\documentclass[twocolumn]{aastex63}%
\usepackage{verbatim} 

\newcommand{\tsys}{$T_{\rm sys}$}
\newcommand{\lsim}{~\rlap{$<$}{\lower 1.0ex\hbox{$\sim$}}}
\newcommand{\gsim}{~\rlap{$>$}{\lower 1.0ex\hbox{$\sim$}}}

\newcommand{\pas}{$\rlap{.}^{\prime\prime}$}

\newcommand{\kms}{km~s$^{-1}$}
\definecolor{darkgreen}{rgb}{0,.3,0}

\shorttitle{Phasing ALMA at 345 GHz}
\shortauthors{Crew, Goddi, Matthews, et al.}

\begin{document}

\title{A Characterization of the ALMA Phasing System at 345 GHz}

\correspondingauthor{Ciriaco Goddi}
\email{cgoddi@usp.br}

\author[0000-0002-2079-3189]{G. B. Crew}
\affiliation{Massachusetts Institute of Technology Haystack Observatory,
  99 Millstone Road, Westford, MA 01886 USA}

\author[0000-0002-2542-7743]{C. Goddi} 
\affil{
Universidade de São Paulo, Instituto de Astronomia, Geofísica e Ciências Atmosféricas, Departamento de Astronomia, São Paulo, SP 05508-090, Brazil}
\affiliation{Dipartimento di Fisica, Universit\'a degli Studi di Cagliari, SP Monserrato-Sestu km 0.7, I-09042 Monserrato,  Italy}
\affiliation{INAF - Osservatorio Astronomico di Cagliari, via della Scienza 5, I-09047 Selargius (CA), Italy}
\affil{
INFN, Sezione di Cagliari, Cittadella Univ., I-09042 Monserrato (CA), Italy}

\author[0000-0002-3728-8082]{L. D. Matthews} 
\affiliation{Massachusetts Institute of Technology Haystack Observatory,
99 Millstone Road, Westford, MA 01886 USA}

\author[0000-0000-0000-0000]{H. Rottmann}
\affiliation{\protect{Max Planck Institut f\"ur Radioastronomie},
Auf dem H\"ugel 69, 53121 Bonn, Germany} 

\author[0000-0000-0000-0000]{A. Saez}
\affiliation{ALMA Observatory,
Av. Alonso de C\'ordova 3107, Vitacura,
\protect{Regi\'on Metropolitana}, Chile} 

\author[0000-0003-3708-9611]{I. Mart\'\i-Vidal}
\affiliation{Departament d'Astronomia i Astrof\'{\i}sica, Universitat de Val\`encia, C. Dr. Moliner 50, E-46100 Burjassot, Val\`encia, Spain}
\affiliation{Observatori Astronòmic, Universitat de Val\`encia, C. Catedr\'atico Jos\'e Beltr\'an 2, E-46980 Paterna, Val\`encia, Spain}

\begin{abstract}
The development of the Atacama Large Millimeter/submillimeter Array
(ALMA) phasing system (APS) has allowed ALMA to function as an
extraordinarily sensitive station for very long baseline
interferometry (VLBI)
at frequencies of up to 230~GHz ($\lambda\approx$1.3~mm). 
Efforts are now underway to extend use of the APS
to 345~GHz ($\lambda\approx$0.87~mm). Here we report a
characterization of APS performance at 345~GHz based on
a series of tests carried out between 2015--2021, including 
a successful global VLBI test campaign conducted in 2018 October
in collaboration with the Event Horizon Telescope (EHT).
\end{abstract}
\keywords{instrumentation: interferometers – methods: observational - techniques: high angular resolution}

\section{Introduction}
In addition to operating as a connected element interferometer, the Atacama Large
Millimeter/submillimeter Array (ALMA) can function as the equivalent of a single very large aperture antenna if the data
from its individual antennas are phase-corrected and coherently added. 
The development of a phased-array capability for ALMA \citep{Doeleman2009,APPPaper} has allowed ALMA to play a
transformational role in the technique of
very long baseline interferometry (VLBI) at millimeter (mm) wavelengths. By boosting the sensitivity of
VLBI baselines in previously existing arrays operating at 230~GHz
($\lambda\approx$1.3~mm) by up to an order of
magnitude,  phased ALMA was crucial to the 
achievement of the first horizon scale images of the supermassive black hole at the center of the M87 Galaxy, M87*  \citep{EHTC2019_1,EHTC2019_2,EHTC2019_3,EHTC2019_4,EHTC2019_5,EHTC2019_6} and the one at the center of our own Milky Way, Sgr A* \citep{EHTC2022_1,EHTC2022_2,EHTC2022_3,EHTC2022_4,EHTC2022_5,EHTC2022_6}.
At 86~GHz ($\lambda\approx$3~mm), phased ALMA was also key to achieving
the first 
scatter-corrected images of SgrA*, the supermassive black
hole candidate at the Galactic Center \citep{Issaoun2019} and
the first ALMA detection of pulsed emission from
radio pulsars \citep{Liu2019,Liu2021}. 

The ALMA phasing system (APS) has been offered to the community for VLBI science observations in ALMA Bands 3 ($\lambda\approx$3 mm; $\nu\approx86$~GHz)
and 6 ($\lambda\approx$1.3 mm; $\nu\approx230$~GHz), where the  Global mm-VLBI Array (GMVA) and the Event Horizon Telescope (EHT),
respectively, serve as partner networks. The first science observations that included the APS in this capacity
were conducted in 2017 April as part of ALMA Cycle 4 \citep{QA2Paper}.
To further expand scientific possibilities, there is now growing
motivation to push VLBI techniques to still shorter wavelengths, i.e.,
$\lambda\approx$0.87~mm or $\nu\approx$345~GHz 
\citep[e.g.,][]{Falcke2001, MiyoshiKameno2002,Weintroub2008, Krichbaum2008, Doeleman2009,
EHTC2019_1,EHTC2019_2}. Not only will this enable even higher
angular resolution ($\lesssim20\mu$as for Earth-sized baselines), but it
will help  to improve $uv$ coverage by enabling the combination of 230~GHz and 345~GHz observations (thus enabling higher fidelity imaging), and will minimize the effects of interstellar scattering on achievable image quality. The latter is particularly important for  imaging SgrA*, where interstellar ionized gas along the line-of-sight causes
significant blurring of images at longer radio wavelengths
\citep{JohnsonGwinn2015, Johnson2016, EHT2022}.

A key component of the effort to extend VLBI capabilities into the
sub-mm regime is
the extension of ALMA's phased array capabilities to 345~GHz.
While this had been an envisioned application of the APS since its
conception \citep[e.g.,][]{Doeleman2010,Fish2013}, commissioning of ALMA's phasing
capabilities was initially limited to the 86~GHz and 230~GHz bands
(ALMA Band 3 and 6, respectively) owing to
time constraints and to the limited availability of suitably equipped VLBI partner sites \citep[see][]{APPPaper}.

While the  APS itself is agnostic to observing
frequency, there are practical
considerations that impact the use of the APS at
$\nu\gsim$345~GHz and the optimization of phasing efficiency at higher frequencies. 
One of the most important is
the shorter coherence timescales at higher frequencies
owing to the effects of tropospheric water vapor, 
which become increasingly significant with increasing baseline length. This in turn will impact
choices such as the
maximum baseline length to include in the phased array and whether or
not to apply ``fast'' phasing corrections---derived from water vapor
radiometer (WVR) data at each ALMA antenna---in addition to the nominal
``slow'' phasing corrections derived by the
phasing engine within the (TelCal) telescope calibration software  \citep{APPPaper}.
Additionally, effects such as pointing errors and wind
speeds will have increasingly important impacts on phased array
performance at higher frequency owing to the smaller beam size of the antennas \citep[see, e.g.,][]{Smith2000}.

Here we present a characterization of the performance and phasing
efficiency of the APS at 345~GHz
based on test sessions conducted between 2015 March and 2021 September. The data sets include observations
obtained in 2018 as part of a multi-day global VLBI test campaign 
that produced for the first time VLBI fringes in the 345~GHz band 
(Event Horizon Telescope Collaboration et al., in preparation; hereafter Paper~II).

\section{Observations}

 Testing and characterization of the performance of the APS for use in Band~7 were done using a combination of ALMA standalone tests 
and a global VLBI  campaign.  
These tests are described in detail in the next two subsections.
In the discussions of phased ALMA that follow, we adopt the following nomenclature:
the {\it reference} antenna is a designated antenna relative to which the phasing corrections are computed for all other antennas;
the {\it sum} antenna is a virtual antenna containing 
the phased signals of all of the phased-array antennas summed together;
a {\it comparison} antenna is an ALMA antenna that is participating in the observations, but is not being phased and
is not included in the phased sum. 

\subsection{Initial Testing: ALMA Standalone Observations}
%

Initial testing  of the APS  at 345~GHz began in 2015 and continued throughout  2021 with a series of short ALMA standalone tests. 
These observations were conducted during ALMA Extension and Optimization of Capabilities (EOC) time or Engineering time,
and typically lasted from a few minutes up to 40 minutes.
A summary of these tests is provided in Table~\ref{tab:b7_tests}, including array and weather parameters. 

Selected observing targets comprised bright ($\gsim$1~Jy at 345~GHz)
quasars and other compact extragalactic sources  that are unresolved on intra-ALMA baselines.
Most of the target quasars are routinely observed at ALMA as part of the flux-density monitoring program  with the ALMA Compact Array (ACA).
This program includes measurements, mostly in Band 3 and Band 7, of bright reference sources, referred to as "Grid Sources" \citep{Remijan2019}).  

Data from these standalone APS tests allowed us to demonstrate the feasibility of phased ALMA operations at sub-mm wavelengths.
Although in several instances the test data were taken in conditions that were sub-optimal for Band 7 observing,
such data enable exploration of the impacts of weather conditions on phasing performance and help to establish guidelines
on the parameter space for scientifically useful phasing operations at higher frequencies (e.g. maximum baseline length in the phased array, maximum wind
speed). More details on the analysis of these test datasets are given in 
Section~\ref{aps_perform}.  

\subsection{The 2018 Global VLBI Test Campaign}
%
2018 October marked the first time that  ALMA's 345~GHz phasing capability was tested
over a sustained observing session, as well as during a global VLBI
campaign, with the goal of obtaining 345~GHz VLBI fringes on global baselines (see Paper~II). 
During this campaign, 
APS operations at 345~GHz were characterized during a series of four
 observing windows from October 17--21. 
A total of six ALMA scheduling blocks  were built and executed, including four  blocks in Band~7 (each spanning $\sim$90~min) and  two  blocks in Band~6 (each spanning $\sim$35~min) for comparison purposes. 
A summary of the observations is reported in  Table~\ref{tab:obslog}. 

  \begin{deluxetable*}{cclcccccc}
    \tablenum{1}
\tablecaption{Standalone ALMA Band~7 Phasing Tests}
\label{tab:b7_tests}
\tabletypesize{\tiny}
\tablehead{
\colhead{UTC Start$^a$}     & \colhead{UTC End$^a$}      & \colhead{Archive UID$^{b}$} & \colhead{$N_{\rm phased}$}  & \colhead{Baselines$^c$} &  \colhead{PWV$^d$} & \colhead{Wind Speed$^d$} & \colhead{$\eta_{v}$$^{e}$} & \colhead{Qual.$^{f}$} \\
(YYYY MMM DD/hh:mm:ss.s)   & (YYYY MMM DD/hh:mm:ss.s)    &  &  & (m) & (mm) & (m s$^{-1}$) & & \\
}
\startdata
2015 Mar 30/02:49:26.4 & 2015 Mar 30/02:51:46.9  & \texttt{uid\_\_\_A002\_X9cdda2\_X42c}      & 9  & 15--193   & 0.53$\pm$0.03 & 8.5$\pm$2.5 & 0.86 & 0.91\\
2015 Aug 02/14:37:58.7 & 2015 Aug 02/14:45:01.4  & \texttt{uid\_\_\_A002\_Xa73e10\_X28dc}     & 35 & 15--1492  & 0.50$\pm$0.04 & 7.5$\pm$4.5 & 0.46 & 0.94\\
2016 Jul 10/08:51:25.1 & 2016 Jul 10/09:38:54.0  & \texttt{uid\_\_\_A002\_Xb53e10\_Xa7a}      & 9  & 19-396   &  2.0$\pm$0.5  &   7$\pm$5 &  0.15 & 0.66\\
2017 Jan 30/21:47:33.5 & 2017 Jan 30/21:51:58.4  & \texttt{uid\_\_\_A002\_Xbd3836\_X4ba}      & 37 & 15--260   &  5.0$\pm$1.5  &  12$\pm$5 & 0.07 & 0.36\\
2017 Jan 30/21:56:56.8 & 2017 Jan 30/22:08:59.2  & \texttt{uid\_\_\_A002\_Xbd3836\_X579}      & 37 & 15--260   &  5.0$\pm$1.5  &  13$\pm$6 & 0.10 & 0.50 \\
2017 Jan 30/22:18:57.0 & 2017 Jan 30/22:34:29.2  & \texttt{uid\_\_\_A002\_Xbd3836\_X739}      & 37 & 15--260  &  4.5$\pm$1.2  &  14$\pm$6 &  0.16 & 0.66\\
2017 Jan 30/22:39:27.4 & 2017 Jan 30/22:51:29.2  & \texttt{uid\_\_\_A002\_Xbd3836\_X87c}      & 37 & 15--260   &  4.5$\pm$1.5  &  13$\pm$4 & 0.11 &0.55\\
2017 Feb 01/03:19:27.0 & 2017 Feb 01/04:02:59.7  & \texttt{uid\_\_\_A002\_Xbd3836\_X4363}$^g$ & 41 & 15--331   &  1.6$\pm$0.6  &   9$\pm$5 & 0.75 & 0.93\\
2019 Mar 08/04:12:43.8 & 2019 Mar 08/04:19:14.4  & \texttt{uid\_\_\_A002\_Xd9435e\_X2859}     & 45 & 15--314   &  1.0$\pm$0.1 & 3.5$\pm$3.5 & 0.74 &0.95\\
2021 Mar 23/23:13:24.1 & 2021 Mar 23/23:21:11.3  & \texttt{uid\_\_\_A002\_Xea64a8\_X321}      & 33 & 15--1232  & 2.7$\pm$0.15  &  10$\pm$4 & 0.49 & 0.88\\  
2021 Mar 24/21:37:20.6 & 2021 Mar 24/21:44:11.3  & \texttt{uid\_\_\_A002\_Xea6cf9\_X1c1}      & 33 & 15--1214   & 2.15$\pm$0.25 &  12$\pm$5 & 0.21 &0.84\\  
2021 Mar 25/01:07:24.7 & 2021 Mar 25/01:15:11.3  & \texttt{uid\_\_\_A002\_Xea6cf9\_Xb5a}      & 35 & 15--1231   & 2.55$\pm$0.15 &   4$\pm$3 & 0.38 &0.90 \\  
2021 Mar 25/19:12:42.6 & 2021 Mar 25/19:20:11.3  & \texttt{uid\_\_\_A002\_Xea6cf9\_X1d9e}     & 25 & 22--969   & 2.0$\pm$0.2   &  13$\pm$5 & 0.11 & 0.69\\  
2021 Aug 26/19:55:06.2 & 2021 Aug 26/20:02:18.0  & \texttt{uid\_\_\_A002\_Xefb0d3\_X7c2}      & 31 & 92--6855 & 1.2$\pm$0.2   &  12$\pm$8 & 0.21 & 0.72\\ 
2021 Sep 02/19:37:24.1 & 2021 Sep 02/19:45:11.6  & \texttt{uid\_\_\_A002\_Xf02179\_X1ea}      & 25 & 237--6855 & 0.45$\pm$0.15 &  10$\pm$6 & 0.26 & 0.87\\  
2021 Sep 03/02:07:24.7 & 2021 Sep 03/02:15:29.9  & \texttt{uid\_\_\_A002\_Xf02179\_X10a0}     & 29 & 237--6855  & 0.4$\pm$0.1  &   4$\pm$3 & 0.65 & 0.91\\
\enddata

\tablenotetext{a}{%
Start times and end times include observations in APS-mode only
(i.e.  standard ALMA-mode calibration scans are excluded).}
 \tablenotetext{b}{Unique identifier (UID) of the data set in the ALMA Archive.}
\tablenotetext{c}{%
Approximate range of baseline lengths in the phased array
(excluding unphased comparison antennas).}
\tablenotetext{d}{%
Weather data (including precipitable water vapor or PWV and wind speed) reported by meteorological stations on the Chajnantor plateau.  The $\pm$ values refer to the range of values
reported by different stations.}
\tablenotetext{e}{Phasing efficiency, averaged over polarizations and basebands, computed according to Eq.~\ref{eqn:etav}.}
\tablenotetext{f}{Phasing quality, averaged over polarizations and basebands, (see Section~3).}
\tablenotetext{g}{%
  This block also included Band~6 observations. }
\end{deluxetable*}
 \setcounter{table}{1}
\begin{table*}
 
  \caption{Observations Log for 2018 October Band\,7 Test Campaign}
\tiny
\begin{center}
\begin{tabular}{cclcccccc}
\hline\hline                  
\noalign{\smallskip}
UTC Start                & UTC End                    & Archive UID &  $N_{\rm phased}$ & PWV & Wind Speed   &  $\eta_{v}$ & Qual. & Band\\
(YYYY MMM DD/hh:mm:ss.s) & (YYYY MMM DD/hh:mm:ss.s)   &                  &                & (mm) & (m~s$^{-1}$) &       &        &        \\
\noalign{\smallskip}
\hline
\noalign{\smallskip}   
2018 Oct 16/23:42:52.4 & 2018 Oct 17/01:00:26.0 & \texttt{uid\_\_\_A002\_Xd3607d\_X6f14} &  23 & $2.0\pm0.3$  & $9\pm5$ & 0.13 &0.42 & 7 \\ 
2018 Oct 17/01:05:24.6 & 2018 Oct 17/01:40:54.3 & \texttt{uid\_\_\_A002\_Xd3607d\_X70fe} &  23 & $2.5\pm0.5$  & $7\pm3$  &0.10 &0.37 &6 \\ 
2018 Oct 17/09:31:07.1 & 2018 Oct 17/11:02:16.5 & \texttt{uid\_\_\_A002\_Xd36f86\_X24dd} &  25 & $1.8\pm0.8$  & $12\pm4$ & 0.07 &0.28 &  7\\ 
	2018 Oct 18/23:23:08.3 & 2018 Oct 19/00:52:37.4 & \texttt{uid\_\_\_A002\_Xd37ad3\_X7ef1} &  25 & $1.1\pm0.3$  & $6\pm4$  & 0.37 &0.91 & 7 \\ 
2018 Oct 19/00:58:00.0 & 2018 Oct 19/01:32:54.5 & \texttt{uid\_\_\_A002\_Xd37ad3\_X82a2} &  25 & $1.0\pm0.1$  & $4\pm3$ & 0.51 & 0.96 & 6\\ 
2018 Oct 21/09:12:54.4 & 2018 Oct 21/10:59:18.3 & \texttt{uid\_\_\_A002\_Xd395f6\_Xd41f} &  29 & $0.85\pm0.10$ & $3\pm3$ & 0.93 &0.97 & 7 \\ 
\hline
\end{tabular} 
\label{tab:obslog}
\end{center}
\tablecomments{See footnotes to Table~\ref{tab:b7_tests} for an explanation of the columns. The final column indicates the ALMA observing band.}
\end{table*}

\begin{table*}
\caption{VLBI Sources Observed During the 2018 October Band\,7 Test Campaign}            
\label{table:sources}       
\begin{center}
\small
\begin{tabular}{cllcc} 
\hline\hline                  
\noalign{\smallskip}
Source & UTC Start$^{a}$    & UTC End$^{a}$    &  Band & Calibration Intent$^{b}$ \\
       &  (YYYY MMM DD/hh:mm:ss) &  (YYYY MMM DD/hh:mm:ss.s) & & \\
\noalign{\smallskip}
\hline
\noalign{\smallskip}  
CTA102  & 2018 Oct 18/23:43:25 &2018 Oct 18/23:58:05 & 7 & \ldots \\
3C454.3 & 2018 Oct 19/00:06:25 &2018 Oct 19/00:20:15 & 7 & Flux \\
BL Lac  & 2018 Oct 19/00:29:25 &2018 Oct 19/00:43:15 & 7 & \ldots \\
BL Lac  & 2018 Oct 19/01:02:25 &2018 Oct 19/01:23:58 & 6 & \ldots \\

J0423--0120   & 2018 Oct 21/09:21:25 & 2018 Oct 21/09:43:15 & 7 & Bandpass \\
J0510+1800  & 2018 Oct 21/09:52:25 & 2018 Oct 21/10:06:15 & 7 & Polarization \\
J0510+1800  & 2018 Oct 21/10:16:25 & 2018 Oct 21/10:28:41 & 7 & Polarization \\
J0522--3627  & 2018 Oct 21/10:36:25 & 2018 Oct 21/10:59:18 & 7 & \ldots \\ 
\noalign{\smallskip}
\hline    
\end{tabular}
\end{center}
\tablenotetext{a}{Only VLBI targets observed on October 18/19 and 21 are listed. Observations on  October 16/17 did not produce good-quality data owing to poor weather conditions and were not included in the analysis (see Table~\ref{tab:obslog} and Appendix~\ref{app:calibration}). }
\tablenotetext{b}{See Appendix~\ref{app:calibration}  for additional information on the calibration of these data. }
\end{table*}
\begin{table*}
\caption{ALMA Frequency Settings}
\centering  
\small
\begin{tabular}{cccccclc} 
\hline\hline                  
\noalign{\smallskip}
 Band & \multicolumn{4}{c}{ Central Freq. (GHz)} &   Chan. Width & No. Spec.  & Integ. time \\
  ($\lambda$)            &    SPW\,0  &  SPW\,1 &   SPW\,2 &  SPW\,3 &    (MHz) & Chans. &  (s) \\
\noalign{\smallskip}
\hline
\noalign{\smallskip}  
6 (1.3\,mm)      &   213.1       & 215.1        &  227.1       &  229.1       &   7.8125  & 240$^{a}$ & 4.03    \\
7 (0.85\,mm)      &   335.5       & 337.5        &  347.7      &  349.7       &   15.625  & 120 & 4.03    \\
\noalign{\smallskip}
\hline
\end{tabular}
\tablecomments{The SPW designations correspond to those in the calibrated CASA measurement set rather than those in the original raw data files.}
\tablenotetext{a}{For bandpass calibration purposes the  Band~6 scans were rebinned in frequency to 120 channels for consistency with Band~7 (see Appendix~\ref{app:calibration}).}
\label{table:freq_b7}
\end{table*}

\begin{table*}
 \caption{ALMA Source Flux Densities from the 2018 October  Band\,7 Test Campaign}
\small
\begin{center}
\begin{tabular}{lcccccccc}
\hline
Source & $S_0$ (Jy) & $S_1$ (Jy) & $S_2$ (Jy) & $S_3$ (Jy)& $\overline{S}$ (Jy) & $S_{\rm arch}$ (Jy) & Ratio & $\Delta t_{S}$ (days) \\
\hline
\multicolumn{9}{c}{Flux Calibrator = 3C454.3, S$_{\rm 343 GHz}$ = 3.53 Jy, $\alpha$=$-$0.69}\\ 
\hline
(1) & (2) & (3) & (4) & (5) & (6) & (7) & (8) & (9) \\ 
\hline
CTA102 & 1.42 & 1.41 & 1.40 & 1.40 & 1.41 & 1.68  &0.84 &  0 \\ 
BL Lac (Band~7) & 1.06 &   1.06 & 1.05 & 1.05 &   1.06  & 1.11 &0.95 & +71 \\ 
BL Lac (Band~6) & 1.26 &   1.25 & 1.22 & 1.24 & 1.24 & ... &  ... & ... \\ 
J0423-0120 & 2.48 & 2.50 & 2.44 & 2.43 & 2.46 & 2.24 & 1.1 & 0 \\ 
J0510+1800 & 1.23 & 1.24 & 1.22 & 1.21 & 1.22 & 1.28 &  0.95 & 0 \\ 

J0522$-$3627 & 4.91 & 4.94 & 4.87 & 4.84 & 4.89 & 4.32 & 1.13 & 0 \\ 
\hline
\end{tabular}
\end{center}
 \tablecomments{Tabulated flux densities include values measured during the 2018 October Band\,7
 VLBI   test campaign  and values retrieved from the ALMA GS calibrator archive.  
Explanation of columns: (1) source name;
(2)--(5)  flux density in Jy, measured in SPW=0,1,2,3,  respectively (see Table~\ref{table:freq_b7} for their central frequencies)
using CASA's \texttt{fluxscale} task and corrected for \tsys\ (see Appendix~\ref{app:fluxscale});   
(6) flux density in Jy, derived at the mean frequency over the four SPWs (342.6~GHz in Band~7 and 221.1~GHz in Band~6);
(7) expected flux density in Jy  at 343~GHz from the ALMA archive (when an entry is available); 
(8)  ratio between the measured and the archive-predicted flux density;
(9) time difference in days between the APS test observation and the archival entry. 
}
\label{tab:fluxes} 
\end{table*}


\subsubsection{Observed Targets}

As for the standalone phasing tests in Table~\ref{tab:b7_tests}, selected observing targets comprised bright quasars and other compact extragalactic sources. 
An effort was made to select sources that would be point-like on the angular scales sampled by intra-ALMA baselines
(to maximize phasing efficiency), while still having sufficient correlated flux density to allow high
signal-to-noise ratio (SNR) fringe detections with short integration times on VLBI baselines. This is particularly important in Band~7, where coherence
timescales are expected to be only $\sim$10~s \citep[e.g.,][]{Doel2011}.

A list of observed sources and their calibration intent is given in Table~\ref{table:sources}; only VLBI targets observed on October 18/19 and 21 are listed (targets observed on previous days were not included in the analysis owing to poor weather conditions; see Sect.~\ref{weather2018} and Appendix~\ref{app:calibration}). 
In most cases,  recent flux density measurements at 230 and 345~GHz
were available from the ALMA Calibrator Source Catalogue.\footnote{https://almascience.eso.org/sc/}
These can be used for cross-comparison and validation of the APS performance and calibration in Band~7 (see Section~\ref{app:fluxcomp}). 

\bigskip
\subsubsection{Observational Setup\protect\label{setup}}
\begin{figure}
\hspace{-5mm}
\includegraphics[width=0.5\textwidth]{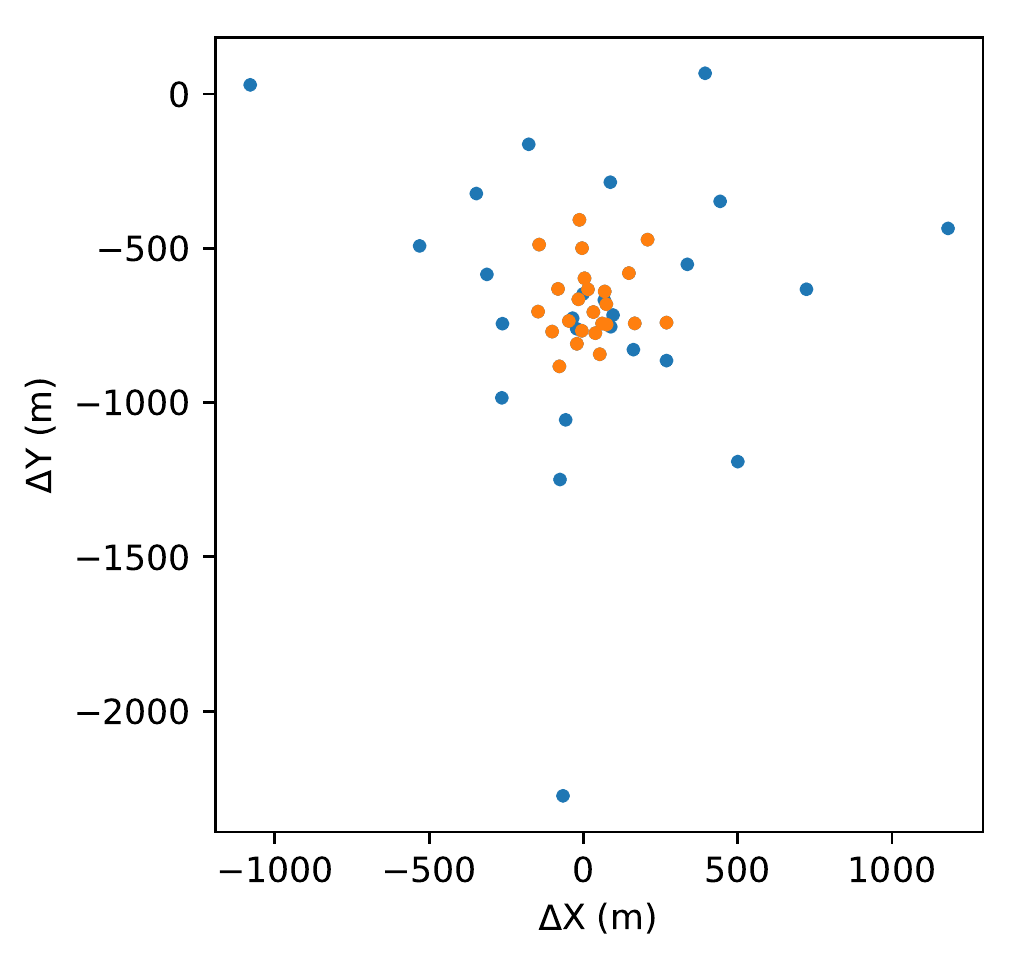} 
\caption{ALMA antenna locations for  the phased array (orange points) and the unphased comparison antennas (blue points) during
  the Band~7 phasing tests in 2018 October.  
Positions are plotted with positive values of $X$ toward local east and positive values of $Y$ toward local north.
}
\label{fig:plotants}
\end{figure}

During the 2018 October test campaign the ALMA antennas were in transition from configuration C43-6 (maximum baseline 2500~m) to C43-5 (maximum baseline 1400~m). 
There were  23--29 12~m antennas included in the ALMA phased array (depending on the session), with a phasing radius  limited to 300~m. 
An additional 16--22 outlying antennas  (with maximum baselines between 1400~m and 2500~m, depending on the day) were withheld for comparison purposes.
Antenna locations are plotted in Figure~\ref{fig:plotants}. 
The observing array was more extended than during previous  science observations in VLBI mode, where only antennas within a radius of 180~m were  phased  \citep[e.g.,][]{QA2Paper}. 
Only 12~m antennas were included in the array;
ALMA's 7~m CM antennas can also be used with the APS, but
are typically excluded from phased-array operations for a variety of practical reasons. 

The spectral setup included four spectral windows (SPWs), each with a bandwidth of 1875~MHz, that were processed by
the  ALMA Baseline Correlator.
Two SPWs were in the lower sideband and two in the upper sideband. In Band~7 the data outputted by the ALMA Baseline Correlator
were averaged in frequency to produce 120 channels per SPW  (corresponding to a channel spacing of 15.625~MHz). 
In Band~6 there were 240 channels per SPW (resulting in a channel spacing of 7.8125~MHz).
The frequency setup is summarized in Table~\ref{table:freq_b7}. 
The spectral data from the ALMA Baseline Correlator are available with a time resolution of  4.032~s. 
In parallel, VLBI recordings of all four basebands, each corresponding to one of the four SPWs, were recorded in dual
linear polarizations, thus exercising the full 64~Gb s$^{-1}$ VLBI recording capability at ALMA (see Paper~II).

\subsubsection{Weather Conditions}
\label{weather2018}
During the 2018 October campaign there were no significant technical issues at ALMA and all aspects of its phasing and VLBI systems appeared
to be performing nominally. However, with the exception of the last observing night where conditions were exceptional, weather conditions
did not meet the usual requirements for Band~7 observing at ALMA. As discussed below, these weather issues 
significantly impacted the quality of the phased array data, but at the same time provided valuable insights into the range
of weather conditions where scientifically useful phased array operations in Band~7 are likely to be possible.

In the first two sessions of the campaign there was
high and variable precipitable water vapor (PWV$\gtrsim$2~mm) and high wind speeds 
 ($\gtrsim$10--15~m s$^{-1}$; see Table~\ref{tab:obslog} and Figure~\ref{fig:weather} in Appendix~\ref{app:weather}). 
 These factors led to unstable atmosphere conditions over timescales of a
few seconds, which  compared unfavorably with the $\sim$~18~second
loop time of the ``slow'' APS phasing solutions \citep[see][]{APPPaper,QA2Paper}. 
The phasing efficiency, $\eta_{v}$ (see Appendix C and Eq.~C2), was consequently rather low:  
typical values  reported by TelCal during the
observations ranged from 5\%--20\%, and
for portions of the session ALMA appeared to be effectively 
unphased (see Section~\ref{aps_perform2018}). 
 
At the onset of the third session (on the night of October 18/19),
atmospheric stability was significantly improved compared with the
previous two VLBI sessions. 
Finally, during the fourth and final VLBI session (corresponding to the fifth day of the VLBI observing window), 
weather at ALMA was excellent, with  precipitable water vapor (PWV) $\sim$0.8~mm and wind speeds of only a few~m s$^{-1}$ (see Table~\ref{tab:obslog} and  the bottom panel of Figure~\ref{fig:weather} in Appendix~\ref{app:weather}). 
Throughout the latter session, the estimated phasing efficiency
reported by TelCal was consistently $>$90\%, and frequently above 95\% 
 (see Section~\ref{aps_perform2018}).

\section{APS Performance Metrics}
\label{aps_perform}

One effective way  to visualize the APS performance is through plots of the phasing efficiency, $\eta_{v}$
(see Eq. 3 in Appendix~\ref{app:efficiency}).
For monitoring purposes,
this quantity is computed by TelCal for a designated comparison antenna and can
be extracted from the archival science data model (ASDM) file metadata \citep[see also][]{QA2Paper}.
Some details about how and why this is done  are discussed in
Appendix~\ref{app:efficiency}.

An additional figure-of-merit computed by TelCal is a `quality' metric, which is a figure of merit intended to provide a sense of the goodness-of-fit of the phasing calculations.  It is constructed from the RMS of the phase residuals, $\sigma_{\rm RMS}$, and assumes values ranging between 0 (no solution) to 1 (excellent fit).  Noting that in the case of pure noise this value is  $\sigma_{\rm RMS, max}~ = ~ \pi / \sqrt{3}$ \citep{TMS2017},  a quality metric for each fit may be constructed as $q ~ = ~ ( \sigma_{\rm RMS, max} - \sigma_{\rm RMS} ) /   \sigma_{\rm RMS, max}$. 

\begin{figure*}[ht!]
\hspace{-5mm}
\includegraphics[width=\textwidth]{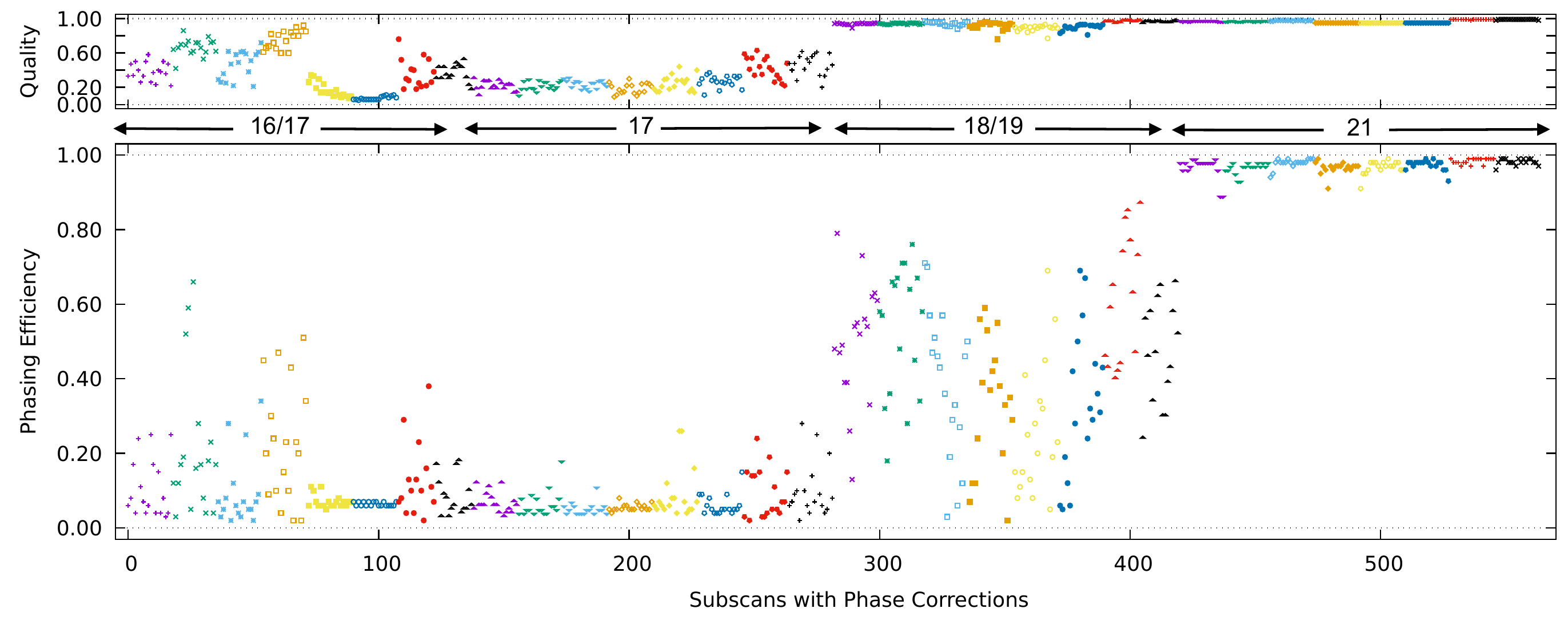} \hspace{-10mm}
\caption{ 
Phasing {\it efficiency} ($\eta_v$, as defined in
Eq.~\ref{eqn:etav}; lower panel) and {\it quality} (top panel) during the phased-array test in
Band 7 as part of the 2018 October VLBI campaign. Calendar dates and their respective time ranges are indicated by the arrows between
the two panels.
All scans are plotted and colored by science target for each day.  
The phasing ``quality''  is a ``goodness-of-fit" parameter derived from the fitting process, which  is scaled so that unity corresponds to perfect phasing.
 The phasing efficiency
ranges from 0 (totally un-phased) to 1 (perfect phasing).
The large departures of efficiency below 0.8 (on days from 16 to 19) correspond
to poor atmospheric conditions (see Section~\ref{aps_perform2018}).
}
\label{fig:pheff_2018}
\end{figure*}

Both are plotted in Figure~\ref{fig:pheff_2018} for the 2018 October data, arranged by correlator subscan
(i.e. the interval of the phasing solution), with the data for each time interval averaged over all basebands.
This plot shows that in the 2018 October test, the APS has achieved
90\% phasing efficiency on October 21, which was the goal specified in
the original operational requirements  \citep{APPPaper}.
On October 18/19
the phasing efficiency was rather modest, 
which is ascribable to poor atmospheric conditions
(see Section~\ref{aps_perform2018}), but still of good {\it quality}. This situation occurs in cases where the phase-solving
algorithm is able to find good-quality solutions, but atmospheric conditions are varying sufficiently rapidly
that the 16-second time delay in the application of these
``slow'' phasing corrections to the data renders them ``stale'' and no longer optimal.
The contrast between these days and the first two
makes clear that the quality metric is a useful discriminator
between different causes of low-phasing efficiency. During the first two observing sessions where wind
speeds were high (October 16/17 and
October 17) $\eta_{v}$ is low and the quality metric is $<<1$. On the other hand, for October 18/19, the
quality metric is consistently $\sim$1 despite periods of low $\eta_{v}$, suggesting that rapid variations in
water vapor rather than wind effects were the dominant source of efficiency loss.

\begin{figure*}[htbp]
\center{
\includegraphics[width=0.45\textwidth,angle=0]{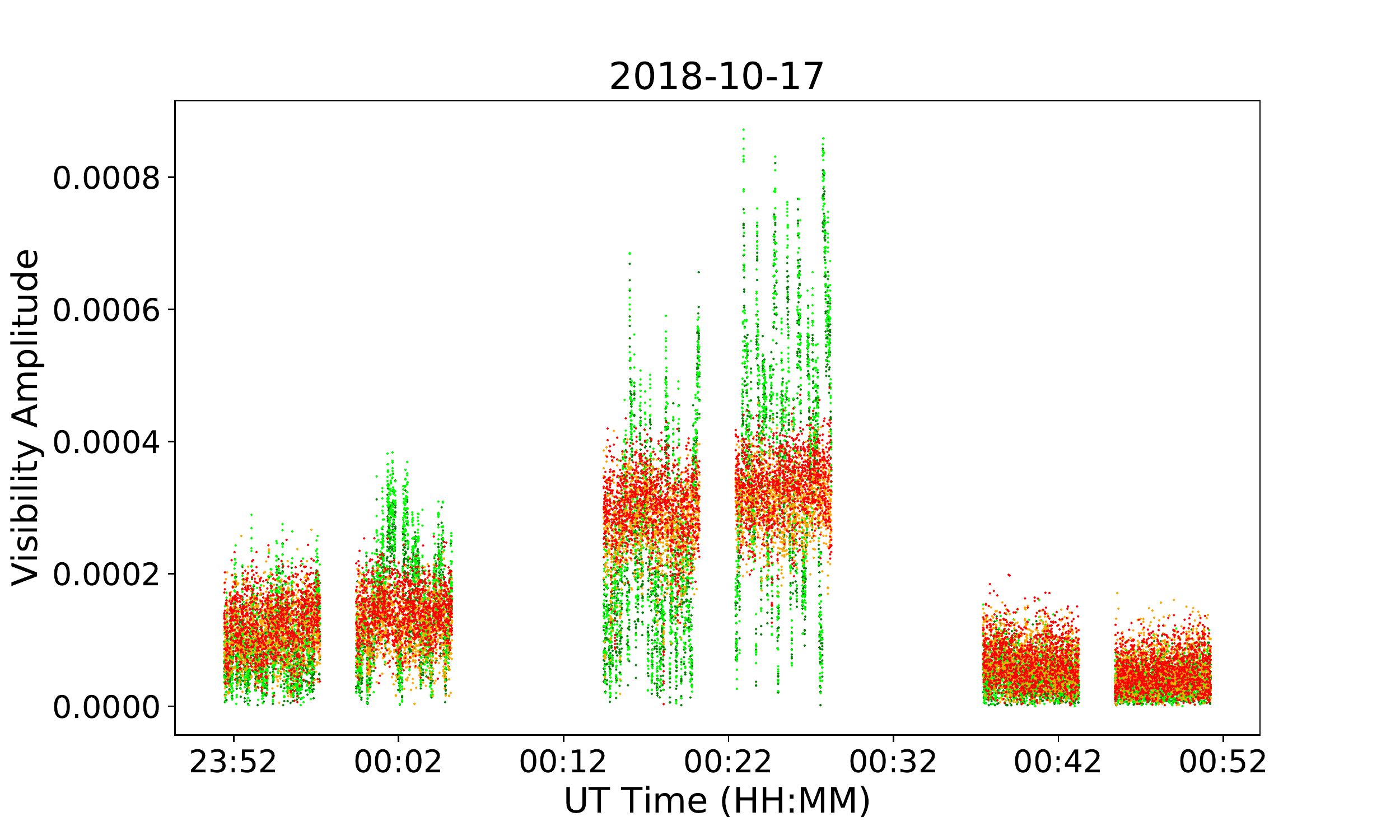}
\includegraphics[width=0.45\textwidth,angle=0]{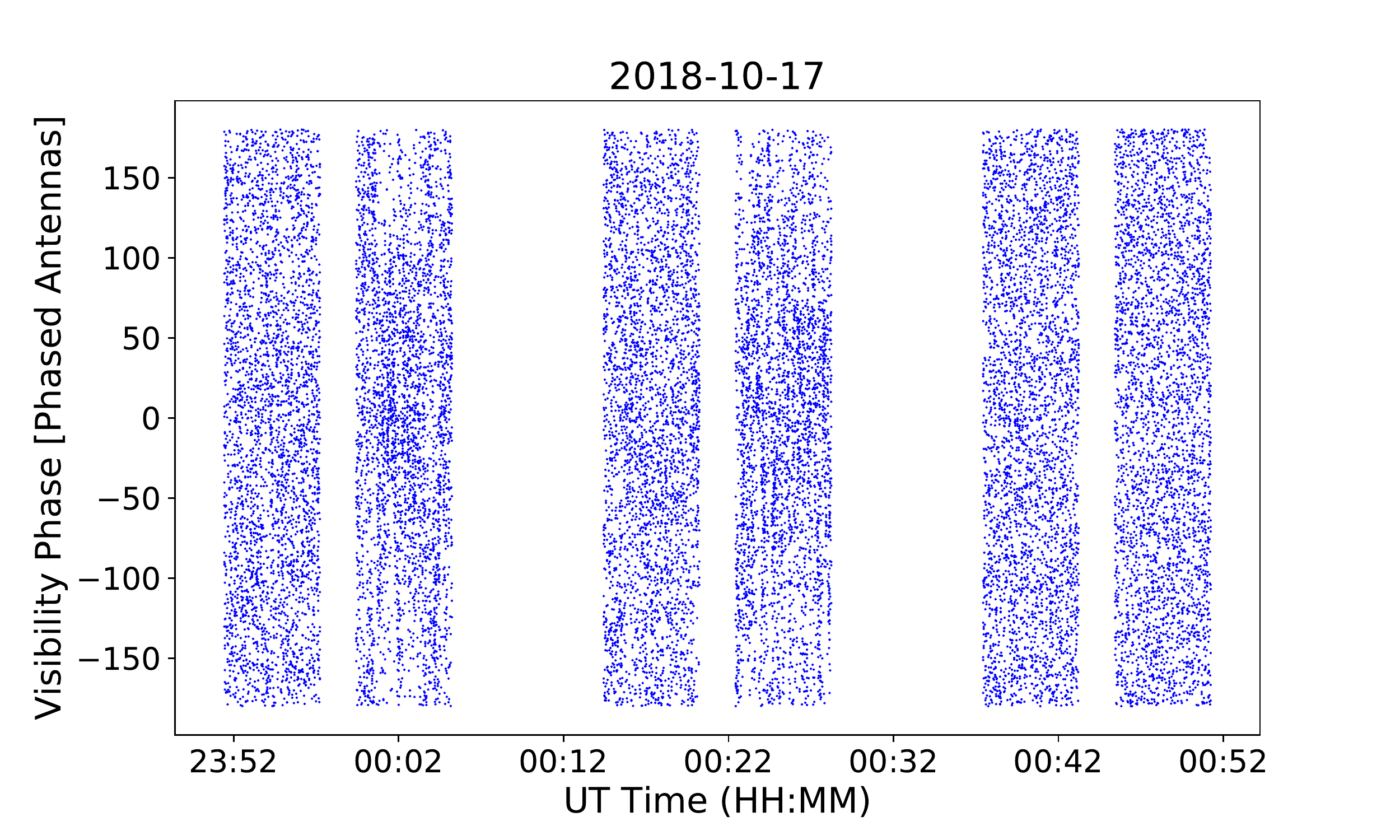}
\includegraphics[width=0.45\textwidth,angle=0]{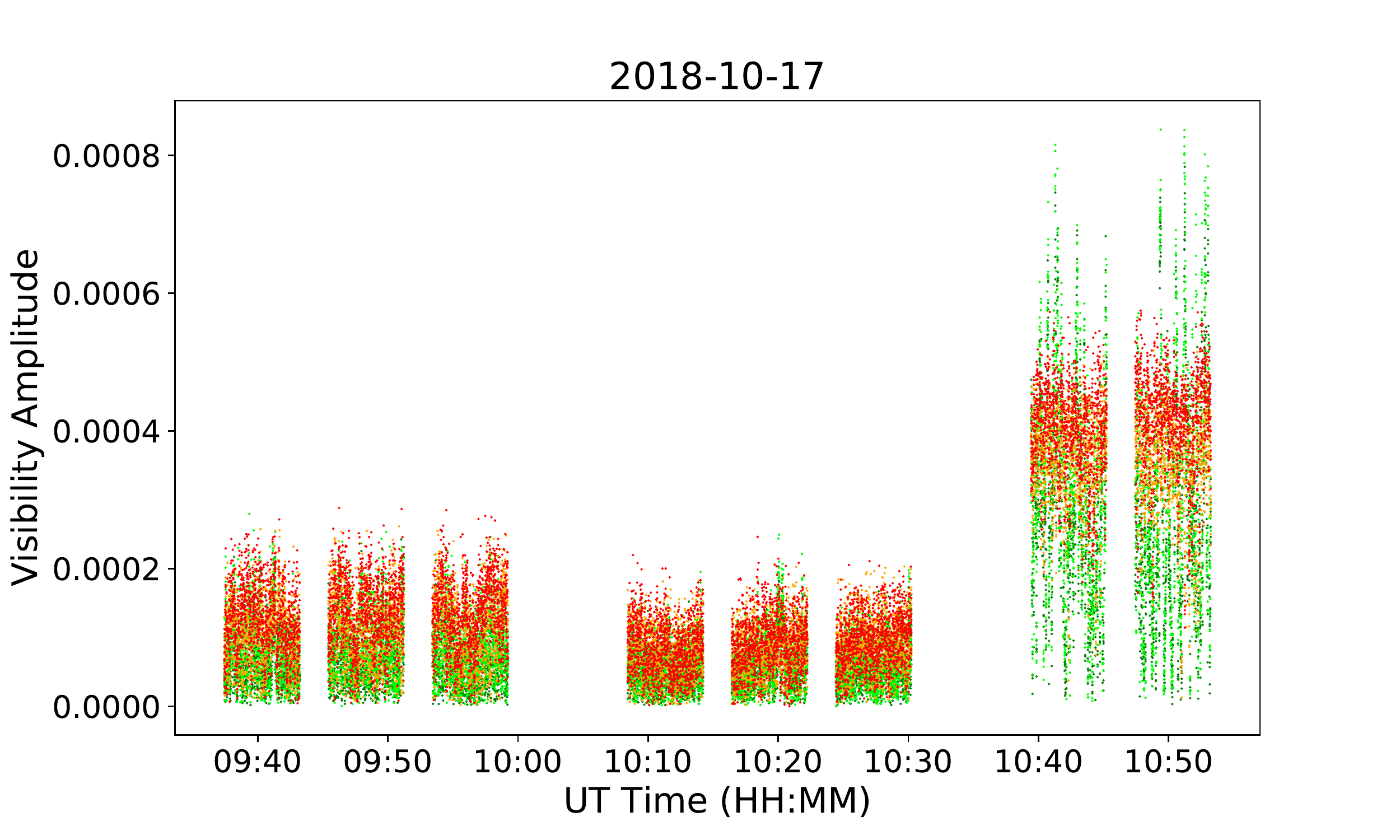}
\includegraphics[width=0.45\textwidth,angle=0]{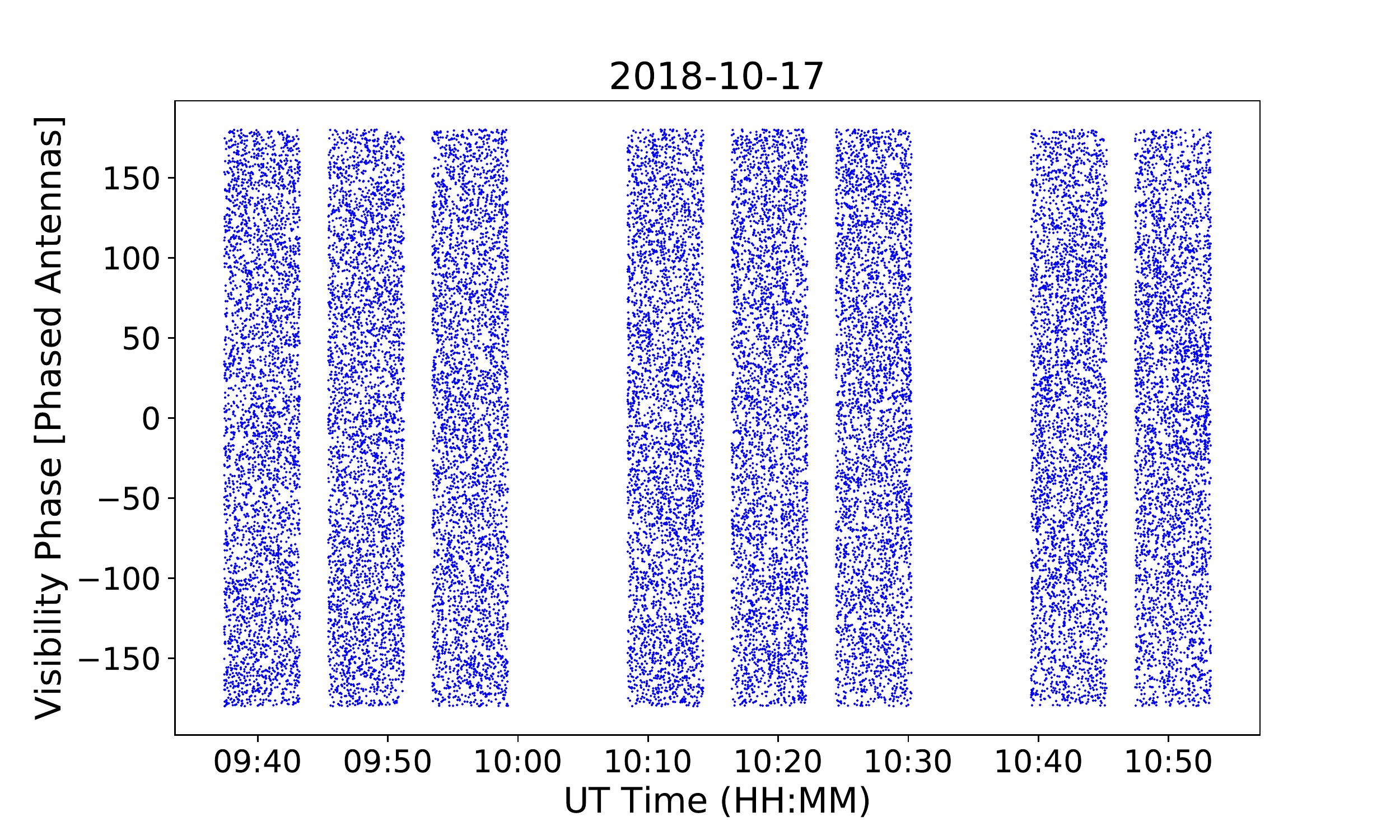}
\includegraphics[width=0.45\textwidth,angle=0]{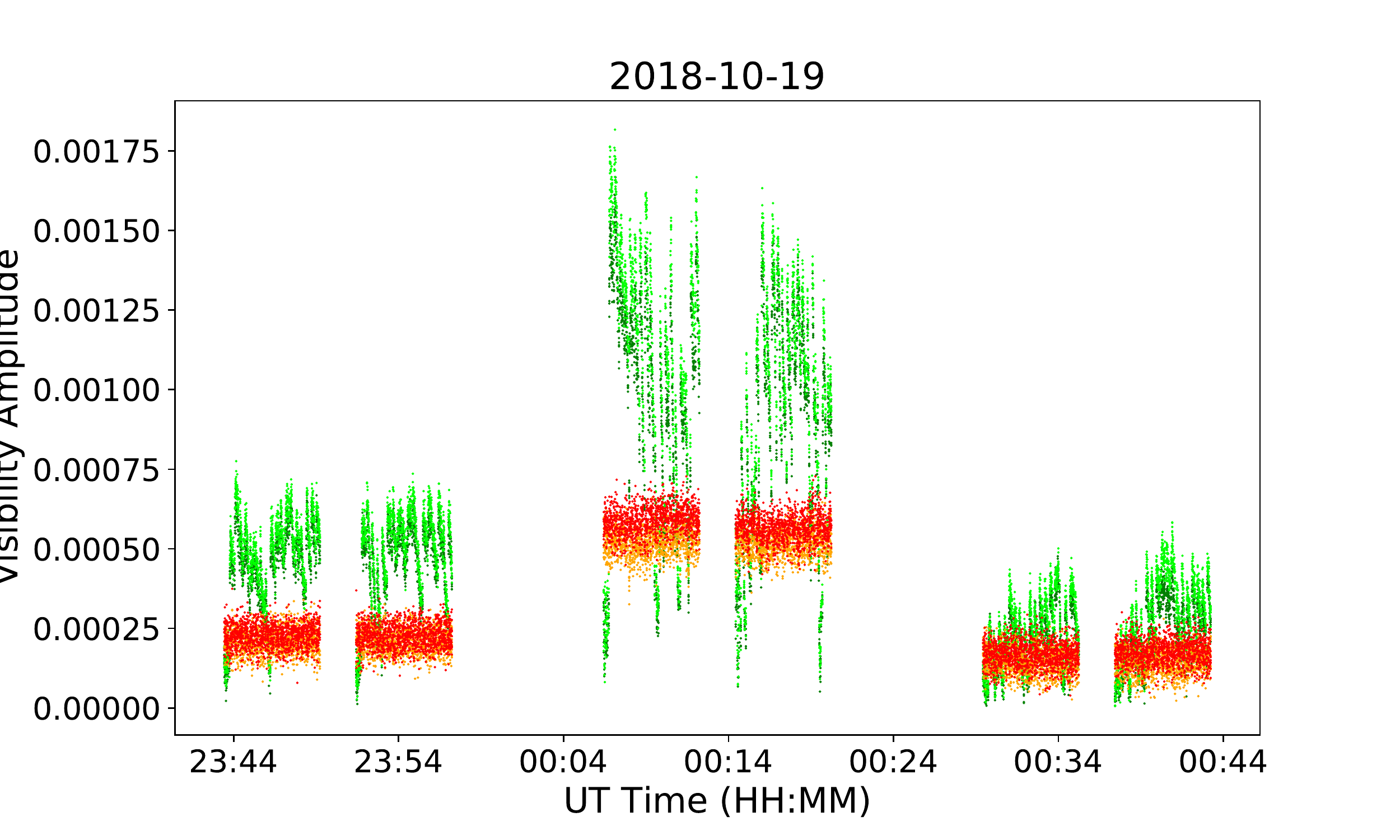}
\includegraphics[width=0.45\textwidth,angle=0]{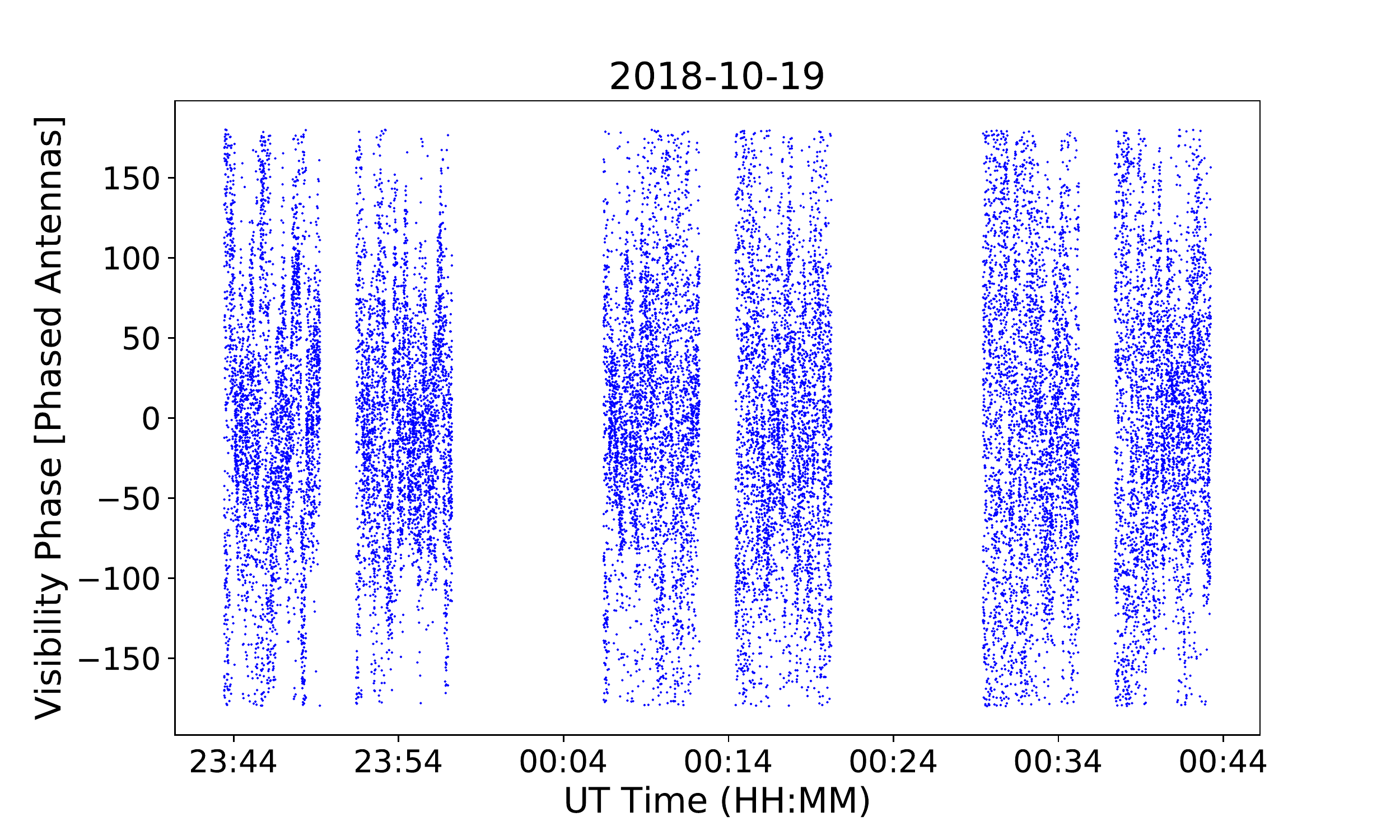}
\includegraphics[width=0.45\textwidth,angle=0]{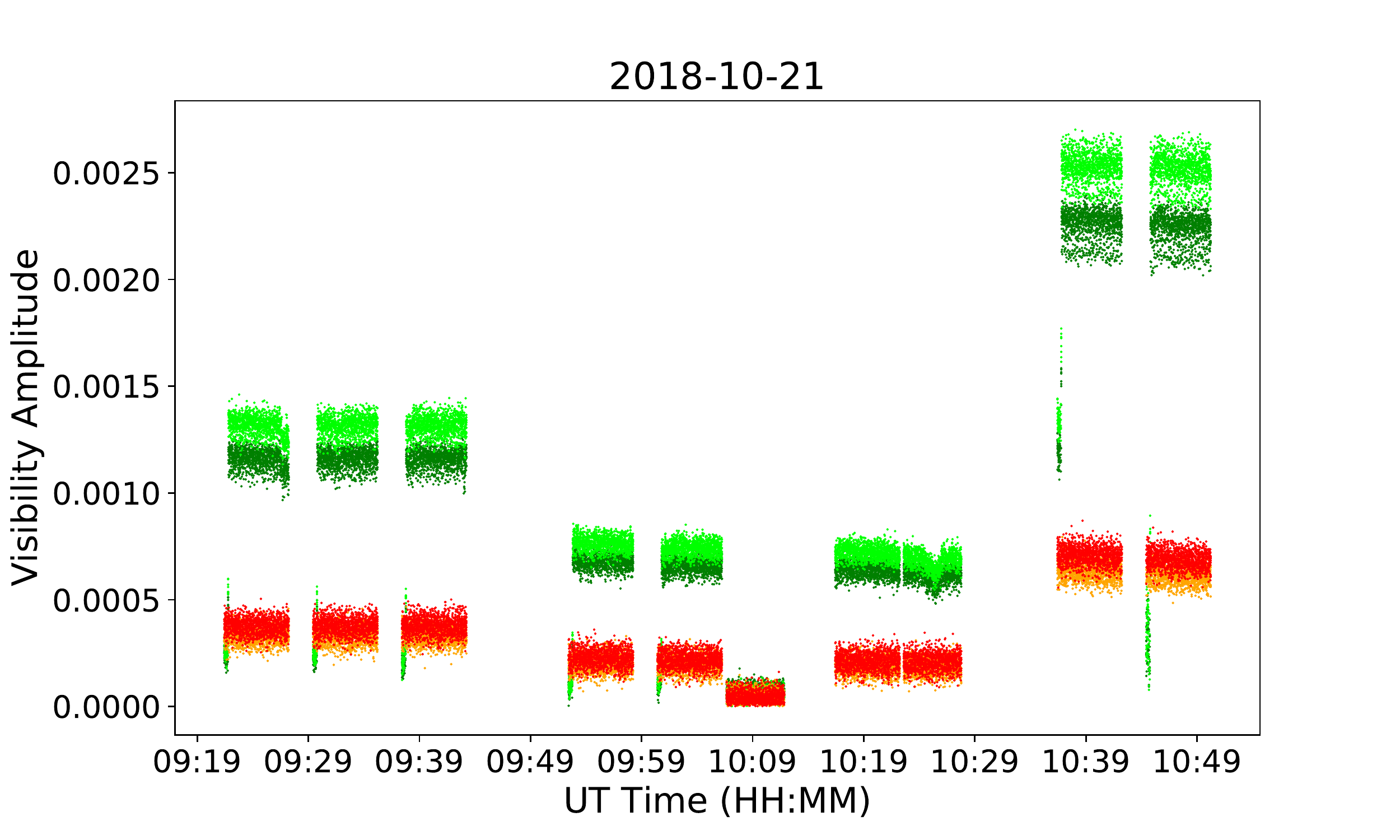}
\includegraphics[width=0.45\textwidth,angle=0]{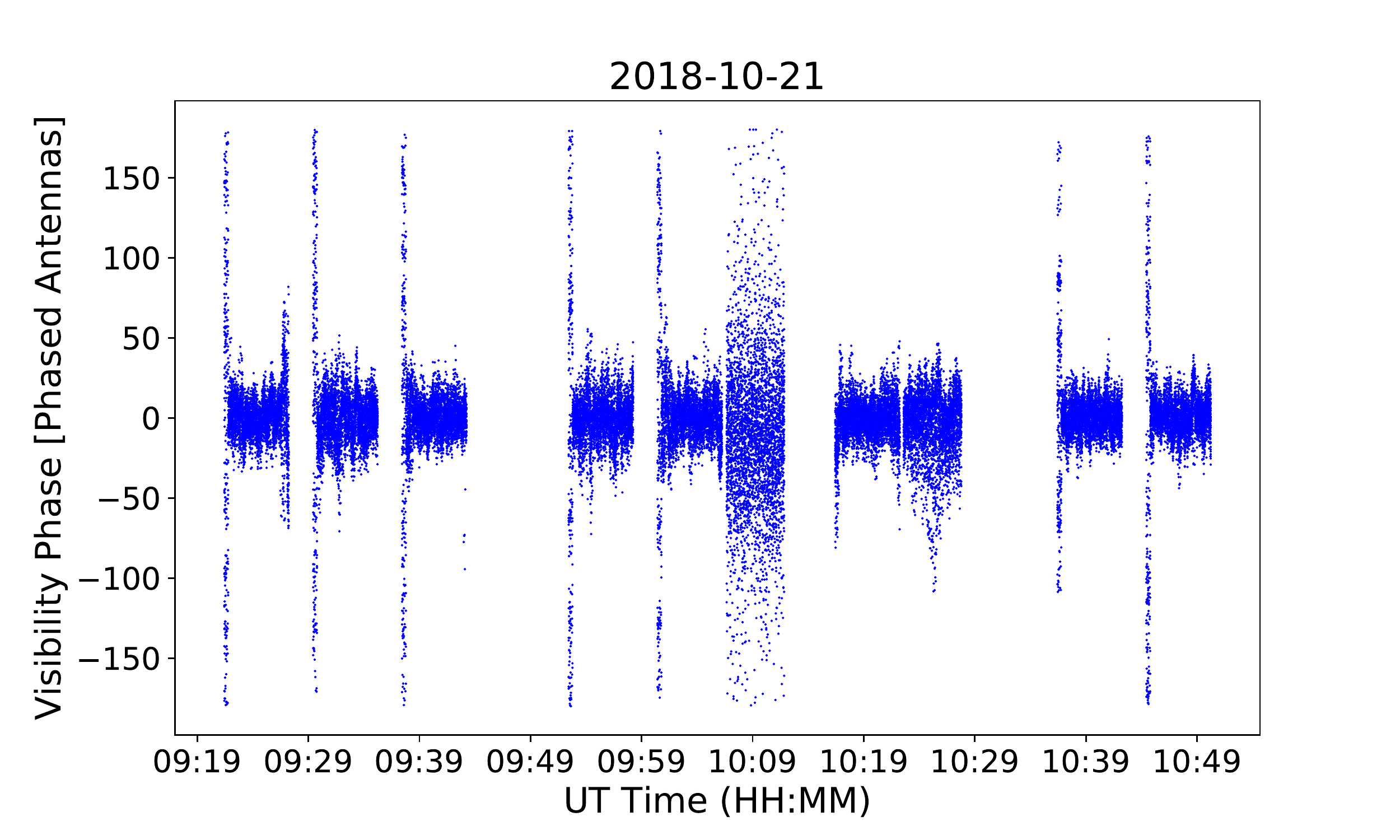}
}
\caption{\footnotesize 
Illustration of the performance of the APS during  the  345~GHz (Band~7) VLBI experiment on 2018 October 16/17, 17, 18/19, 21 (from top to bottom). 
Left panels: 
correlated amplitude is plotted as a function of time on two sets of baselines: 
(1) the phasing reference antenna with two  unphased comparison antennas (red points); 
(2) the phased sum antenna with the  same comparison antennas (green points).  
 Right panels:
phase versus time is plotted on baselines between the ALMA reference antenna and
the other  phased ALMA antennas (blue points). 
The uncorrelated phases  during the first few integrations of each observing block (lower-right panel), are due to the fact that phases are still being adjusted (i.e. the array is unphased; see \S3).
In both columns, data from a single correlator quadrant (baseband 3, corresponding to SPW = 2 in Table~\ref{table:freq_b7}) and a single polarization (XX) are shown. 
Data in the other SPWs and polarization YY show similar behaviors.}
\label{fig:amp+pha_vs_time}
\vspace{-0.2in}
\end{figure*}

An alternative way to display the APS performance during observations is to plot directly amplitudes and phases of
the interferometric visibilities including the sum and the reference antennas, on baselines to one or more comparison antennas. 
In Figure~\ref{fig:amp+pha_vs_time} (left panels) we show a
comparison of the correlated amplitude as
a function of time for the 2018 October data on: 
(i) baselines between the phasing reference
antenna and each of two different unphased comparison antennas; and 
(ii)  baselines between the phased sum antenna 
and the same comparison antennas. 
For an optimally phased array, the correlated amplitude of (ii)
should ideally improve by a factor of
$\sim\sqrt{N_{\rm phased}}$ when phasing is active.
Plots of (i) are useful in making this assessment. We see that for October 16/17 and October 17, the correlated
amplitude for a baseline with the phased sum is comparable to that on a baseline with a single antenna,
implying that the array is effectively unphased as a result of the poor weather conditions. On October
19, the phased sum shows a significant improvement in correlated amplitude (green points), though the data are noisy
and the improvement does not match the ideal $\sqrt{N_{\rm phased}}$ scaling. Finally on October 21 nearly
ideal phasing performance is seen.

In Figure~\ref{fig:amp+pha_vs_time} (right panels)
we show phase versus time on baselines which are part of the phased sum for the same four data sets.
On October 21 the phases in individual scans show a low RMS dispersion and are tightly clustered near 0, 
except for a few seconds near the start-up of each scan, when the phases are still being adjusted\footnote{The APS scans are started two subscans (18-s each) prior to the start of the VLBI recording to allow the APS to calculate and apply the phase adjustments. The ``phase-up'' occurs during the first 22~s of each scan (where typical scan lengths are several minutes); these intervals are routinely flagged to prevent using poorly phased data. See \citet{APPPaper,QA2Paper} for details.}.
(Note that the scan
at 10:06~UT was passively\footnote{The APS supports a “passive” phasing mode where a bright calibrator  located within a few degrees of the fainter target is used to phase up the array \citep{APPPaper}.} rather than actively phased, hence the higher noise level).
On October 19 some hints of phase coherence are seen, but the data are much noisier. Finally on October 16/17 and 17 the phases appear nearly random, consistent with an unphased array.

The performance of the APS in Band 7 relative to Band 6 is discussed in Section~\ref{aps_B6-B7}. Although the weather conditions were sub-optimal during the test observations, we nonetheless see comparable RMS phase fluctuations in the two bands, suggesting that there is no systematic degradation in phasing performance in Band 7 compared to Band 6.

\section{Variables Impacting Phased Array Performance in Band~7}
\label{aps_perform2018}

Because the data recorded during the Band~7 APS tests presented in Section~2 were acquired using different arrays of ALMA antennas
and across a range of  weather conditions, this allows us to begin to
investigate how different array parameters, weather conditions, and other variables affect APS performance in Band~7.
These effects are discussed in the following subsections.

\subsection{Impact of Weather Conditions}
The four dates of the 2018 October campaign were conducted with a phased array with a fixed radius (300-m)
and all included a similar number of phased antennas ($\sim$25). However, weather conditions varied on the different days.
We can therefore use the 2018 data to assess how weather conditions affect the APS performance. 

In Section~\ref{weather2018} we point out that observations on October 16 and 17 were plagued by variable PWV and high winds (see also Appendix~\ref{app:weather}).
Under these conditions, it was not possible to successfully phase the array, with the phased sum antenna performing no better than a
single antenna.
Under conditions of moderate wind but still relatively high PWV fluctuations (October 18/19), the array could be successfully phased, but the variable conditions reduced the duration of the validity of the solution, resulting in a lower than
expected improvement in the correlated amplitude and a phasing efficiency of only 20\%--80\%.
Finally, under low-wind and low-PWV conditions (October 21), the correlated amplitude of the phased antennas reaches the expected 
square root of the  number of phased dishes  (29 in this case), once the known efficiency losses are considered (Appendix~\ref{app:efficiency}), indicating  an optimally phased array
  (overall phasing efficiency, $\eta_{v}$ of $\gtrsim$90\%;  Figure~\ref{fig:pheff_2018}).

  In addition to the standard ``slow'' phasing corrections computed by TelCal, the APS has an option to apply in real time ``fast'' phasing corrections
  (with $\sim$1.6~s cadence)\footnote{The underlying measurements
    are currently made every 1.152~s, and
it takes an additional $\sim$0.5~s to apply the correction.}  computed
  from the WVR data available at each antenna \citep{APPPaper}. These real-time fast corrections were not used
  during the 2018 October observations, but we have investigated the expected impact of such corrections
  by applying WVR-derived corrections 
  off-line  to the individual elements of the phased array, via the {\tt{wvrgcal}} task in CASA.
Since the phased sum is formed in real time, it is not possible to use the fast corrections in post-processing to improve the SNR,
    as they apply to the individual antennas used to form the sum.
  Nonetheless we can gauge the expected impact by computing the improvement in the phase
  coherence of the individual baselines in the phased array. In the case that
rapid phase fluctuations (such as those observed on October 16/17) result from atmospheric water vapor varying on
timescales more rapid than the computed ``slow'' phasing solutions, we should expect to see an improvement in the phase coherence on individual
baselines after application of the WVR corrections. 
For example, the analysis of
ALMA data with $v_{\rm wind}\lsim$10~\kms\ by \citet[]{Maud2017}  and \citet[]{Matsushita2017}
  suggest that such corrections are
typically helpful in reducing phase fluctuations and coherence loss
for baselines $<$500~m when PWV$>$1.
We find, however, that for the October 16/17 data the
WVR-based corrections do not improve the overall phase  coherence; instead they appear to add additional noise. 
This suggests that  the rapid phase fluctuations, which lead to a systematic degradation of phasing efficiency towards the beginning of the VLBI campaign
(as displayed in Figures~\ref{fig:pheff_2018}   and~\ref{fig:amp+pha_vs_time}),  are not induced by variations in
tropospheric water vapor alone,
but most likely arise instead from a combination of water vapor and
wind-induced atmospheric turbulence  \citep[e.g.,][]{Nikolic2013,Maud2017}.

\begin{figure}[htbp]
\center{
  \includegraphics[width=0.48\textwidth,angle=0]{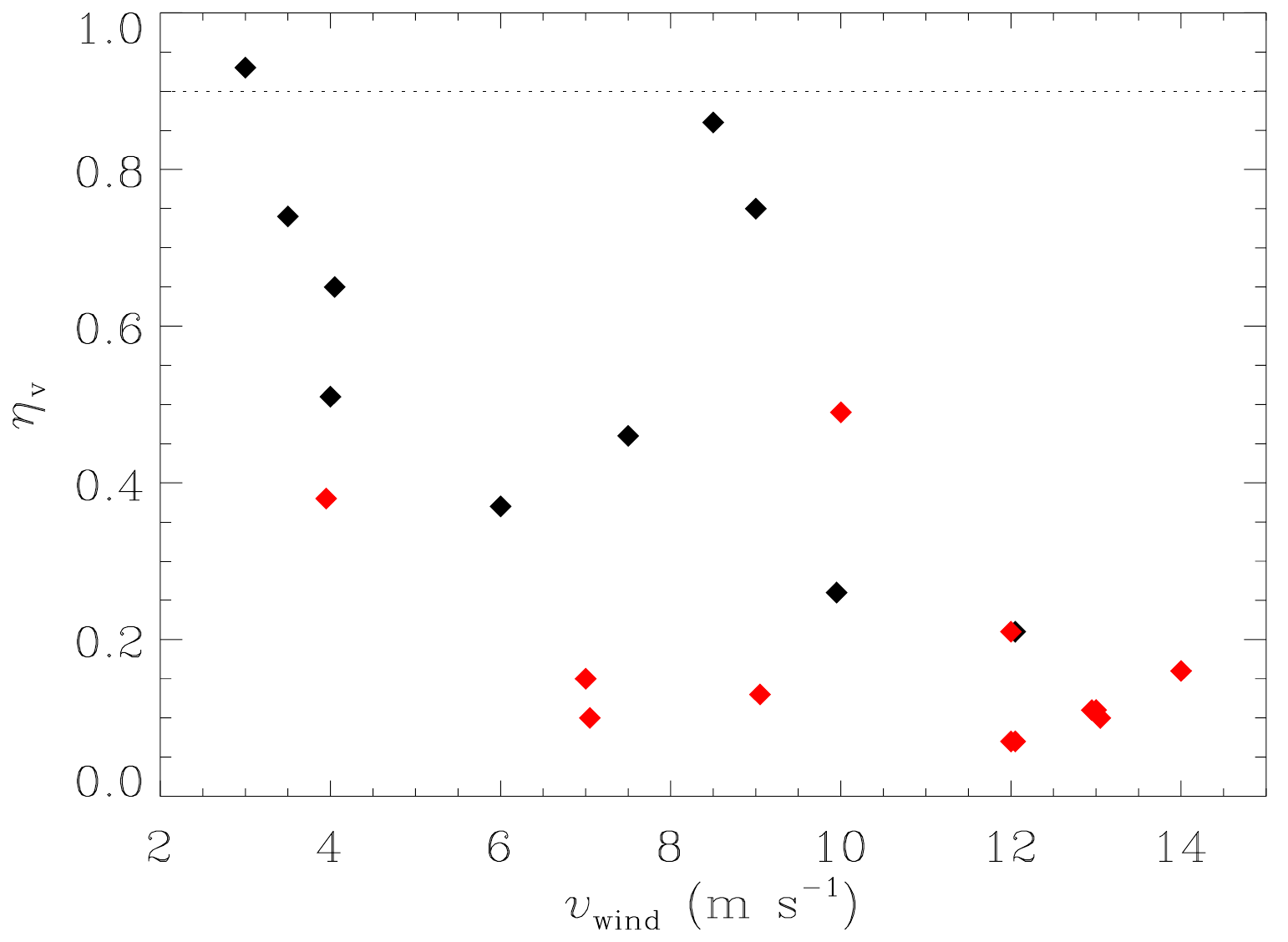} 
 \includegraphics[width=0.48\textwidth,angle=0]{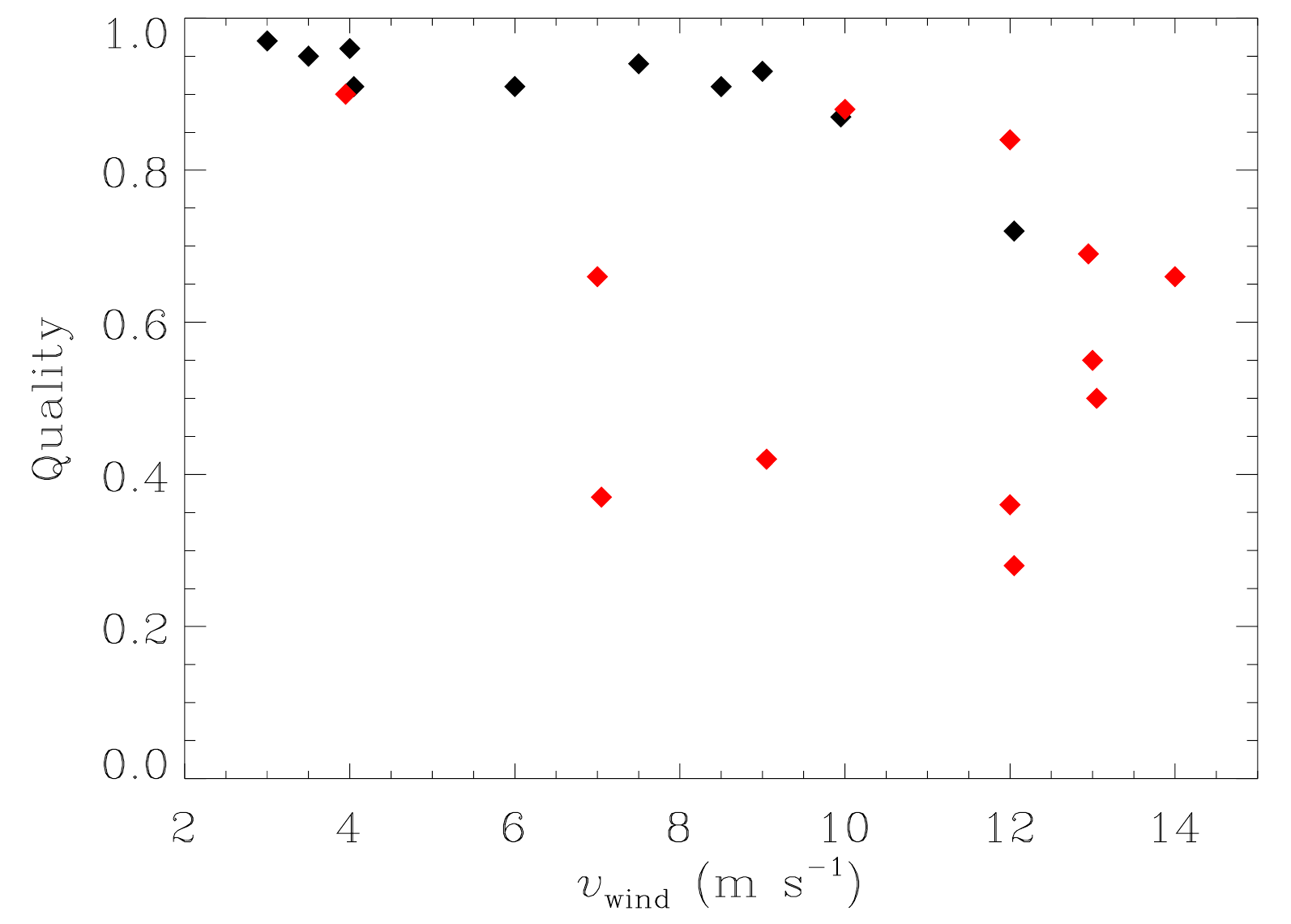}
  \caption{
  Phasing efficiency (top) and phasing quality  (bottom) as a function of wind speed for the data sets presented in the current paper (Tables~1 \& 2).
  Data sets with PWV$\ge$2.0~mm are indicated in red; data with PWV$<$2.0~mm are plotted in black. The horizontal dashed
  line in the upper panel indicates the nominal APS efficiency goal of $\ge$90\%. All of the data sets plotted here were
  taken without the use of WVR-based fast phasing corrections.
}}
\protect\label{fig:windspeed}
\end{figure}

To obtain a preliminary assessment of how the combination of wind speed and PWV affects phasing performance
at 345~GHz, we plot in Figure~4 the phasing efficiency $\eta_{v}$ and the phasing quality as a function of wind speed $v_{\rm wind}$
for each of the data sets presented in the current paper (Tables 1 \& 2). Data points with PWV$\ge$2.0~mm are shown in red
and data with PWV$<$2.0~mm are shown in black. 

Figure~4 shows that irrespective of wind speed,  when PWV$>$2.0~mm, phasing efficiency in Band~7 generally falls below $\sim$50\%.
Thus operation of the APS in Band~7 in conditions with PWV$>$2.0~mm is not recommended in general, although it may be possible to
relax this restriction with future use of the fast phasing mode (see above). 

We also see in Figure~4 that when wind speeds exceed $\sim$10~m s$^{-1}$, phasing efficiency is consistently quite low ($\lsim$20\%), even 
in one case with PWV$<$2~mm. Furthermore, phasing solution {\it quality} is seen to decline systematically for such high wind speeds. This suggests
that for the high wind-speed regime, the use of fast phasing corrections is unlikely to improve the overall phasing performance. 
It is thus recommended that phased array observations in Band~7 are strictly avoided in conditions with $v_{\rm wind}>$10 m s$^{-1}$.

For intermediate wind speeds ($3\le v_{\rm wind}<10$~m s$^{-1}$) the situation is more complex. We find that (in absence of fast phasing corrections)
one generally does not meet the nominal
APS efficiency goal of $\eta_{v}\ge$0.9. However, for VLBI, the sensitivity and strategic importance of phased ALMA mean that
even data with lower phasing efficiency may be scientifically useful.  For example, when PWV is low ($<$2.0~mm), in many cases
$\eta_{v}\gsim$0.5; assuming 37 phased 12~m antennas, this still provides the sensitivity comparable to a 25~m diameter parabolic dish.
Furthermore, the generally good phasing {\it quality} seen for data sets with $3\le v_{\rm wind}<10$~m s$^{-1}$ and PWV$\le$2.0~mm
suggests that the fraction of experiments achieving $\eta_{v}>$0.5 under this combination of conditions
is expected to grow significantly with the use of the fast phasing mode. 
We are currently in the process
of acquiring additional regression test data to explore how much improvement the fast mode provides under a range of observing
conditions, including moderate wind speeds ($<$10 m s$^{-1}$) and moderate PWV values ($\sim$2--3~mm).

\subsection{Impact of Array Size and Maximum Baseline Length}
\label{aps_perform_array}
\begin{table*}
\caption{Comparison of phase RMS as a function of baseline length.}
\centering  
\small
\begin{tabular}{ccccccc} 
\hline\hline                  
\noalign{\smallskip}
Data set & Max. baseline & Date & Target & Flux & Phase RMS &Phase RMS \\
&&&& density&all baselines &baselines$<$200 m \\
&(m) &(YYYY MMM DD) & & (Jy) & (deg) &  (deg)\\
\noalign{\smallskip}
\hline
\noalign{\smallskip}  
B180    &   180       & 2015 Mar 30    &  3C273          &  4.0  & 16  &  16    \\
B1500  &   1500     & 2015 Aug 02    &  J0522--3627 &  4.5  &  69 &  35    \\
B6900  &   6900     & 2021 Sep 03    &  J1924--2914  &  3.0 &  55 &  39    \\
\noalign{\smallskip}
\hline
\end{tabular}
\label{table:phvsbsl}
\end{table*}

\begin{figure}[htbp]
\center{
\includegraphics[width=0.5\textwidth,angle=0]{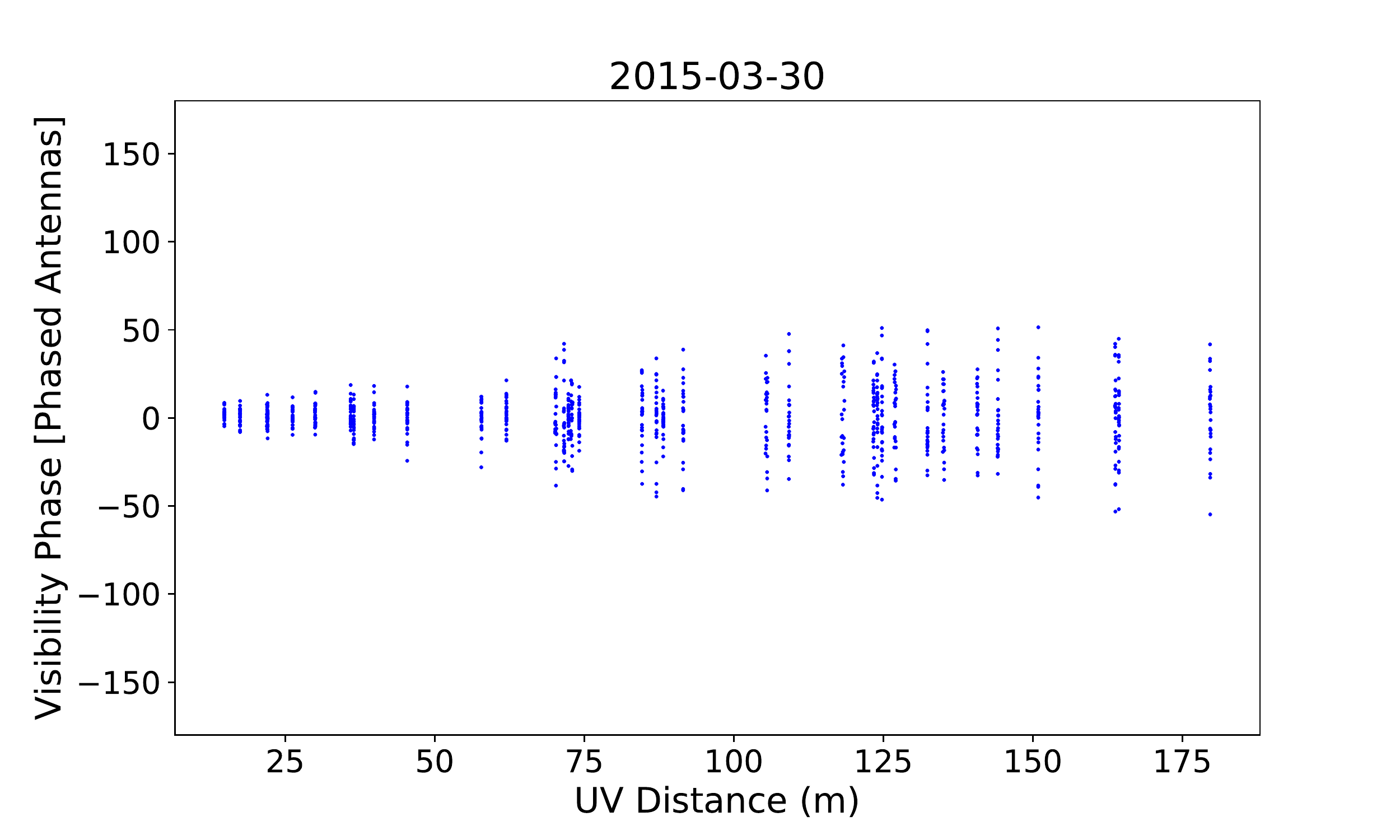} 
\includegraphics[width=0.5\textwidth,angle=0]{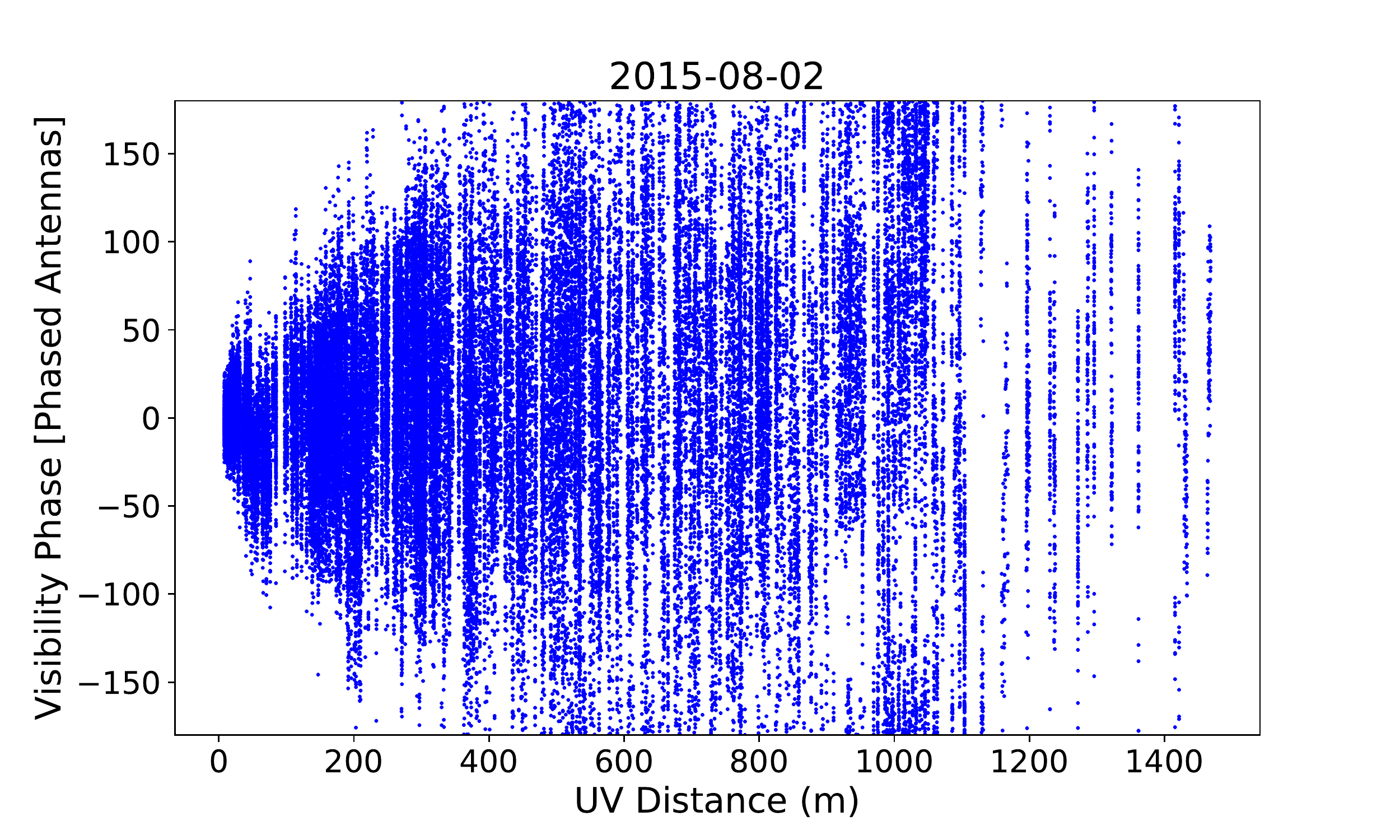} 
\includegraphics[width=0.5\textwidth,angle=0]{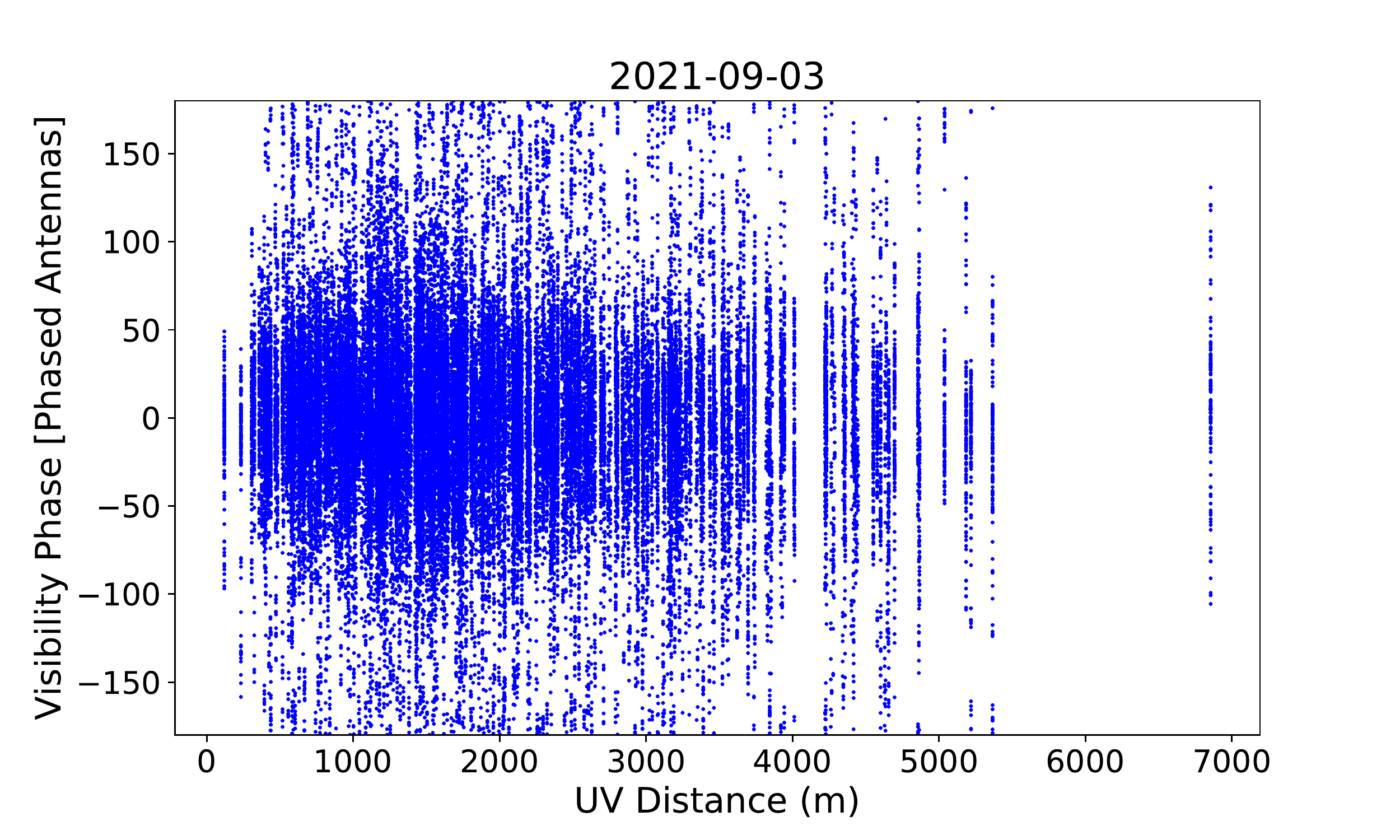} 
}
\caption{\footnotesize 
  Phase (in degrees) as a function of projected baseline length (in meters)
  for Band~7 phasing tests on 2015 March 30 (top), 2015 August 2 (middle),
  and 2021 September 3 (bottom). 
In each case the target flux density is a few Jy (see text for details). 
The tests were taken under similar weather conditions (see Table~\ref{tab:obslog}).
The mean RMS dispersions in the phases in the 2015 August and 2021 September data (with longest baselines $<$1.5~km and $<$6.9~km,
respectively) are larger than in the 2015 March data where the longest baselines $<$0.2~km, even
on the shortest baselines. 
Data from a single correlator quadrant (SPW=0, averaged over all channels) and a single polarization (XX) are shown in all panels.
The phased array for the 2015 August observations included a number of 7~m CM antennas which are typically not included
in the phased array (see Section~\ref{setup}).
}
\label{fig:uv_test}
\end{figure}

Figure~\ref{fig:uv_test} compares phase as a function of $uv$ distance for  two tests carried out in 2015 (on March 30 and  August 2),
and one carried out in 2021 (on September 3).
The tests were taken under similar weather conditions (PWV$\sim$0.5~mm) but
with different baselines ranges ($<$180~m, $<$1500~m, and  $<$6900~m, respectively).
Table~\ref{table:phvsbsl} provides a summary of these tests, labelled B180, B1500, and B6900, respectively.  

Considering the single correlation quadrant (SPW=0) and polarization (XX) that is plotted for each data set,
we find that
the RMS dispersion in the phases for all baselines in the phased array is significantly higher in the B1500 data (69~deg)
and the B6900 data (55 deg) compared with the  B180 data
(16~deg). This is true even if we limit our comparison to baselines $<$200~m for all three data sets; in this case
the RMS phase dispersions are 35~deg (B1500), 39~deg (B6900) and 16~deg (B180).
We thus see evidence that even under relatively good weather conditions it is advantageous to limit the phased array to short
baselines (less than a few hundred meters) when observing at wavelengths $\lambda\lsim$1~mm. Because the correlated amplitude
scales as e$^{-\sigma_{p}^{2}/2}$ where $\sigma_{p}$ is the RMS dispersion (in radians) of the phase, the sensitivity gained by the
inclusion of antennas on long baselines
will be significantly diminished by an overall decrease in phasing efficiency of the entire phased array. This can be understood as a result of the fact that
the APS phase solver uses a least-squares method to convert baseline phases
to station phases, and too many noisy baselines (typically those longer than a few hundred meters)
will impact the overall quality of the phasing solutions.

\subsection{Comparison between Band~6 and Band~7}
\label{aps_B6-B7}
To begin to assess how the performance of the APS in Band~7 compares with Band~6, we have performed preliminary analysis of test observations where Band~6 and Band~7 measurements were obtained within a single session. 
In particular, during the  2017 February 1 ALMA-only test and the  2018 October 18 VLBI test, 
 data in both Bands 6 and 7 were acquired using comparable arrays, with baseline lengths ranging from $\sim$15~m to $\sim$300~m, while the PWV content varied in the range $\sim$1.5--2.0~mm.

To explore the relative performance in the two bands, in Figure~6 
 we compare the results from scans of a few minutes duration on the source J0522--3627 in each band, acquired on 2017 February 1. 
 The RMS phase fluctuations for all phased baselines were 43 deg in Band 6 and 36 deg in Band 7. These relatively high phase dispersions
 reflect the sub-optimal weather conditions for observing
 in these bands (PWV$\sim$~1.6~mm; wind speed $\sim$9 m s$^{-1}$), but these results nonetheless indicate that the phasing system is capable of comparable performance in Band 7 compared with Band 6.
 In this example, the fluctuations are actually slightly lower in Band 7 relative to Band 6. However, as the observations we compare here were not co-temporal,
 those differences can be ascribed to changes in PWV, coupled with changes in the source elevation. 
 We reach similar conclusions from a preliminary analysis on the 2018 test dataset.
 
Our initial results suggest that for compact array sizes (baselines $\lesssim$300~m) and moderately good or better observing conditions (PWV $\lesssim$2~mm), high-quality and high-efficiency phased array performance will be
possible in both Bands~6 and 7 and that Band~7 does not show any appreciable loss in phasing performance compared with Band~6.
Under conditions where the phase fluctuations are dominated by tropospheric water vapor, some degradation
in Band~7 performance compared with Band~6 is naturally expected to occur as a consequence of the linear wavelength dependence
in the temporal phase fluctuations  \citep[e.g.,][]{Rioja2012} and the slightly lower aperture efficiency of the ALMA antennas in Band~7 compared with Band~6 
\citep{Remijan2019}. 
However, in practice, we did not see any clear
evidence of systematic degradation in phasing performance in Band~7, in part because only limited comparison data are available to date, and also because
such comparisons are complicated by the modest
variations in weather conditions that typically occur over tens of minutes
during available test periods at ALMA.

\begin{figure*}[htbp]
\center{
\includegraphics[width=0.45\textwidth,angle=0]{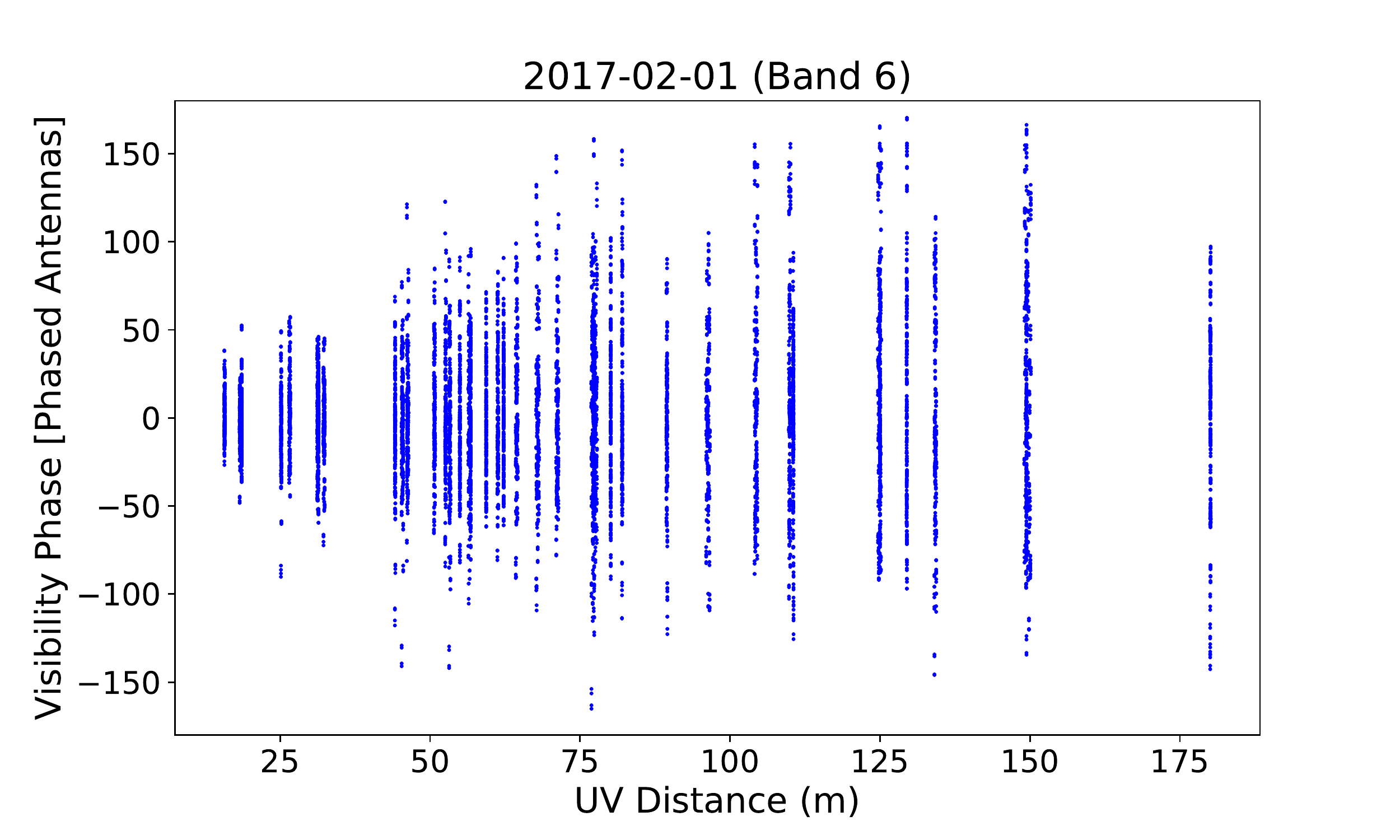} 
\includegraphics[width=0.45\textwidth,angle=0]{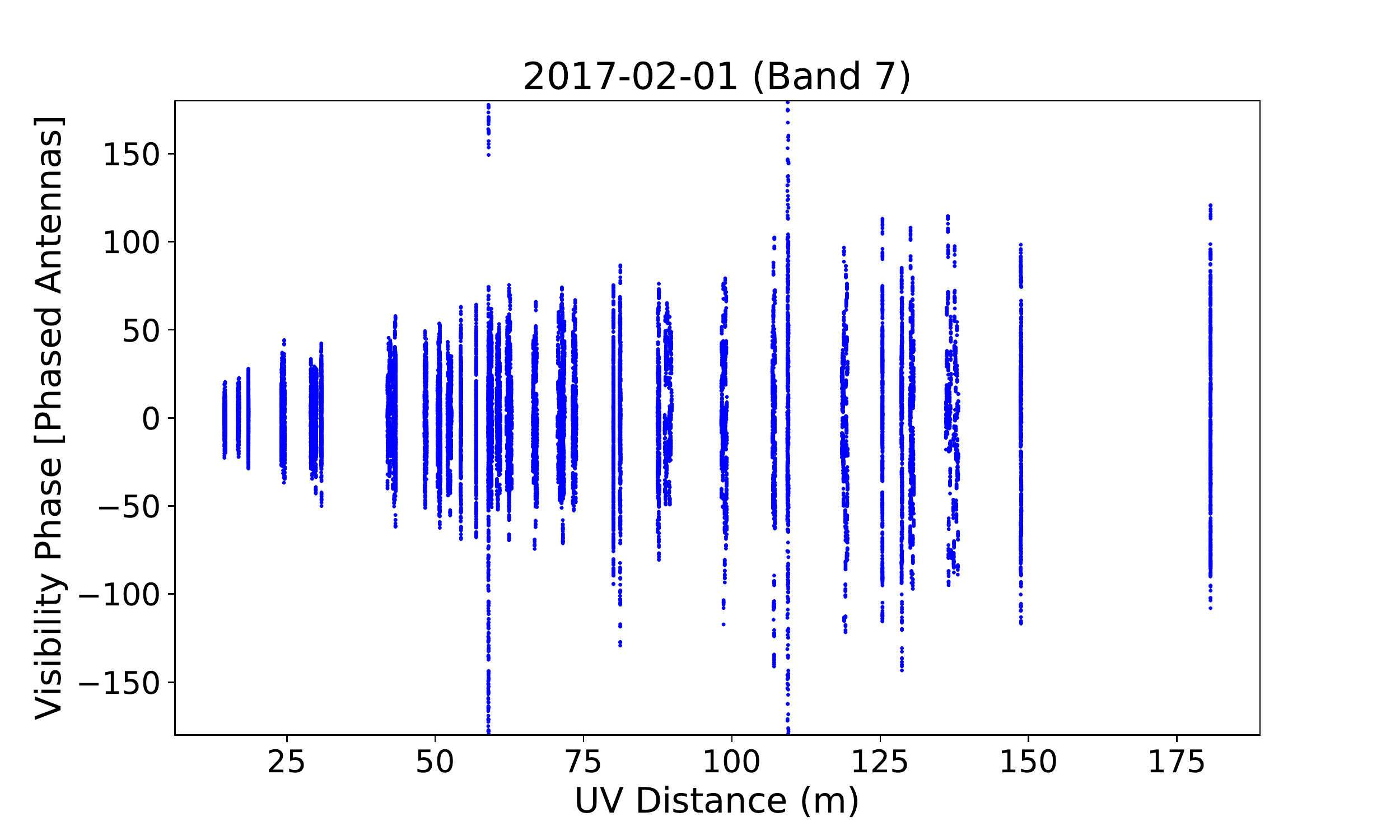} 
\includegraphics[width=0.45\textwidth,angle=0]{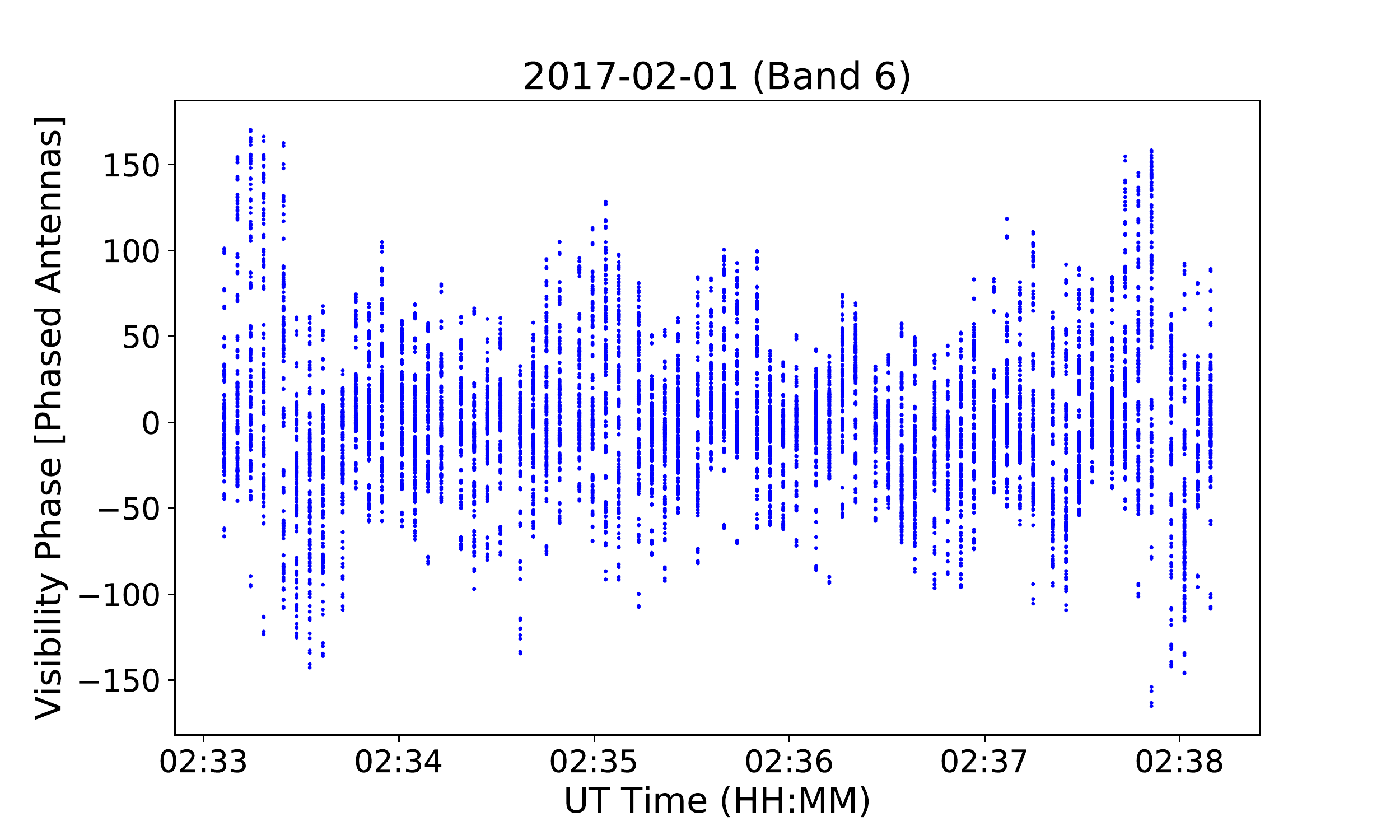} 
\includegraphics[width=0.45\textwidth,angle=0]{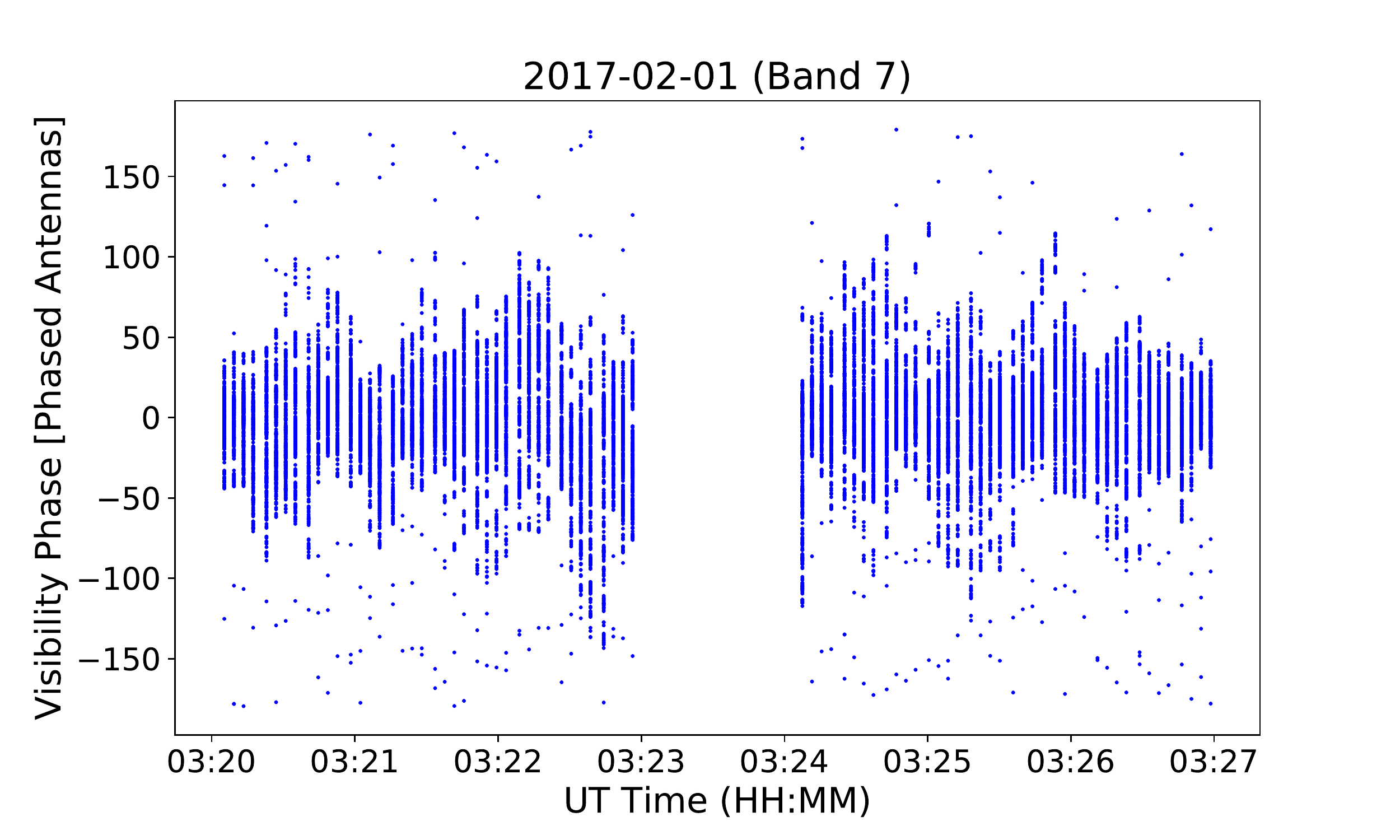} 
\caption{\footnotesize 
  Phase (in degrees) as a function of projected baseline length in meters (upper panels) and observing time (lower panels)
  during scans of a few minutes duration on the source J0522--3627 using the APS on 2017 February 1 in Band 6 (left) and Band 7 (right), respectively,
  under conditions with PWV$\sim$1.6~mm and wind speeds of $\sim$9~m~s$^{-1}$.
  The RMS phase fluctuations are $\sim$43~deg in Band 6 and $\sim$36~deg
  in Band 7, respectively, indicating comparable phasing performance in the two bands.
}}
\label{fig:B6vsB7}
\end{figure*}

\section{Additional Assessments of Phased Array Data Quality}
\label{qualass}
The primary goal of phasing ALMA in Band~7 is to harness the enormous sensitivity and collecting
area of ALMA for use a sub-mm VLBI station \citep[e.g.,][]{Fish2013}. This will be discussed
further in Paper~II. A key part of the process of turning the phased ALMA array into a functional
sub-mm VLBI station will be to first calibrate the interferometric visibilities \citep{QA2Paper}. 
Because the data taken during the 2018 October VLBI test were intended for testing and engineering
purposes only, a full suite of calibrators was not observed. However, as described in Appendix~\ref{app:calibration}, we have
been able to perform a modified calibration scheme to the data to allow these data to
be meaningfully combined with other VLBI stations (Paper~II). This calibration scheme additionally allows us to
perform some further quality assurance checks, as described below.

\subsection{Accuracy of the Absolute Flux-density Scale}
\label{app:fluxcomp}

To assess the accuracy of the flux density calibration in VLBI mode, \citet{QA2Paper} compared the measured flux densities of VLBI targets with values
derived from the independent flux monitoring done with the  ACA,  
taking advantage of the fact that some of the Grid Sources are also observed in VLBI observations.
The analysis in \citet{QA2Paper}  showed that the flux density values estimated from ALMA during VLBI observations  are generally within 5\% in Band~3 and 10\%
in Band~6 when compared with the Grid Sources monitoring values (consistent with the expected  absolute flux calibration uncertainty at ALMA; see \citealt{Remijan2019}).

We have performed a similar analysis for the Grid Sources observed in Band~7. 
Table~\ref{tab:fluxes} reports the measured flux values (per SPW) for all sources observed in Band\,7  during the 2018 October
campaign along with the archival flux values for Grid Sources.
The flux values of the VLBI sources  are estimated in the {\it uv}-plane using the {\sc CASA} task \texttt{fluxscale},
which adopts a point-source model (this assumption is valid since Grid Sources are unresolved on ALMA baselines).
The expected flux density of Grid Sources at a given time and frequency are retrieved from the ALMA archive via the \texttt{getALMAflux()} function implemented in the \texttt{CASA analysis utils}. 
Table~\ref{tab:fluxes} also reports the time difference between the VLBI observations and the archival entry, $\Delta t_{S}$, which is  $<$1 day for all sources
(i.e. they  were observed with the ACA within less than a day of the VLBI observations) except BL Lac. 
 The nominal calibration uncertainty at ALMA in Band~7 is $\sim$10\% 
 (see ALMA  Technical Handbook -- \citealt{Remijan2019}),  
 and  most of our new flux density
 estimates are consistent with the archival values to within this range, with the exception of CTA102 (with a  16\% lower flux) and J0522$-$3627 (with a  13\% higher flux). 

\subsection{Interferometric Test Images}
\label{imaging}

\begin{figure*}
\includegraphics[width=0.5\textwidth, trim={0 2cm 4cm 3cm,clip}]{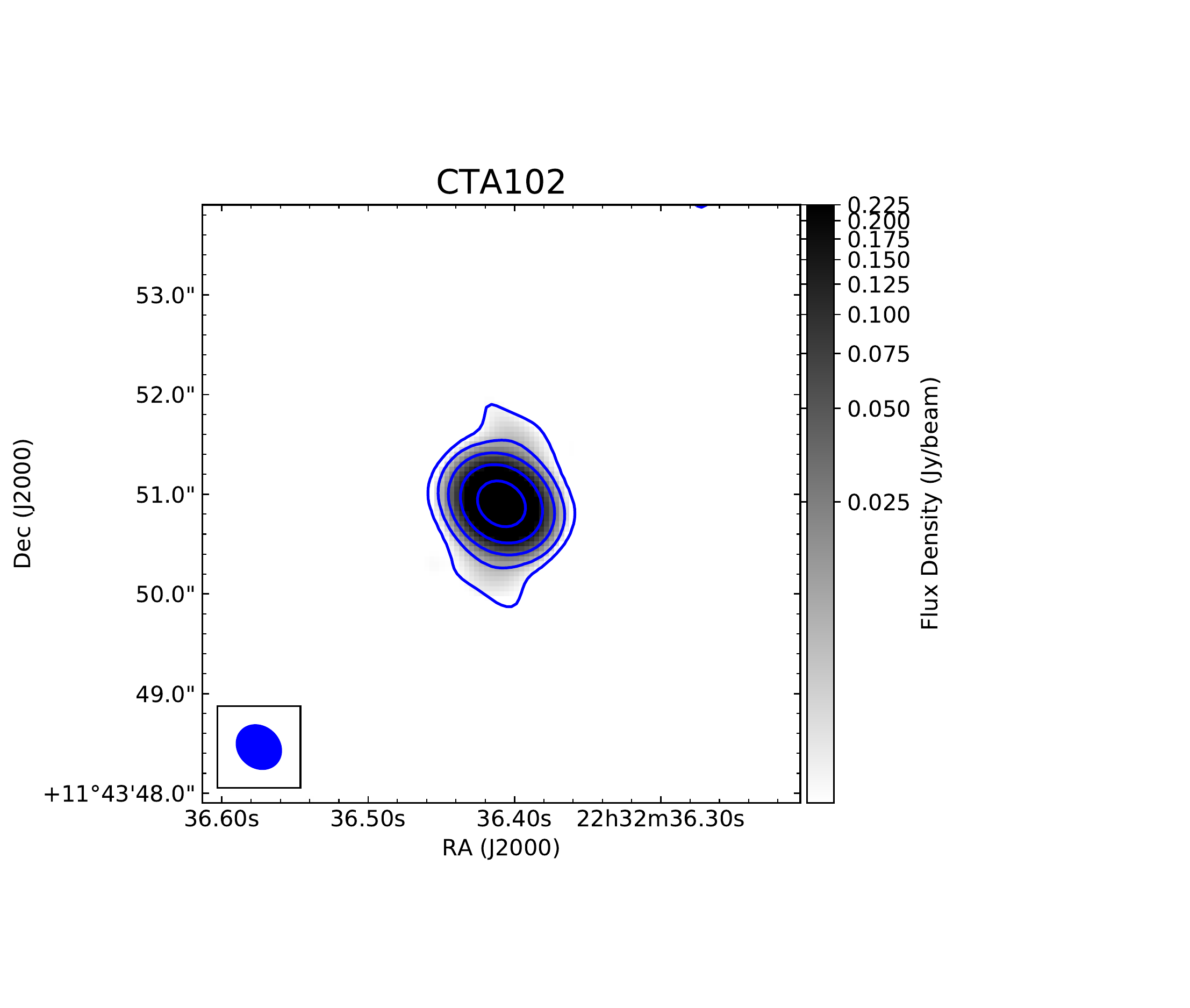}
\includegraphics[width=0.5\textwidth, trim={0 2cm 4cm 3cm},clip]{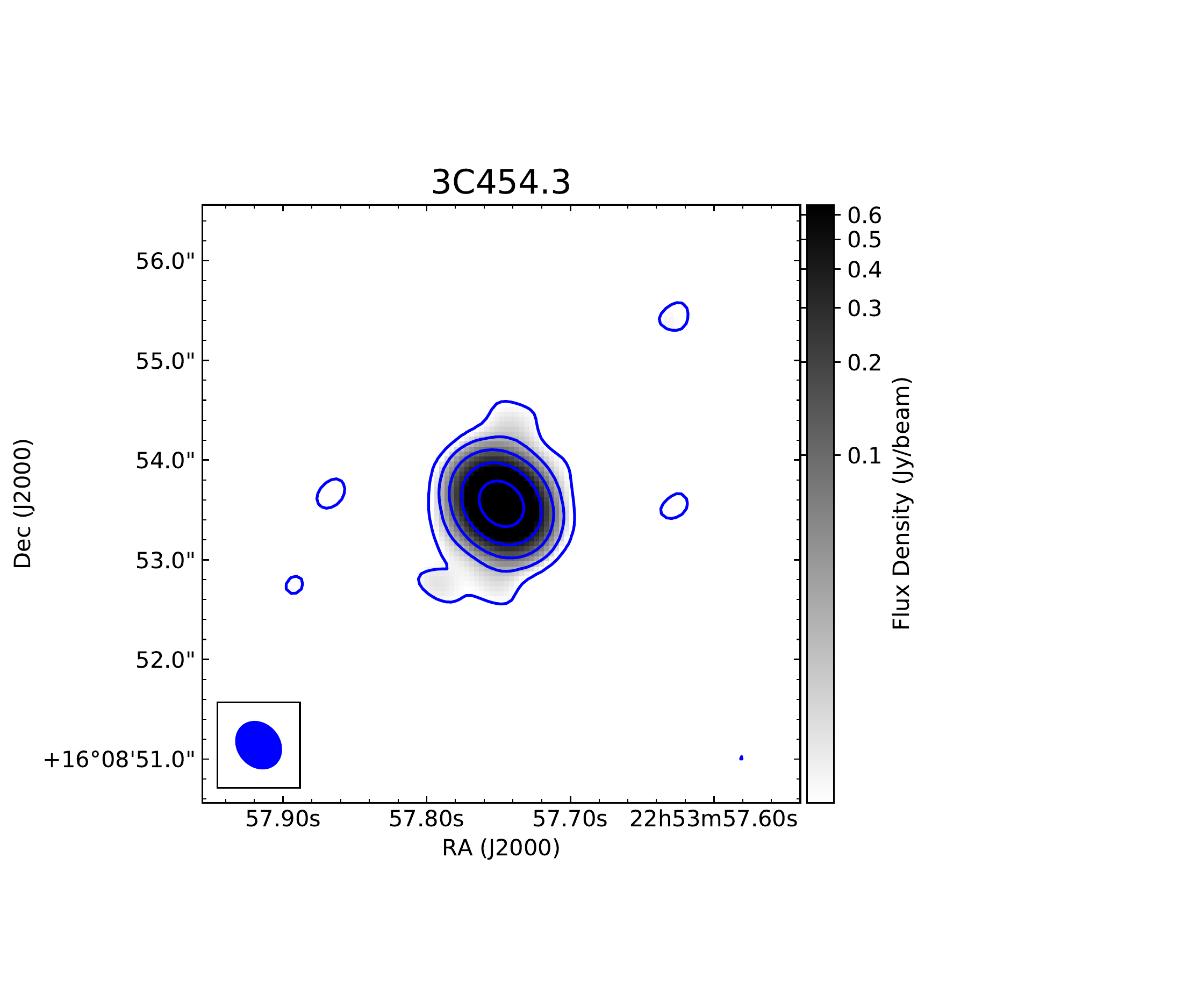}
\includegraphics[width=0.5\textwidth, trim={0 2cm 4cm 3cm},clip]{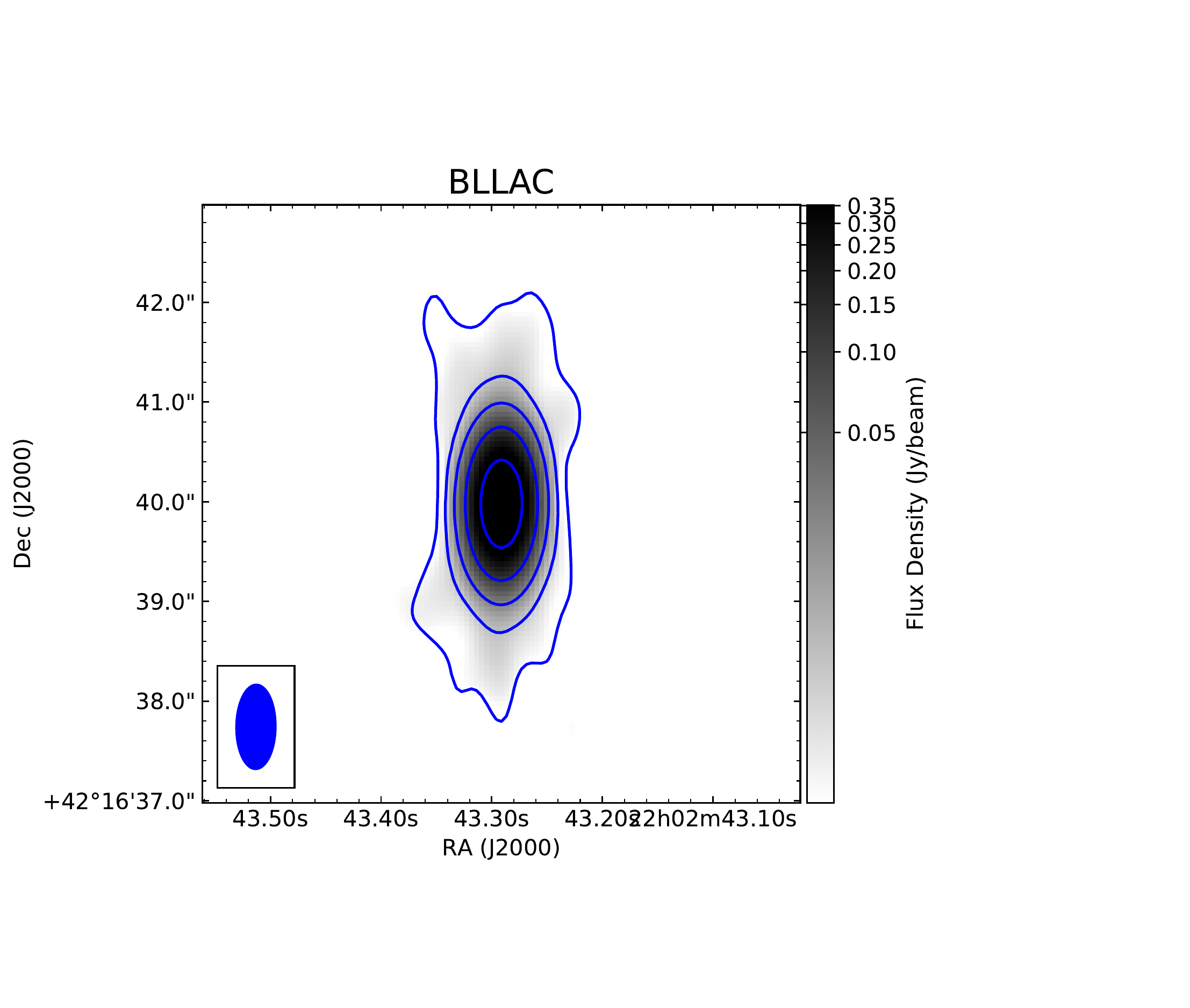}
\includegraphics[width=0.5\textwidth, trim={0 2cm 4cm 3cm},clip]{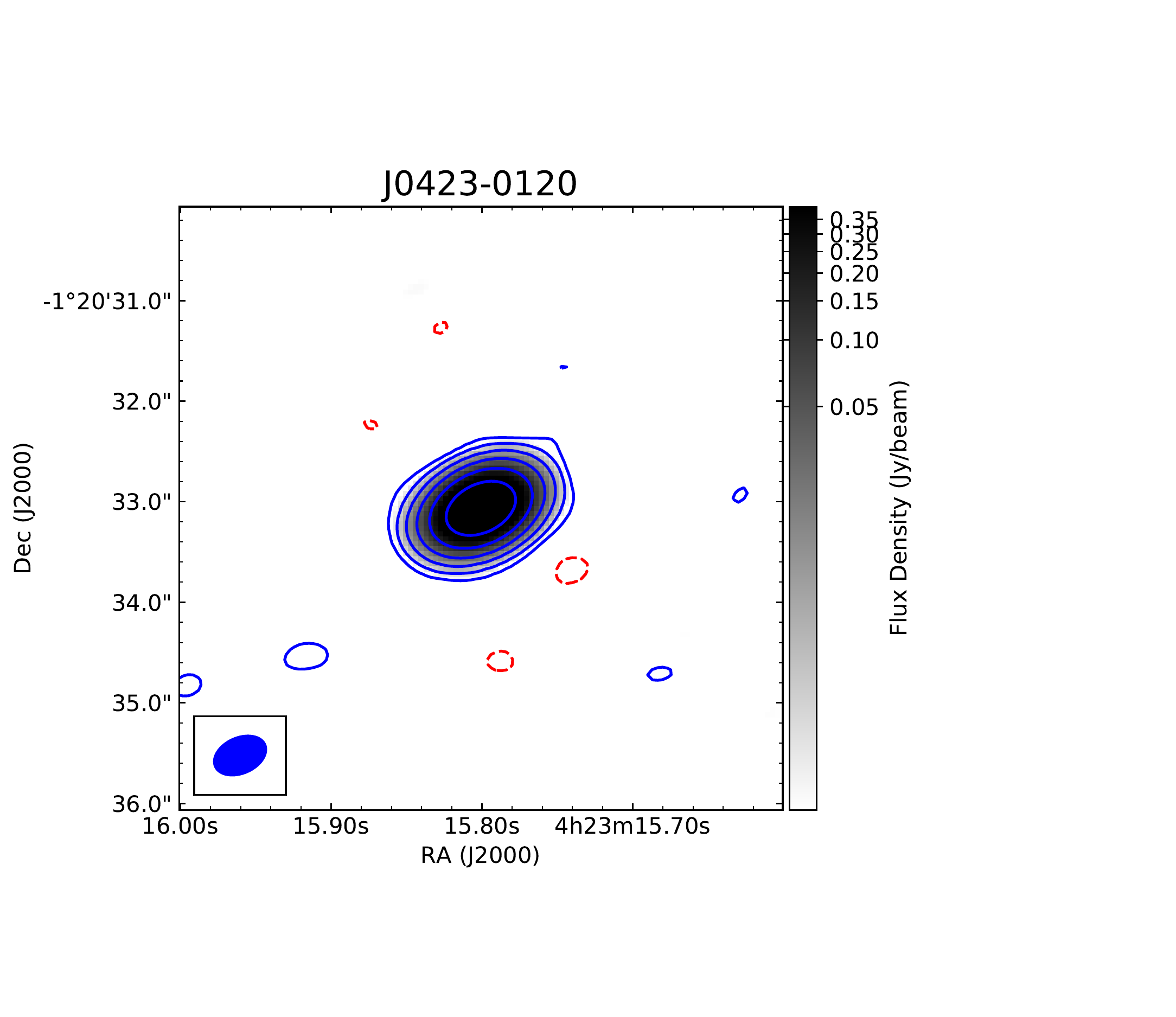}
\includegraphics[width=0.5\textwidth, trim={0 2cm 4cm 3cm},clip]{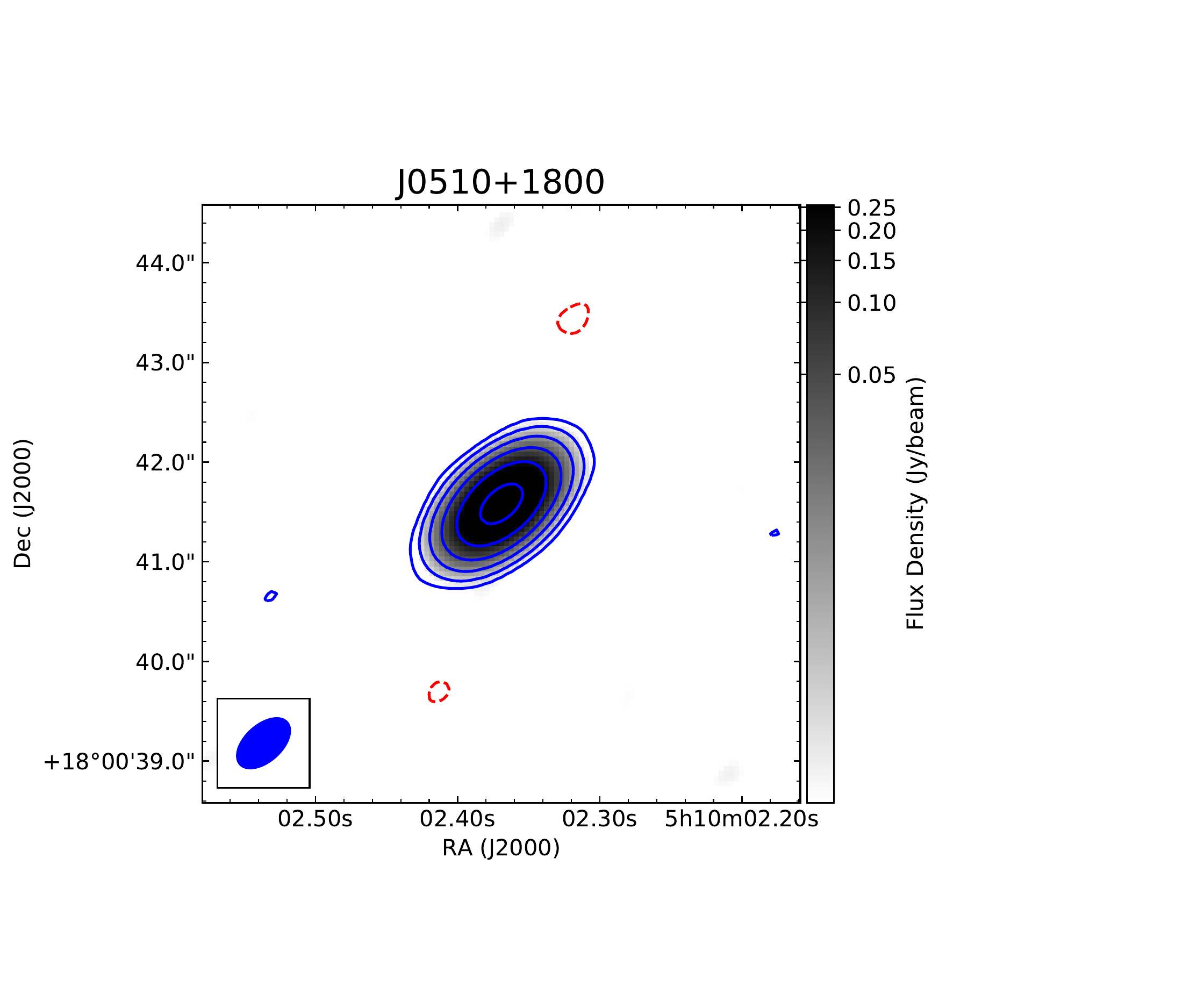}
\includegraphics[width=0.5\textwidth, trim={0 2cm 4cm 3cm},clip]{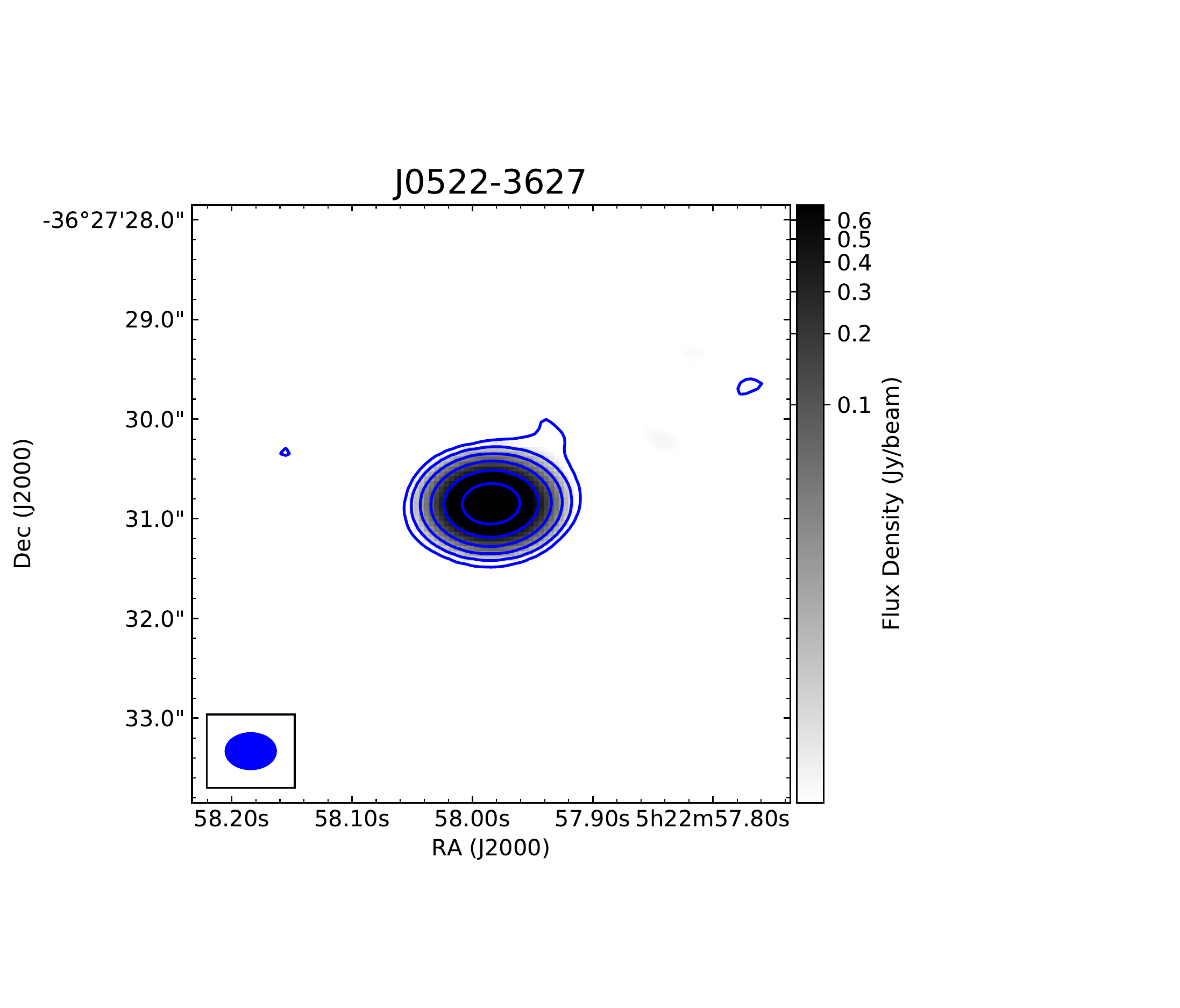}
\caption{
Representative total intensity images of targets observed during the 2018 October VLBI campaign. 
The grey-scale image shows emission at 347.6~GHz  (SPW=2) while the blue  contours show emission centered at 336.6~GHz  (SPW=0);
the red (dashed) contours indicate negative values. 
The contour levels are $\pm 3\sigma \times 2^n$ where $\sigma=$[0.9, 2.5, 0.7, 0.25, 0.27, 0.7]~mJy~beam$^{-1}$ for CTA 102,  3C454.3, BL Lac, J0423--0120, J0510+1800, J0522--3627,
respectively, and $n=0\,,1\,,2\,,3\,\ldots$ up to the peak flux-density.
The intensity brightness is plotted using  a logarithmic weighting function (starting from the $3\sigma$-level). The major axis of the synthesized beam for BL~Lac is $\sim$0\pas62 and
for the remaining sources is 0\pas45-0\pas5. }
\label{fig:maps}
\end{figure*}

As an additional means of assessing
the science readiness of the APS  in Band~7, we have produced images  of the VLBI targets  from fully-calibrated  ALMA interferometric visibilities (see Appendix~A),
following the same  procedures outlined in \citet{Goddi2021}.   We show representative images
in Figure~\ref{fig:maps}. 
The images displayed cover an area slightly smaller than the primary beam of the ALMA antennas (18\arcsec\ at 350 GHz) and
have a synthesized beamsize of roughly 0\pas45. The correction for the attenuation of the primary beam is not  applied to these maps. 

We have conducted a series of quality-assurance self-consistency tests on these images. 
We first assessed that the images are consistent with unresolved point sources (as expected for the selected VLBI targets), indicating that they are not smeared by residual phase errors. 
We then established that the size determined from a Gaussian fit  matches the size of the synthesized beam (within $<1$\%) and the peak flux and integrated flux
have the same value (within $<1$\%), as expected for point-like sources. Finally we confirmed that the peak flux occurs exactly at the phase center.
Besides these self-consistency checks, we also estimated  source flux densities from the
images (following the methods outlined in \citealt{Goddi2021}) and assessed that they are consistent  with the values estimated in
Table~\ref{tab:fluxes} (within $\lesssim$10\% for sources observed on the 18th/19th and within $\lesssim$5\% for sources observed on the
21st, respectively). 

\section{Amplitude Calibration of Phased-ALMA as a Single VLBI Station}
\label{fluxcal}

Traditionally VLBI stations store time-dependent amplitude corrections, $A(t)$, as a combination of  $T_{\rm sys}$ 
 (one value per intermediate frequency and integration time)
 and an instrumental gain given in degrees per flux unit (K/Jy) or DPFU 
 (assumed to be stable over time and frequency): 
 $$
  A(t) = \sqrt{T_{\rm sys}/\mathrm{DPFU}} 
 $$

\noindent In VLBI, one also often defines a {\it system-equivalent flux density} (SEFD) as the total system noise represented
in units of equivalent incident  flux density, which can be written as
\begin{equation}
   \mathrm{SEFD}= \left<T_{\rm sys}\right> /\mathrm{DPFU}.
   \label{eq:SEFD}
   \end{equation} 
   
\noindent In the following, we estimate DPFU, \tsys, and SEFD for phased-ALMA in Band~7 using the data collected on 2018 October 18/19 and 21.
Representative values are reported in Table~7. 

   \begin{table}
\begin{center}
\caption{ALMA Band 7 Antenna Parameters for Observations in 2018 October }
\begin{tabular}{cccccc} 
\hline 
Date & $N_{\rm phased}$ & \tsys$^{a}$  & DPFU$^{b}$  &  \tsys[sum]$^{c}$ &    SEFD$^{d}$ \\
(2018 Oct.) & Ant. & (K) & (K/Jy) & (K) & (Jy) \\ 
\hline
\noalign{\smallskip}
\multicolumn{6}{c}{Band 7} \\
21 &29 & 155 & 0.011 & 2.6 & 238 \\  
18/19 &25 & 200 & 0.011 & 6.4 & 578 \\  
\multicolumn{6}{c}{Band 6} \\
19 &25 & 80 & 0.006 & 0.9 & 150 \\  
\noalign{\smallskip}
\hline
\end{tabular}
\end{center}
\tablecomments{DPFU, \tsys, and SEFD estimates for phased ALMA in Band 7, as derived from observations in 2018 October.}
\tablenotetext{a}{Antenna-wise median of valid \tsys\ measurements.}
\tablenotetext{b}{Antenna-wise average of DPFUs, estimated with Eq.~\ref{eq:DPFU}.}
\tablenotetext{c}{Median phased-array \tsys, estimated with Eq.~\ref{TsysEq}.}
\tablenotetext{d}{Phased-array SEFD, estimated with Eq.~\ref{eq:SEFD}.}
\label{table:dpfu}
\end{table}
\subsection{DPFU}
 
While in a single-dish telescope the DPFU is fixed, in a phased array it scales with the number of phased antennas.
Because the number of phased antennas can change during the observations, the DPFU may also change. In order to
keep the DPFU of phased ALMA constant over a given observation (for calibration purposes), we set the DPFU of phased ALMA
to the antenna-wise {\it average} of DPFUs (instead of the antenna-wise {\it sum}). 
The DPFU of a single antenna $i$ is calculated using the measured \tsys\ and amplitude gains $g_{a,i}$ computed from
self-calibration during QA2 (these are stored in the \texttt{<label>.flux\_inf.APP.OpCorr} table; see Appendix~B):

 \begin{equation}
{\rm DPFU}_i = \langle 1/g_{a,i}^2\rangle /\langle 1/T_{ {\rm sys},i } \rangle
\label{eq:DPFU}
 \end{equation}

\noindent where the average $\langle$~$\rangle$ is computed over all scans where a $T_{\rm sys}$ is measured at the antenna $i$.

Using the data collected on October 21 (which have higher quality), we estimate DPFU = 0.011 K/Jy. 
We assume this value as the single-antenna average DPFU for Band~7 phased ALMA observations in 2018 October. The DPFU for a
phased array of 29 antennas would be: (DPFU)$_{29}$ = 0.32 K/Jy.

\subsection{T$_{\rm sys}$ and ANTAB files}
\label{antab}

The amplitude calibration for phased ALMA is computed 
via a linear interpolation of the ALMA antenna gains 
and  it  is stored in the ``ANTAB" format. This
is the standard file used in VLBI to store amplitude a-priori information, and readable by the Astronomical Image Processing System
(AIPS) task \texttt{ANTAB} \citep{AIPS2003}.

The ANTAB files are generated directly from the antenna-wise average of all the  amplitude gains $g_a$ and phase gains $g_p$ (stored
in the QA2 \texttt{<label>.flux\_inf.OpCorr.APP} and  \texttt{<label>.phase\_int.APP} tables, respectively; see Appendix~B):

 \begin{equation}
   T_{\rm sys} = \frac{{\rm DPFU}}{\langle[g_a e^{i g_p}]^2\rangle}  \frac{1}{N_{\rm phased}}
     \label{TsysEq}
 \end{equation}
 where the average $\langle$~$\rangle$ runs on all time integrations where gain solutions are found and on all phased antennas (see Eq. 16 in \citealt{QA2Paper}).

\begin{figure}[ht!]
\centering
\includegraphics[%
  width=0.5\textwidth]{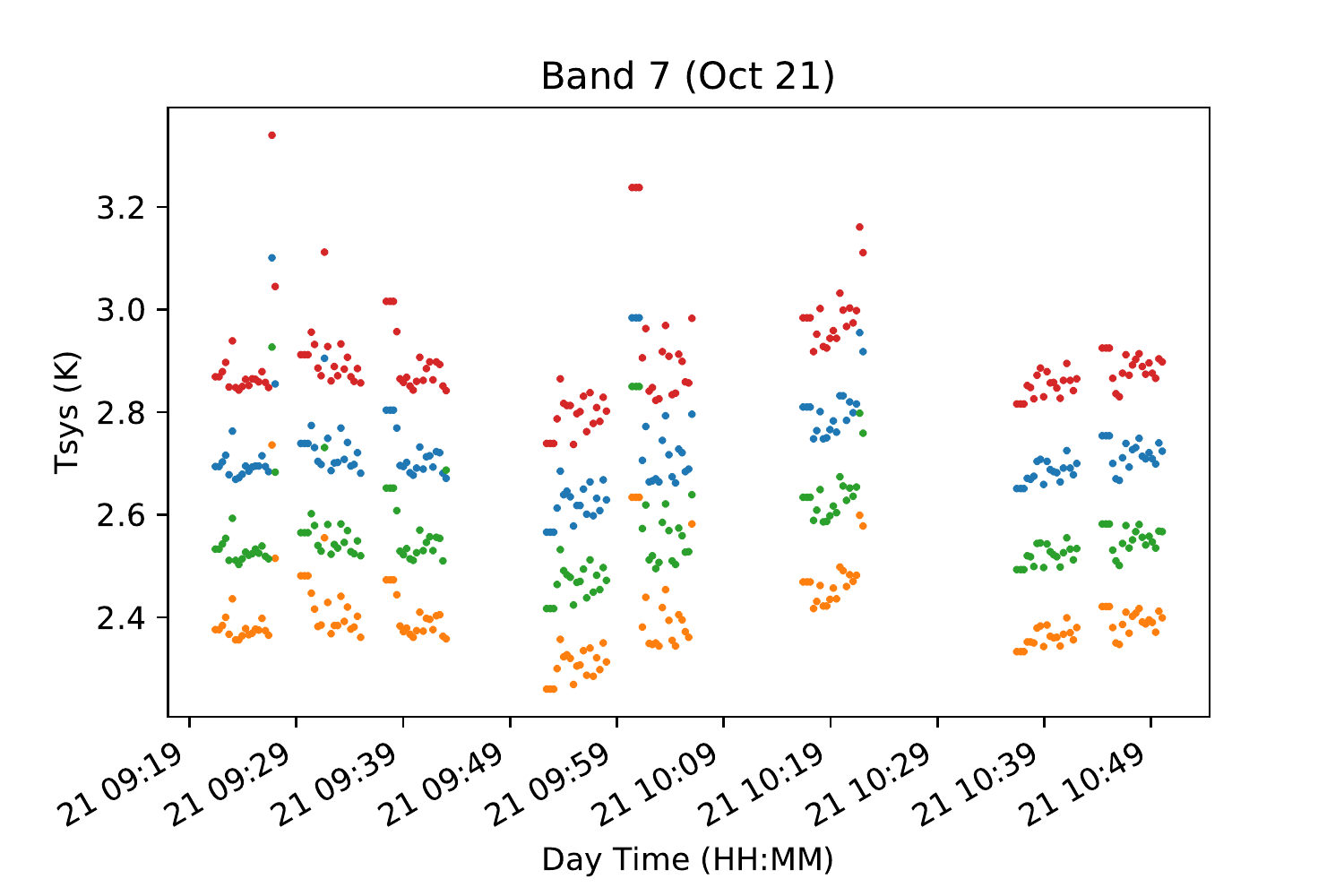}
\includegraphics[%
  width=0.5\textwidth]{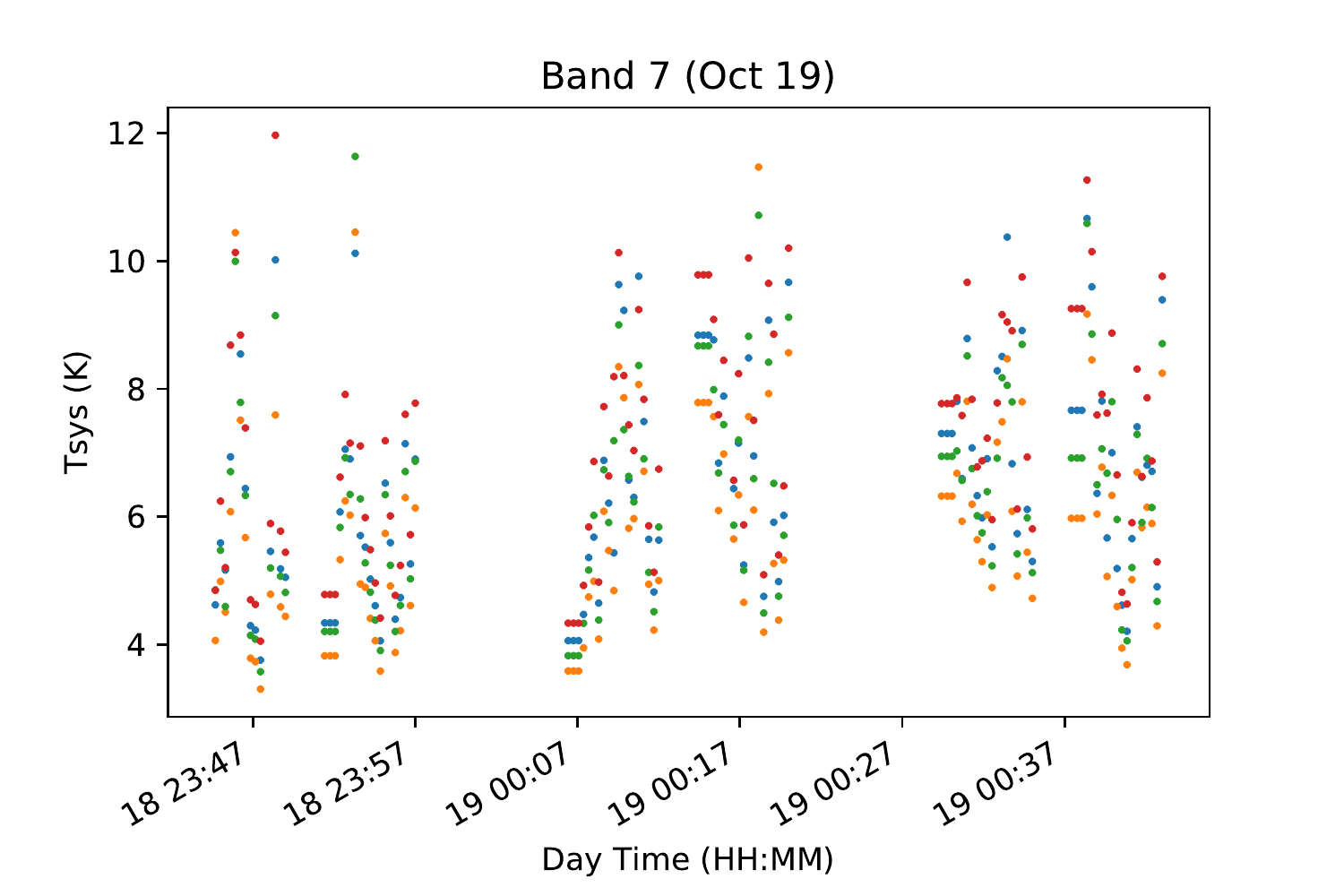}
\includegraphics[%
  width=0.5\textwidth]{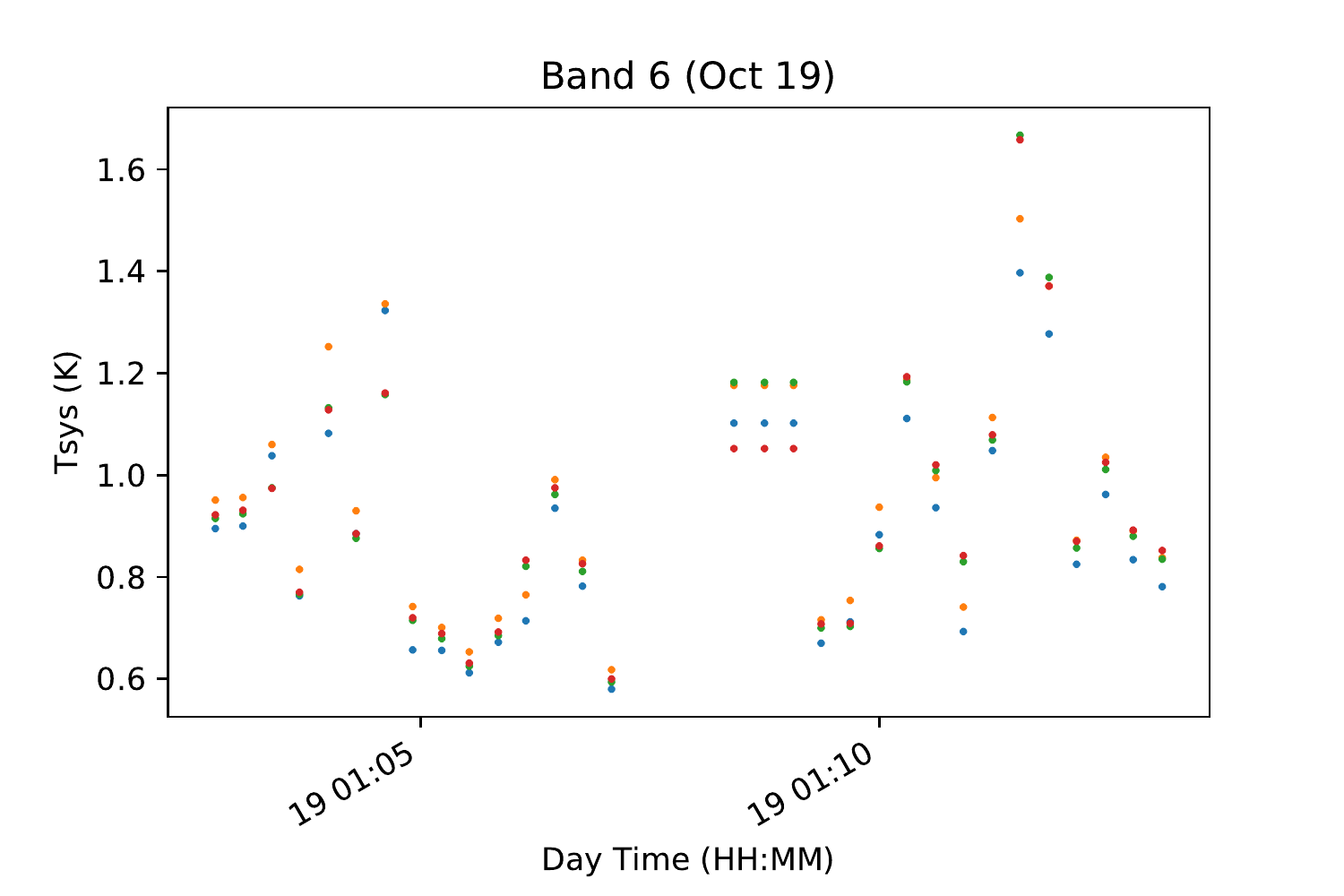}
\caption{
ANTAB \tsys\ values derived as described in Section~\ref{antab} on October 21 (top) and 18/19 (middle) in Band~7 and October 19 in Band~6 (bottom). 
Note the much larger scatter and higher values of the \tsys\ computed on October 18/19 with respect to October 21. 
Colors display different SPWs. 
A DPFU=0.011Jy/K and 0.006 Jy/K  is assumed in Band~7 and 6, respectively. }
\label{fig:antab}
\end{figure}

Using Eq.~\ref{TsysEq} and DPFU $= 0.011$\,K/Jy, one can derive an effective {\it phased-array system temperature} for each scan.
These are shown in Figure~\ref{fig:antab} for 2018 October 18/19 and 21. 
Note that since we set the DPFU of phased-ALMA to the antenna-wise average of DPFUs, there is a factor of $N_{\rm phased}$
that must be absorbed by $T_{\rm sys}$ in order to keep the same amplitude correction.
This explains why the plotted values (varying
in the range $\sim$2-10~K) are much lower than the measured \tsys\  of the individual antennas
in Band 7 ($\sim$100-300~K). 

Note that the specification for the ALMA receiver noise performance is $\sim$80~K in Band~6 and $\sim$150~K in Band~7 \citep{Remijan2019}, very close to  the median \tsys\ values measured on October 19  in Band~6  and on October 21 in Band~7, respectively; the much higher \tsys\ measured on the October 18/19 in Band~7 reflects the rapid changes in the weather conditions on that night (see Table~\ref{table:dpfu}).

\subsection{SEFD}

Using Eq.~\ref{eq:SEFD}, and plugging in  the median of the phased-array \tsys\ values measured on October 21 (2.6~K) and the estimated DPFU (0.011 Jy/K), one  derives SEFD = 238~Jy in Band~7. 
One can also  compare this estimate with the  theoretically expected SEFD$_{\rm th}$ as provided by the ALMA observatory: 
\begin{equation}
 \mathrm{SEFD}_{\rm th}= \frac{\left<T_{\rm sys}\right > \ 2 k_B }{ \eta_A A_{\rm geom}}, 
\end{equation} 
where  \tsys\ is the  opacity-corrected system temperature, $k_B$  is the Boltzmann constant, $\eta_A$ is the aperture efficiency, and $A_{\rm geom}$
is the geometric collecting area. 
For a 12~m ALMA dish in Band~7, $\eta_{A}$=0.63 \citep{Remijan2019}.  
Assuming a phased array of 29 antennas of 12~m diameter and taking a representative value of \tsys= 155~K (corresponding to the mean value of the \tsys measured on October 21), one derives SEFD$_{\rm th}$ = 207~Jy. 
Taking into account the phasing efficiency and defining $ \mathrm{SEFD}=  \mathrm{SEFD}_{\rm th} / \eta_v$, one finds SEFD = 230 Jy for 90\% efficiency, close to  the estimate from the QA2 gain calibration.

A similar analysis on the data acquired on October 18/19  provides SEFD = 150~Jy in Band~6 and SEFD = 578~Jy in Band~7. The latter
is a factor of 2.4 higher than the value estimated on October 21, likely a consequence of the poor observing conditions.

\section{Summary}
\label{summary} 
This paper presents the first test observations of ALMA as a phased array at 345~GHz (Band~7).
These include phased array data acquired during
a series of short ALMA standalone tests and during a global VLBI  test campaign conducted in 2018 October.
We also present a description of the special procedures for calibration and a preliminary analysis of the ALMA observations aimed at
validation of the Band~7  phased-array operations in VLBI mode.

We find that under nominal Band~7 observing conditions at ALMA (PWV$\lsim$2.0~mm; $v_{\rm wind}<$10 m s$^{-1}$), ALMA can perform as a scientifically effective phased array,
with a phasing efficiency $\eta_{v}\gsim$0.5. Typical phasing efficiencies achieved under these conditions are
expected to increase  with future use of fast (WVR-based) phasing corrections in addition
to the TelCal-based phasing corrections used in the present study. 
At 1.3 mm the order-of-magnitude boost in sensitivity of phased ALMA was crucial for enabling the first ever images of SgrA* and M87*. At 345 GHz, ALMA is now poised to provide a comparable leap in capabilities and serve as a vital anchor station for the first VLBI observations at submillimeter wavelengths. 
However, we find that phasing performance in Band~7 diminishes significantly during periods of high winds ($\gsim$12 m s$^{-1}$)
which should thus be avoided, even when PWV is low. 
Under high-wind  conditions, the effective collecting area of the phased array may approach that of a single 12~m antenna.
We also conclude that it is generally advantageous to limit the maximum baseline lengths in the phased array to
a few hundred meters or less
when operating in Band~7.

As will be described in Paper~II, 
the phased ALMA data collected on October 18/19 and 21 led to the detection of the first VLBI fringes in the 345~GHz band between ALMA and other EHT sites. 
The detection of 345~GHz VLBI fringes   represents a crucial first step, not only for commissioning sub-mm VLBI at ALMA, but in general for establishing
that such measurements are both technically and practically feasible.
Pushing the VLBI capability of phased-ALMA to the highest frequencies is crucial
for the success of scientific experiments that require extremely high-angular resolution, 
including studies of black hole physics on event horizon scales 
and accretion and outflow processes around black holes in AGNs.

\acknowledgements
The authors are grateful to Remo Tilanus for valuable comments and suggestions.
We thank the anonymous referee for a thorough and constructive report.
This work was supported by an ALMA Cycle 5
North America Development award and made use of the following ALMA
data: ADS/JAO.ALMA \# 2017.1.00019.CSV, 2018.1.00005.CSV, 2018.1.00006.CSV, 
2018.1.00007.CSV.
This work was partially supported by FAPESP (Fundação de Amparo à Pesquisa do Estado de São Paulo) under grant 2021/01183-8 and by the Generalitat Valenciana GenT Project CIDEGENT/2018/021 and the MICINN Project PID2019-108995GB-C22.
The National Radio Astronomy Observatory is a facility of the National Science Foundation operated under cooperative agreement by Associated Universities, Inc.

\appendix
\section{Calibration of the 2018 VLBI Campaign Data}
\label{app:calibration}

In order to  turn the phased ALMA array into a single mm or sub-mm VLBI station and combine it  with other VLBI stations (e.g. Paper II), 
it is necessary to first calibrate the interferometric visibilities \citep{QA2Paper}. Although the data sets described in this
paper did not include the full suite of calibration measurements that would usually be part of an ALMA VLBI science observation,
we were nonetheless able to perform some basic calibrations of the data sets from the 2018 October global Band~7 VLBI test campaign
(Table~2) using a modified version of the procedure described in \citet{QA2Paper}. The steps are outlined here.

In brief, we have calibrated the ALMA data using the
Common Astronomy Software Applications ({\sc casa}) package \citep{CASA2022} and the special  procedures 
known as ``Quality Assurance Level 2'' (QA2)  described in \citet{QA2Paper}. 
Normally  VLBI observations  are arranged and calibrated in ``tracks'', where one track consists of the observations taken during the same day or session.
Since the observations carried out during each night were insufficient to allow for their independent calibration, we concatenated
the data collected in 2018 October into a single dataset.  
In addition, since the poor weather conditions made the data collected on days 16 and 17 unusable (see Section~\ref{aps_perform2018}),  we flagged them from the concatenated dataset. 
We have therefore based our analysis on  the  data collected on days 18/19 and 21;  the latter yielded the highest-quality data and were 
used to compute the bandpass and polarization calibration tables (see Appendix~\ref{app:validation}). 
We then applied those calibration tables to the full dataset, under the
assumption  that the observed sources are stable over a time interval of 2.5 days. 

\subsection{Absolute Flux-density Scale}
\label{app:fluxscale}

ALMA normally tracks the atmospheric opacity by measuring system temperatures (\tsys) at each antenna. However,
\tsys\  values are  not used in the phased-array data calibration to avoid biasing the calibration of the phased sum antenna  \citep[see][]{QA2Paper}. 
While the bulk of the opacity effects are removed with self-calibration,  $T_{\rm sys}$ (usually measured a few times per hour) can still be used to correct for {\em second-order} opacity effects, 
related to the difference between the opacity correction in the observation of the primary flux calibrator and the (average)
opacity in the observation of any given source.

In \citet{QA2Paper}   the opacity-corrected flux density $S^S_{\tau}$ for a given source $S$
was estimated post-QA2 calibration by computing the products of the  antenna-wise average of the amplitude gains, $g_a^S(t)$,
times the antenna-wise average of valid $T_{\rm sys}$ measurements, $T^S_{\rm sys}(t)$, and then taking the ratio  between
a given source {\it S} and the  primary flux-density calibrator  {\it P} such as:

\begin{equation}
S^S_{\tau} = \left( \frac{\left< g_a^S(t) T^S_{\rm sys}(t) \right>}{\left< g_a^{P}(t) T^P_{\rm sys}(t) \right>}      \right)^2 S^S_{QA2}
\label{TauFluxEq}
\end{equation}

\noindent
Here we have followed a similar approach, except that
the opacity correction is applied to the data right after the gain calibration, yielding opacity-corrected amplitude gains
(contained in the \texttt{<label>.flux\_inf.APP.OpCorr} table in the ALMA archive). 

The absolute flux-density scale was  derived from self-calibration  on  3C454.3. 
This quasar was chosen as the primary flux-density calibrator because it is  the only Grid Source observed in both Bands~6 and 7, allowing
calibration of the amplitude scale in both bands.   
For this purpose, $S^P_{343 GHz} = 3.53$~Jy and $\alpha$=--0.69 was assumed. 
See Section~\ref{app:fluxcomp} for an assessment  of the accuracy of the absolute flux-density scale. 

\subsection{Polarization Calibration}
\label{app:polcal}
Obtaining accurately calibrated VLBI  science data from ALMA requires a full polarization calibration
of the interferometric visibilities to allow  conversion of ALMA's linearly polarized data
products to a circular polarization basis, for consistency with
other VLBI stations \citep{PCPaper,APPPaper,QA2Paper}. 
Observations of a polarization calibrator with sufficient parallactic angle coverage
are required to simultaneously derive a reliable model of the polarization calibrator and an estimate of the XY cross-phase at the reference antenna. 
The latter is computed by running the \texttt{CASA} task \texttt{gaincal} in mode \texttt{XYf+QU} 
 \citep[see \S5.2.3 of][]{QA2Paper}. 
 
 Since the 2018 October campaign was conceived as a VLBI ``fringe test", it was not designed to obtain fully calibrated science data from
 ALMA and, consequently, no frequent observations of a  suitable polarization calibrator were obtained. 
Instead, only a few scans were obtained towards two sources with a high linear polarization fraction ($\gtrsim$5\%), J0522--3627 and J0510+1800 (see Table~\ref{table:sources}). 
 This precluded deriving a reliable polarization model from the observations using \texttt{gaincal}. 
 Fortunately, both sources were observed with the ACA in Band~7 on October 10 and October 21/22  and we could  estimate their Stokes
 parameters using the AMAPOLA\footnote{\url{http://www.alma.cl/~skameno/AMAPOLA/}} polarimetric Grid Sources. 
 We  set J0510+1800 as the polarization calibrator and its  Stokes parameters as
 {\it IQUV} = [1.28, --0.134, 0.079, 0.0]~Jy
 (see Appendix~\ref{app:polcal} for an assessment of the goodness of the polarization model). 
We then ran the  \texttt{CASA} task \texttt{polcal} in mode \texttt{Xf}
to estimate the XY phase at the reference antenna.

\begin{figure*}
\hspace{-5mm}
\includegraphics[width=0.5\textwidth]{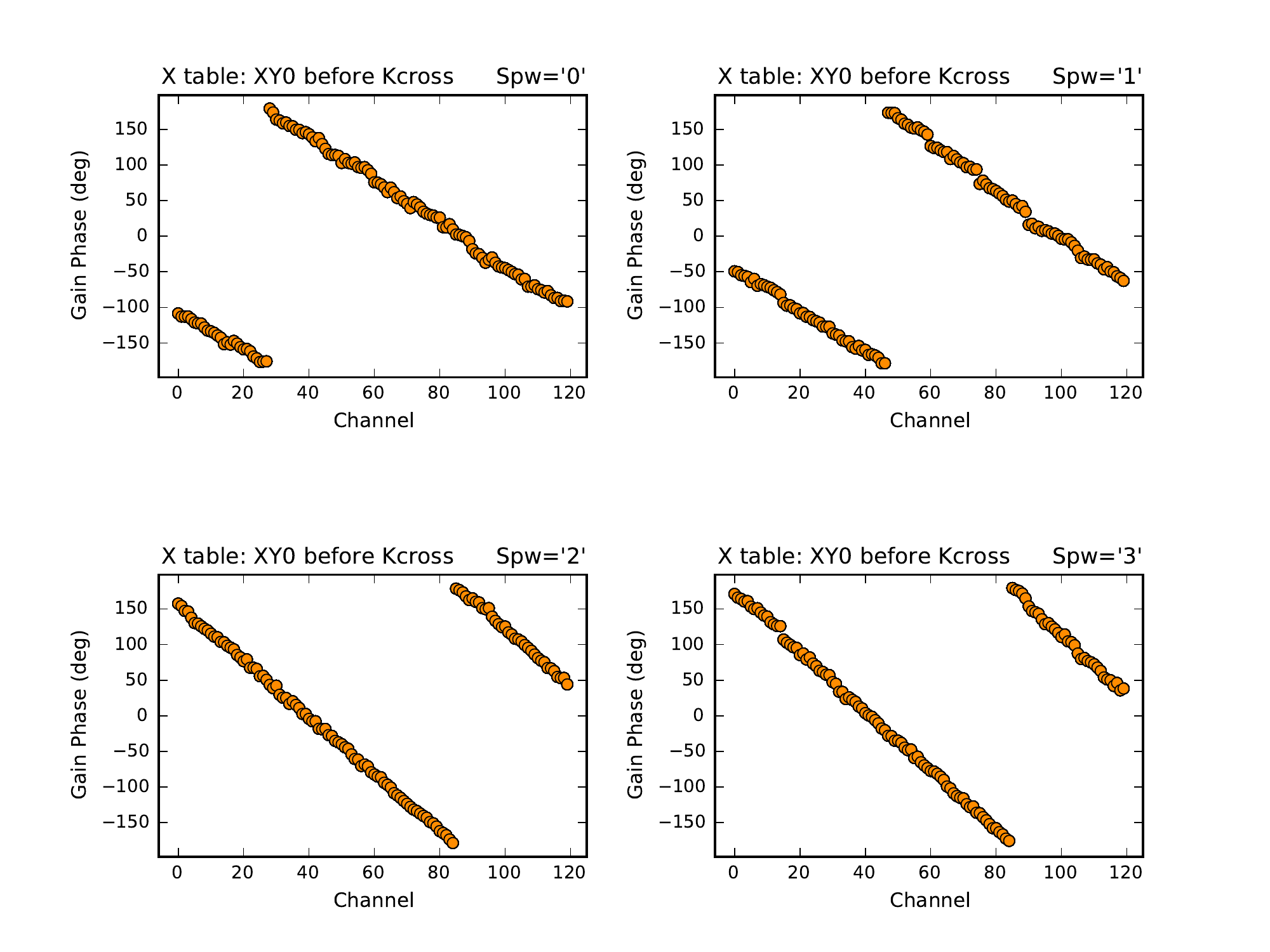} 
\includegraphics[width=0.5\textwidth]{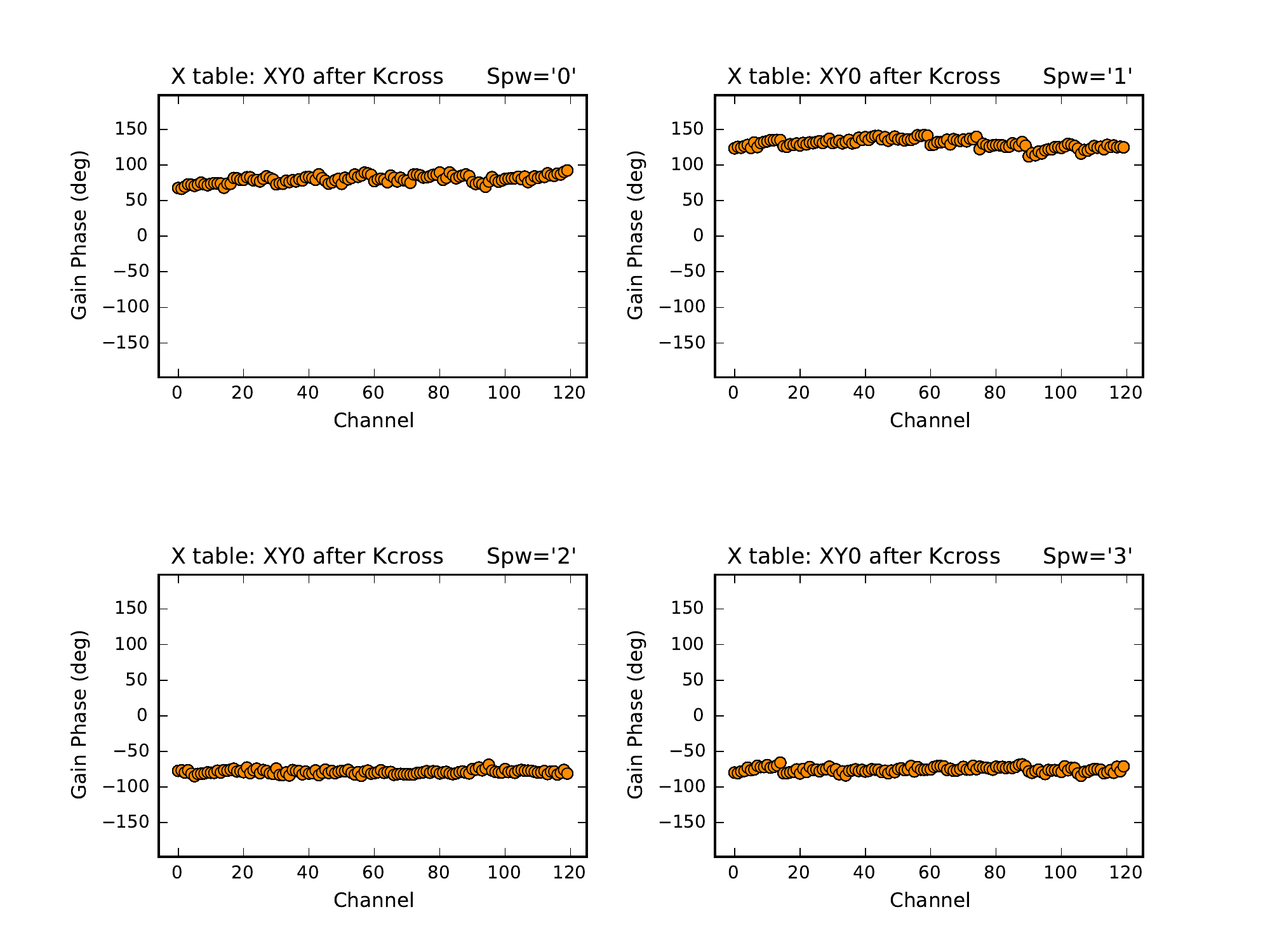} 
\caption{X-Y cross-phase of the reference antenna  in APS mode  for each of the four SPWs 
before (left)  and after (right) removing the cross-polarization phase delay.
}
\label{fig:crossphase}
\end{figure*}

The computed cross-hand phases  as a function of frequency are plotted in Figure~\ref{fig:crossphase}.
The four left panels  reveal a steep phase slope with frequency, which indicates the presence of a significant cross-hand delay
($>$0.5 nanoseconds or a full wrap within the $\sim$2~GHz SPW) intrinsic to  the reference antenna.

This can be a consequence of two factors. 
First, the gain and bandpass calibration  only correct for parallel-hand delay residuals (the two polarizations are referenced independently),
thus leaving a single cross-hand delay residual from the reference antenna.  
Second,  the online system does not apply the static baseband delay corrections when the APS is active \citep{APPPaper}.
In general,
the phasing corrections applied to contiguous frequency chunks are expected to remove most
of the (generally large) baseband delays. (This was previously demonstrated in science observations carried out in
Band~3 and Band~6; see \citealt[][]{QA2Paper}). Despite this,
a significant residual delay  in the XY cross-phases may still be present,
depending on the nature of the observations, as demonstrated in this case. 
  This cross-hand delay can be estimated with \texttt{gaincal} using \texttt{gaintype=`KCROSS'}.
We therefore included this additional correction with respect to the procedures outlined in Section~5 of \citet[][]{QA2Paper}. 
Figure~\ref{fig:crossphase} (right panels) shows plots of cross-hand phases after removing the cross-hand delay,
which indeed reveal a flat slope with frequency.

We explicitly note that the approach to polarization calibration outlined above could be useful in future cases where no suitable polarization calibration is observed,
or only short VLBI observations
can be obtained because of weather or scheduling constraints.

\section{Validation of the APS  Calibration}
\label{app:validation}

The  QA2 data calibration presented in Appendix~\ref{app:calibration}  provides a full set of calibration tables 
 in \texttt{Measurement Set} format (i.e. readable in \texttt{CASA}): 

 \begin{itemize}

\item \texttt{<label>.CSV.phase\_int.APP.XYsmooth}:  phase gains (per integration time).

\item \texttt{<label>.CSV.flux\_inf.APP.OpCorr}:   amplitude gains scaled to Jy units (per scan).

\item \texttt{<label>.CSV.bandpass\_zphs}:   bandpass (with zeroed phases).

\item \texttt{<label>.CSV.XY0.APP}: 
cross-polarization phase at the TelCal phasing reference antenna.

\item \texttt{<label>.CSV.Kcrs.APP}: 
cross-polarization phase delay at the TelCal phasing reference antenna (see Appendix~\ref{app:polcal}).

\item \texttt{<label>.CSV.Gxyamp.APP}: 
amplitude cross-polarization ratios for all antennas. 

\item \texttt{<label>.CSV.Df0.APP}: 
 D-terms at all antennas. 

\end{itemize}

 These tables can be used to fully calibrate the ALMA interferometric visibilities, which in turn can be used to do science analysis,
 e.g. deriving the mm emission properties of the VLBI targets such as integrated fluxes (see Section~\ref{app:fluxcomp}), or imaging them on arcsecond-scales (see Section~\ref{imaging}).
The same tables  can also be processed  to calibrate phased-ALMA as a single VLBI station (see Paper~II). 

\section{Weather Metrics}
\label{app:weather} 

\begin{figure*}[htbp]
\center{
\includegraphics[width=0.4\textwidth,angle=0]{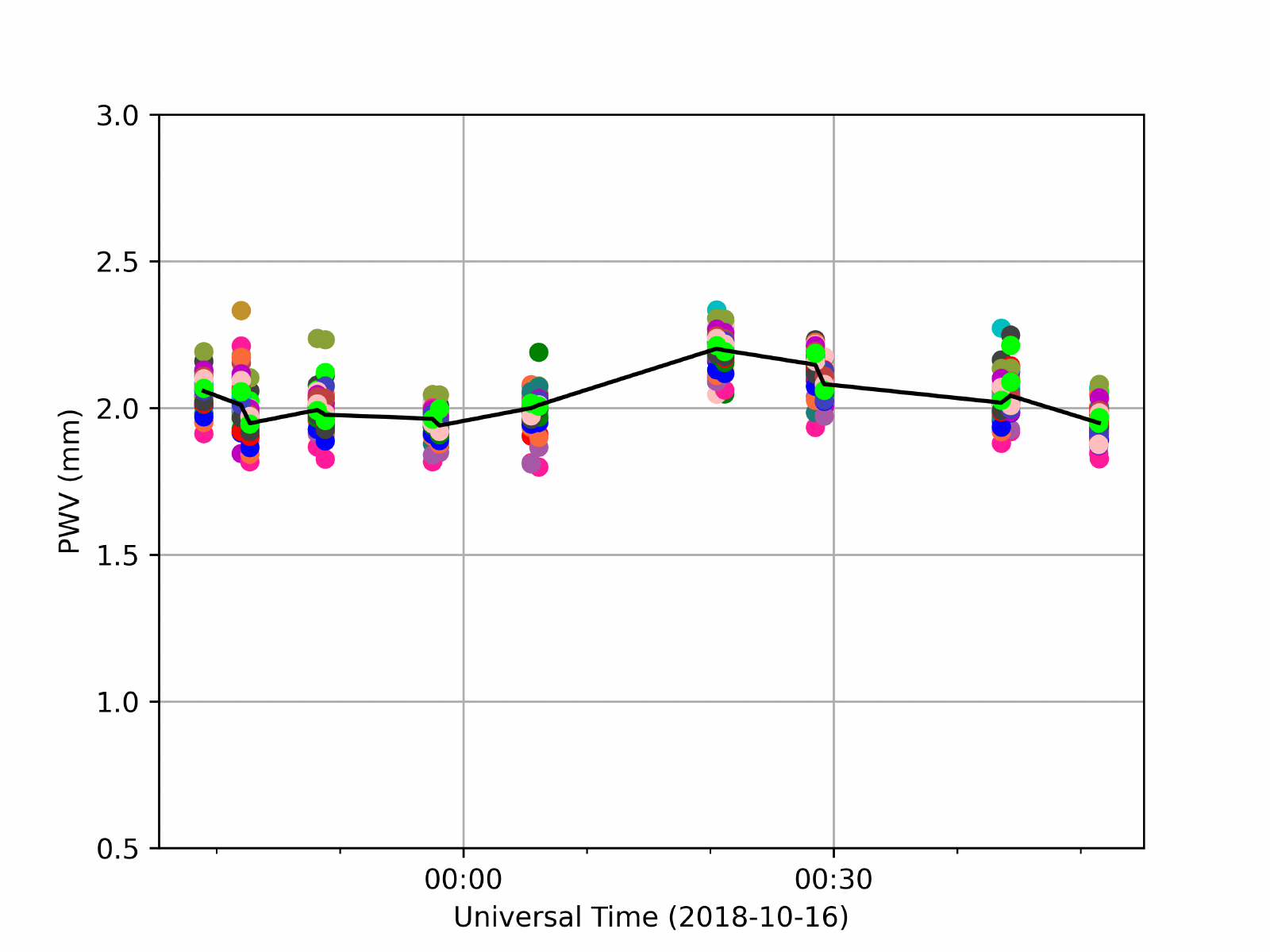}
\includegraphics[width=0.4\textwidth,angle=0]{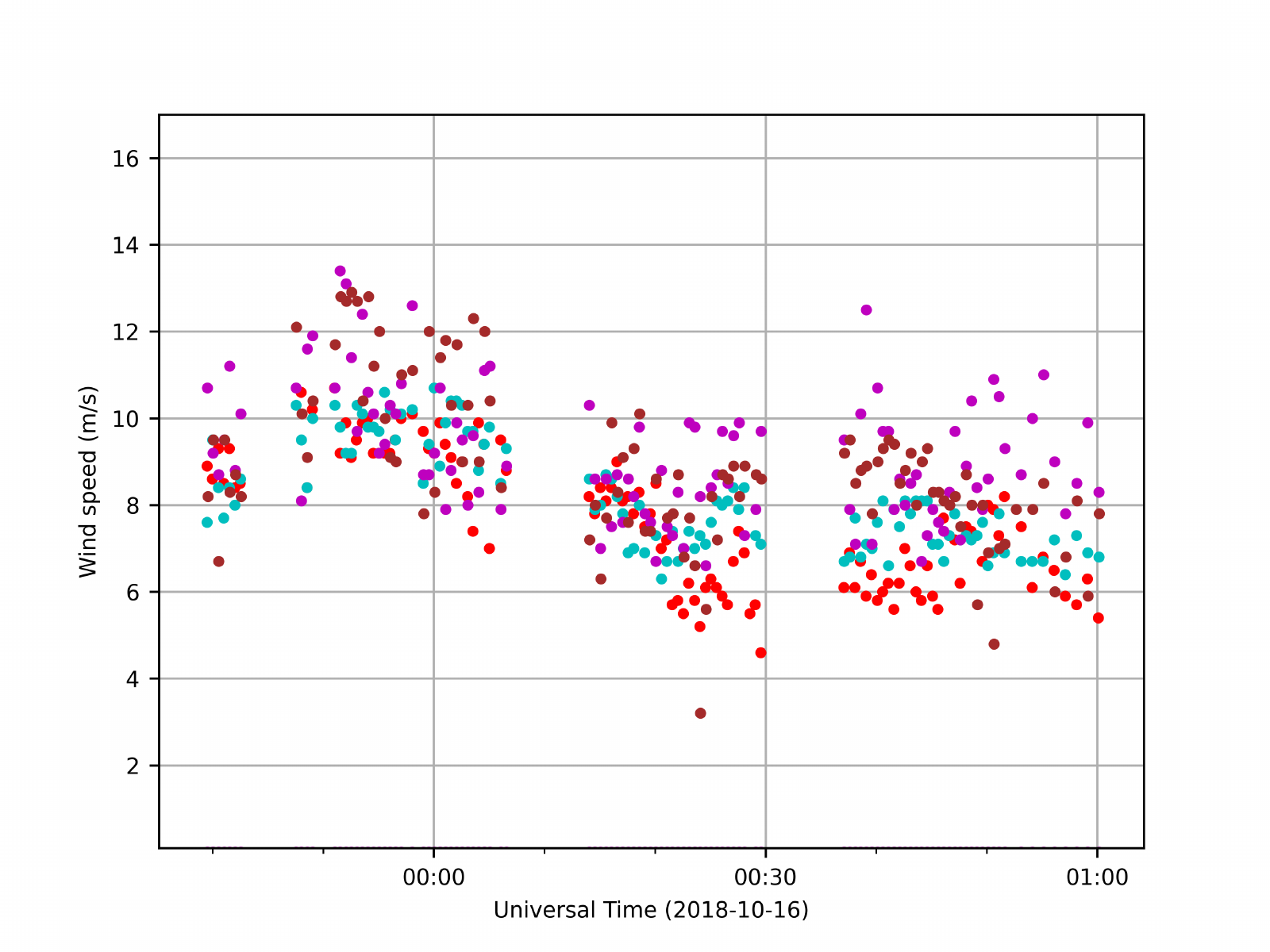}
\includegraphics[width=0.4\textwidth,angle=0]{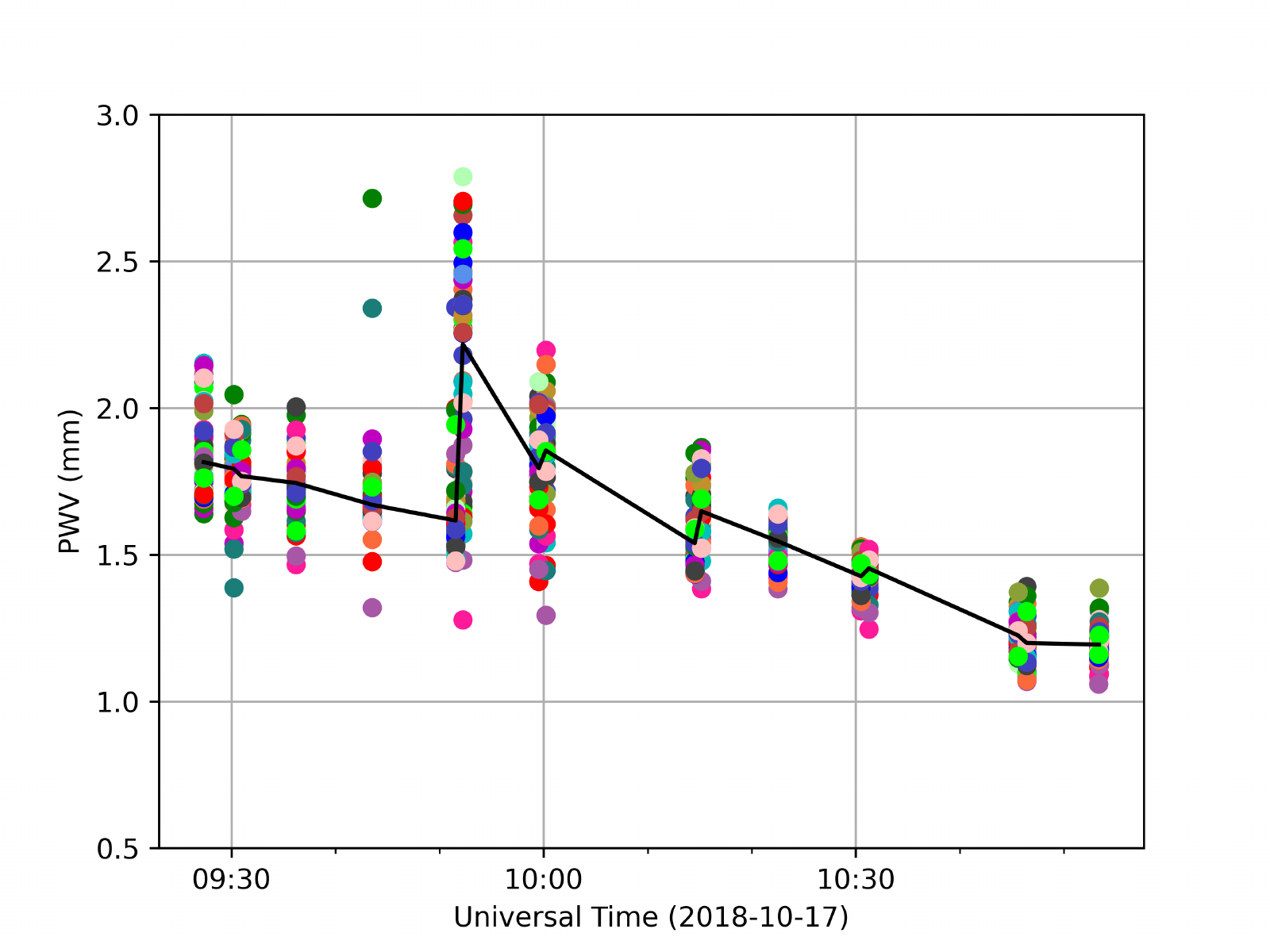}
\includegraphics[width=0.4\textwidth,angle=0]{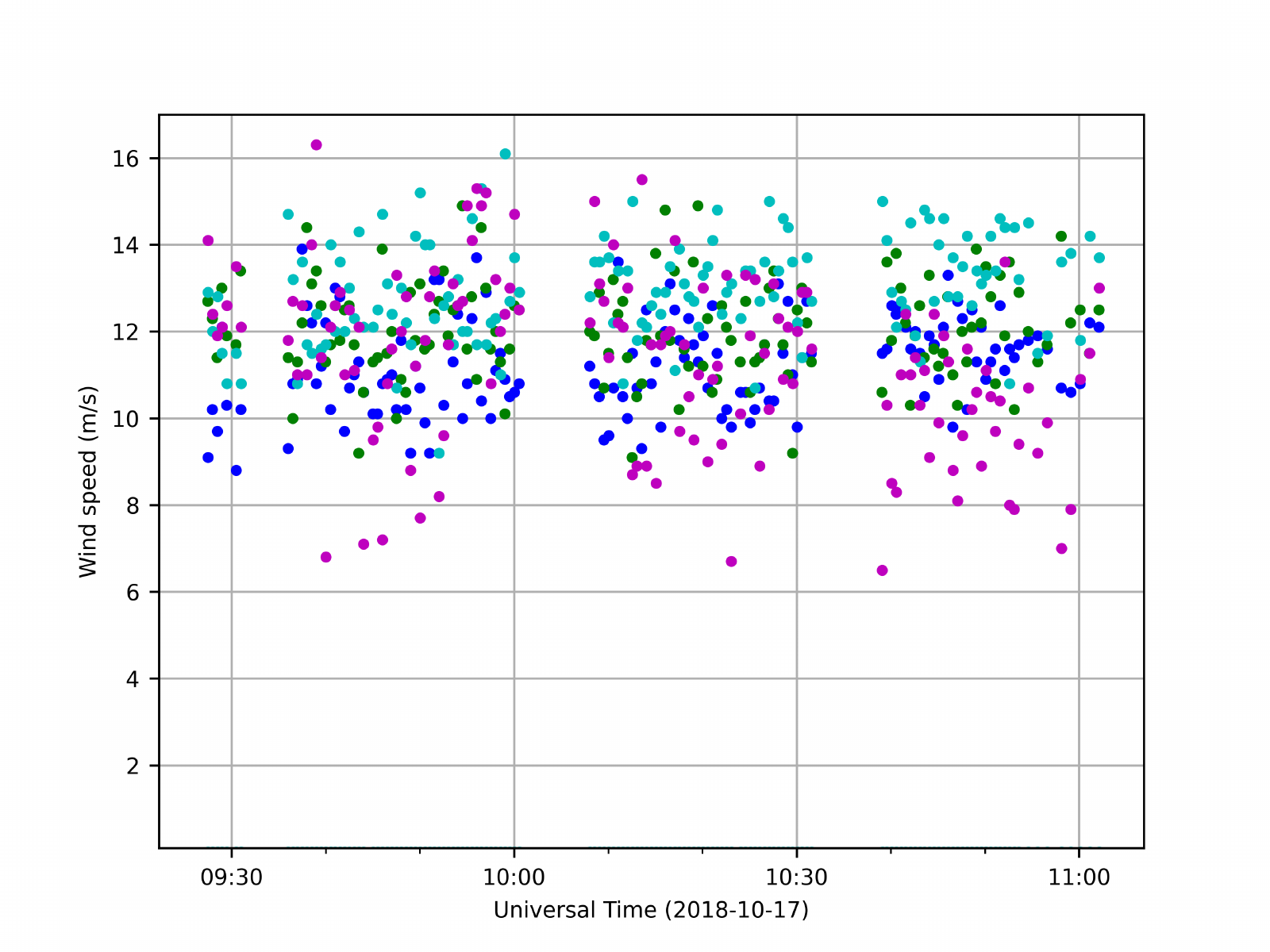}
\includegraphics[width=0.4\textwidth,angle=0]{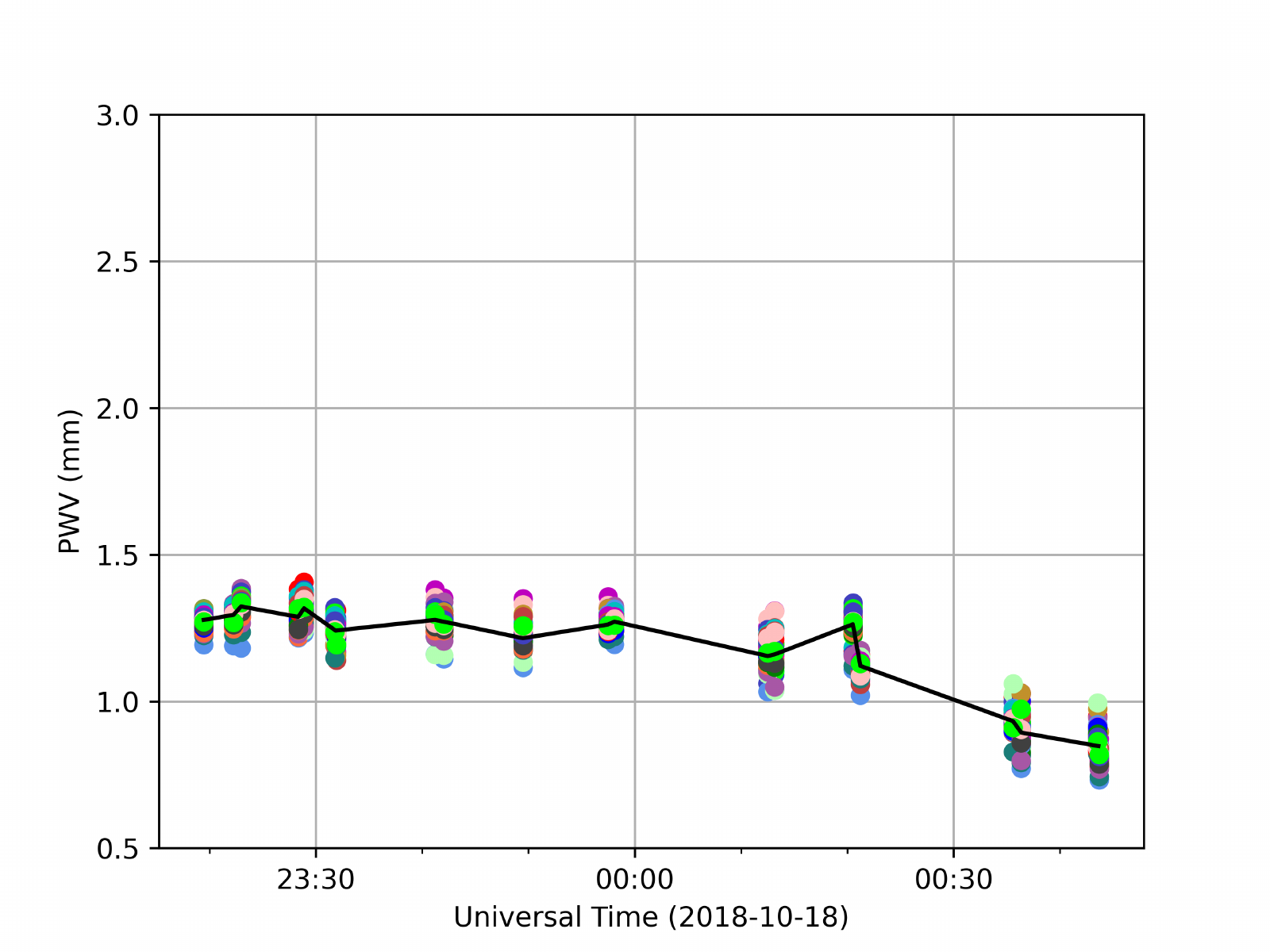}
\includegraphics[width=0.4\textwidth,angle=0]{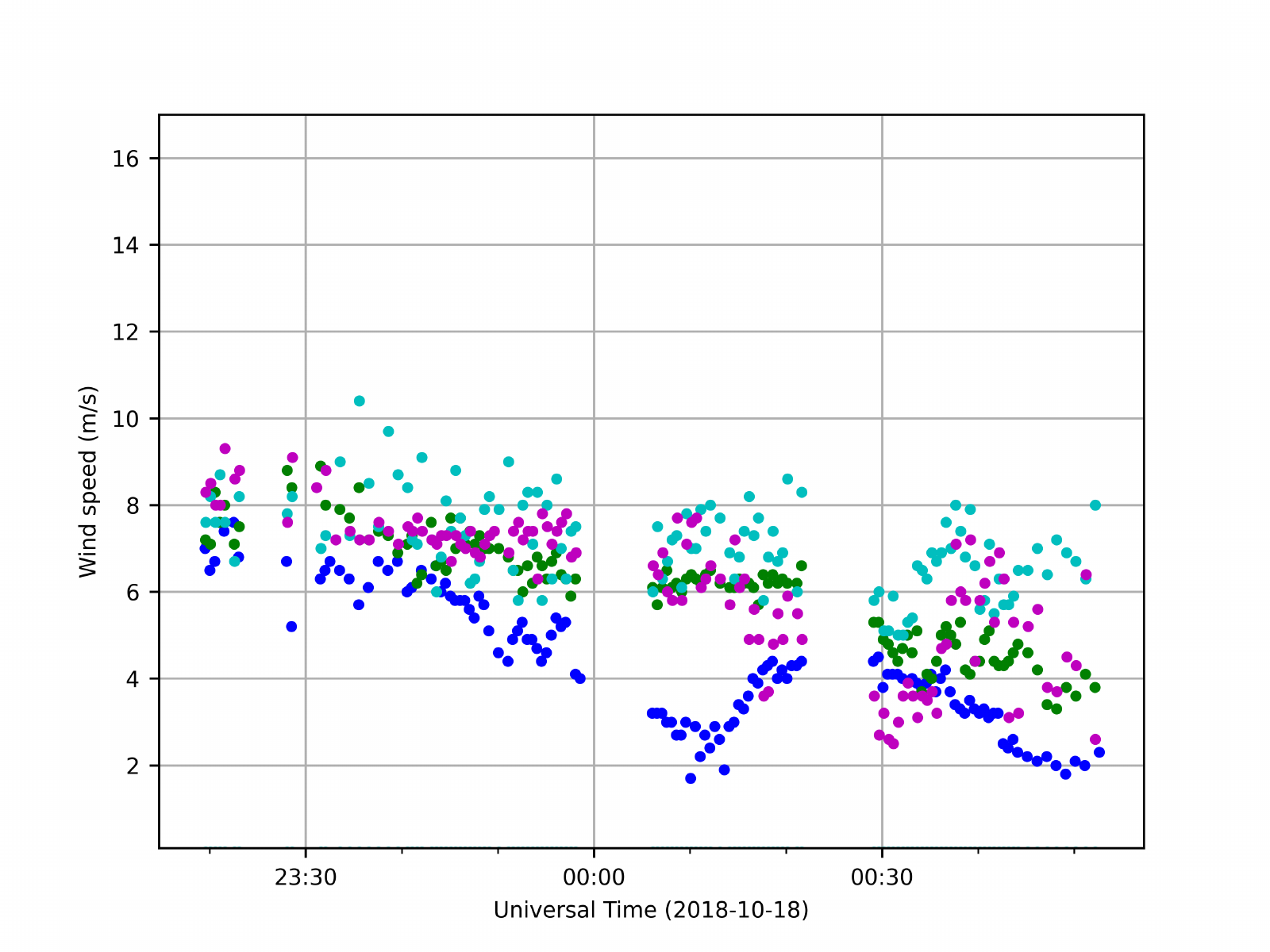}
\includegraphics[width=0.4\textwidth,angle=0]{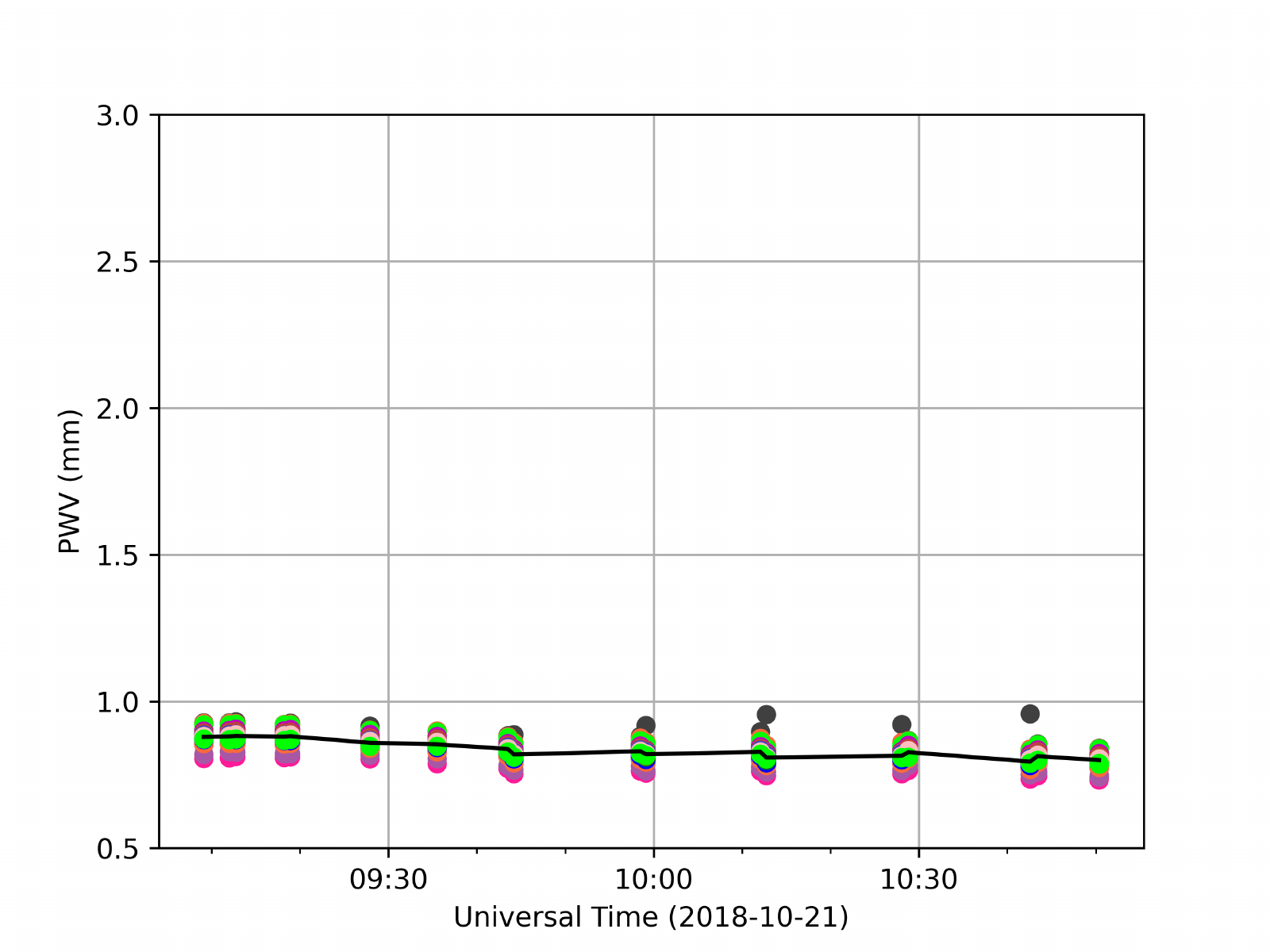}
\includegraphics[width=0.4\textwidth,angle=0]{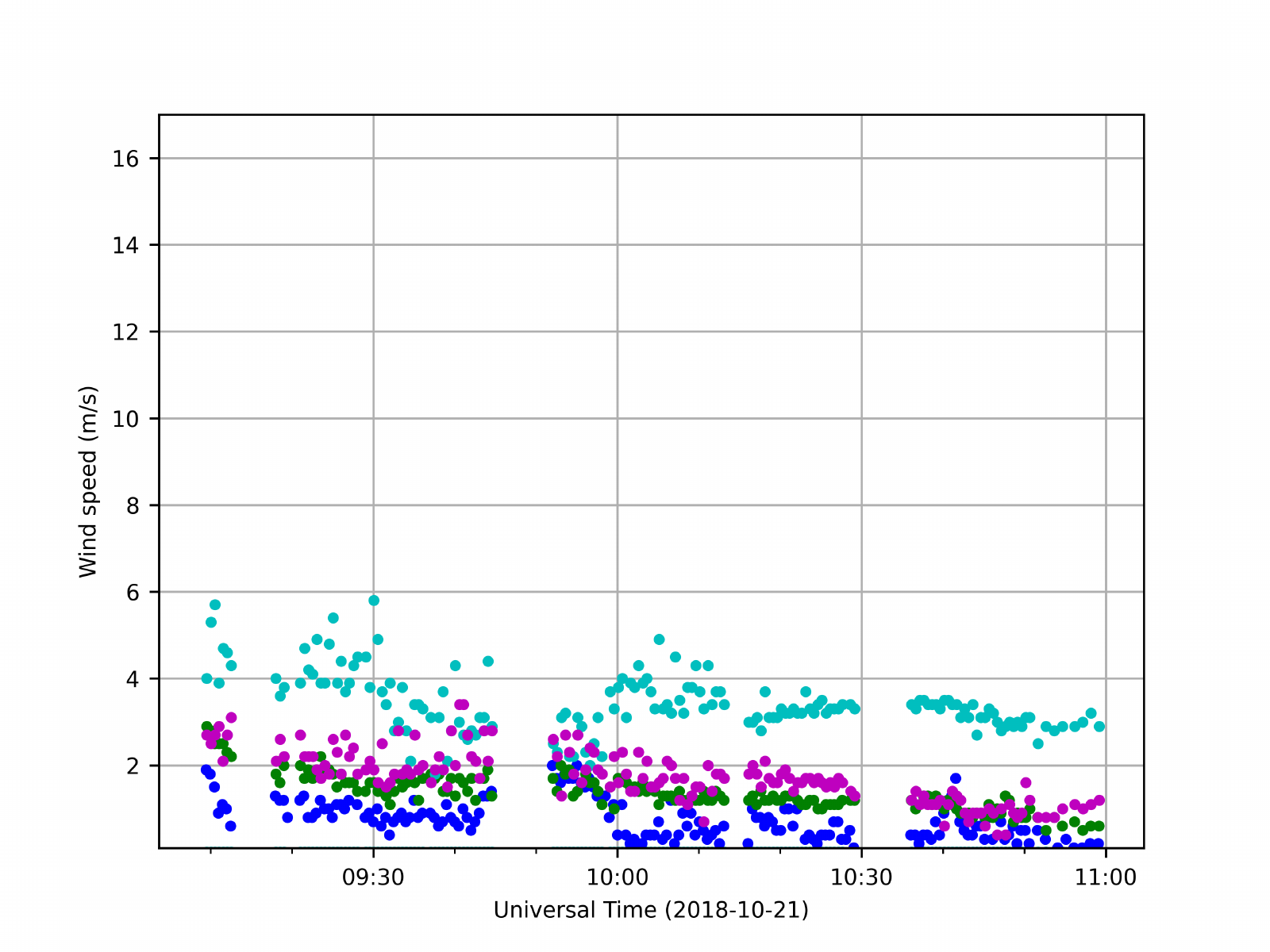}
}
\caption{\footnotesize 
Illustration of the weather conditions during the  345~GHz (Band~7) VLBI test experiment on 2018 October 16/17, 17, 18/19, 21 (from top to bottom). 
Left panels: PWV column in mm as a function of observing time. Colors display different antennas. The black line is the median PWV value computed across all antennas in the array. 
Right panels: wind speed  in m~s$^{-1}$ as a function of observing time. 
Colors indicate data from different weather monitoring stations. 
}
\label{fig:weather}
\vspace{-0.2in}
\end{figure*}

Figure~\ref{fig:weather} illustrates the weather conditions during  the  345~GHz (Band~7) VLBI experiment on 2018 October.
The left and right panels show PWV column in mm and the wind speed  in m~s$^{-1}$, respectively, as a function of observing time.  
Different rows show different days, from October 16 to October 21 (top to bottom). 

At the beginning of the campaign (on the night of October 16/17 and the morning of October 17, respectively),
the conditions on the Chajnantor Plateau were extremely unstable, with moderately high 
 and variable  PWV ($\sim 2\pm0.3$ and $1.8\pm0.8$~mm)  and high wind speeds 
 ($\sim 9\pm5$~m s$^{-1}$ and $\sim 12\pm4$~m s$^{-1}$) compared
with typical conditions at ALMA for Band~7 observing.
Atmospheric stability was significantly improved on the night of the 18/19 
(PWV$\sim 1.1\pm0.3$~mm) but  wind speed was still significant ($\sim 6\pm4$~m s$^{-1}$). 
On the last day of observing (October 21) weather at ALMA was excellent, with  PWV$\sim 0.8\pm 0.1$~mm and wind speed of $\sim 3\pm3$~m s$^{-1}$. 

\section{Assessment of the Polarization Calibration model used for the October 2018 VLBI test}
\label{app:polcal}

To evaluate the goodness of the  AMAPOLA model  for J0510+1800, 
we ran the  \texttt{CASA} task \texttt{polcal} with different input models. 
The latter  included deviations in either
the linearly polarized flux, $P= \sqrt{Q^2+U^2}$,  or polarization position angle, $2\chi=\mathrm{arctan}(U/Q)$,
with respect to the AMAPOLA model (with $P_{\rm amapola}=0.16$~Jy and $\chi_{\rm amapola}=75^{\circ}$). 
 In particular, we considered the following 17 alternative  models: 
 $P = \Delta P \times  P_{\rm amapola}$ where $\Delta P =$ [0.25, 0.50,0.75,1.25,1.50,1.75,2.00]  (these models kept the same $U/Q$ ratio) 
 and
  $\chi =  \chi_{\rm amapola} \pm \Delta {\chi}$ where $\Delta {\chi} = \pm[10,20,30,40,50]$deg. 

\begin{figure*}
\hspace{-5mm}
\includegraphics[width=0.5\textwidth]{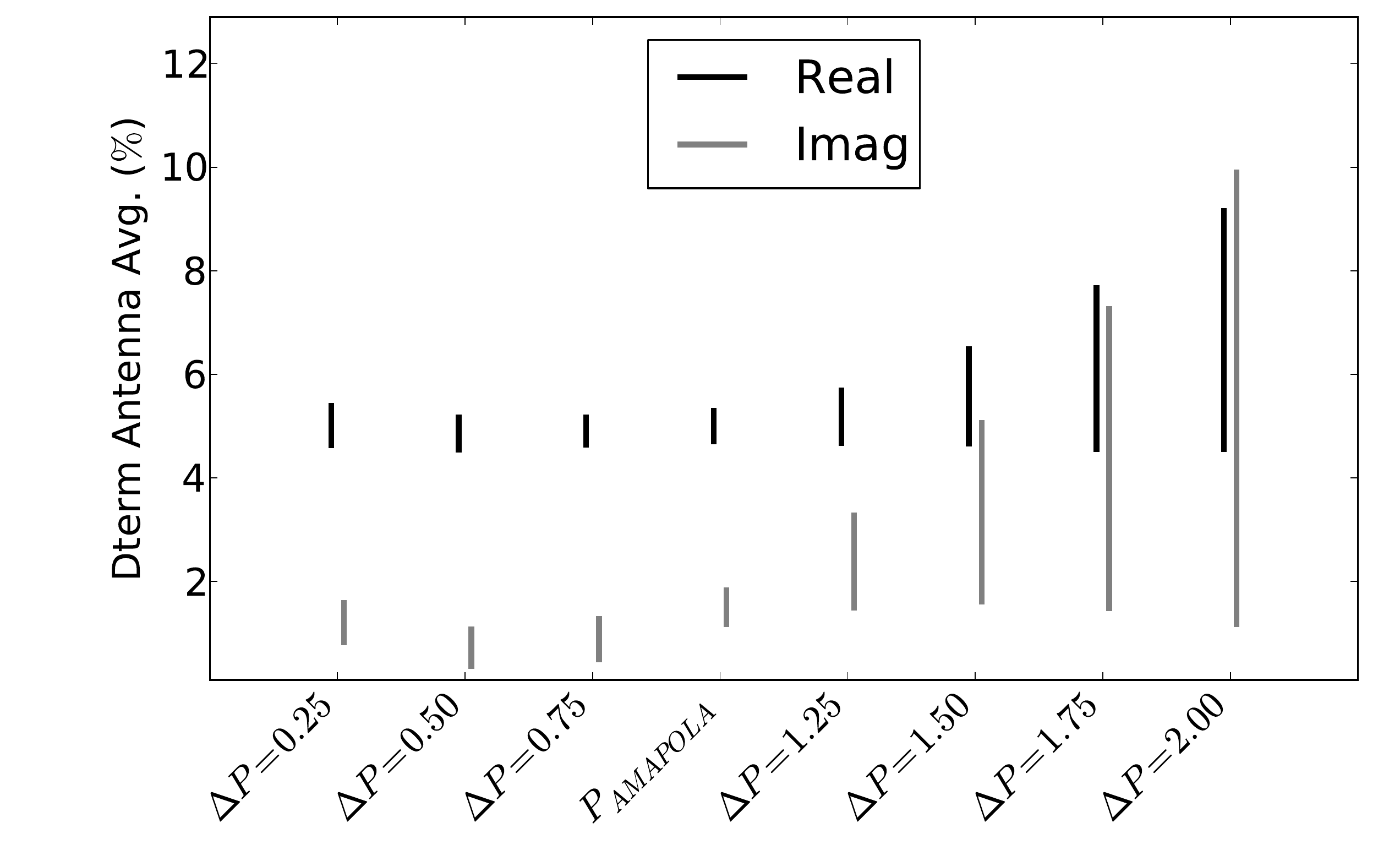} 
\includegraphics[width=0.5\textwidth]{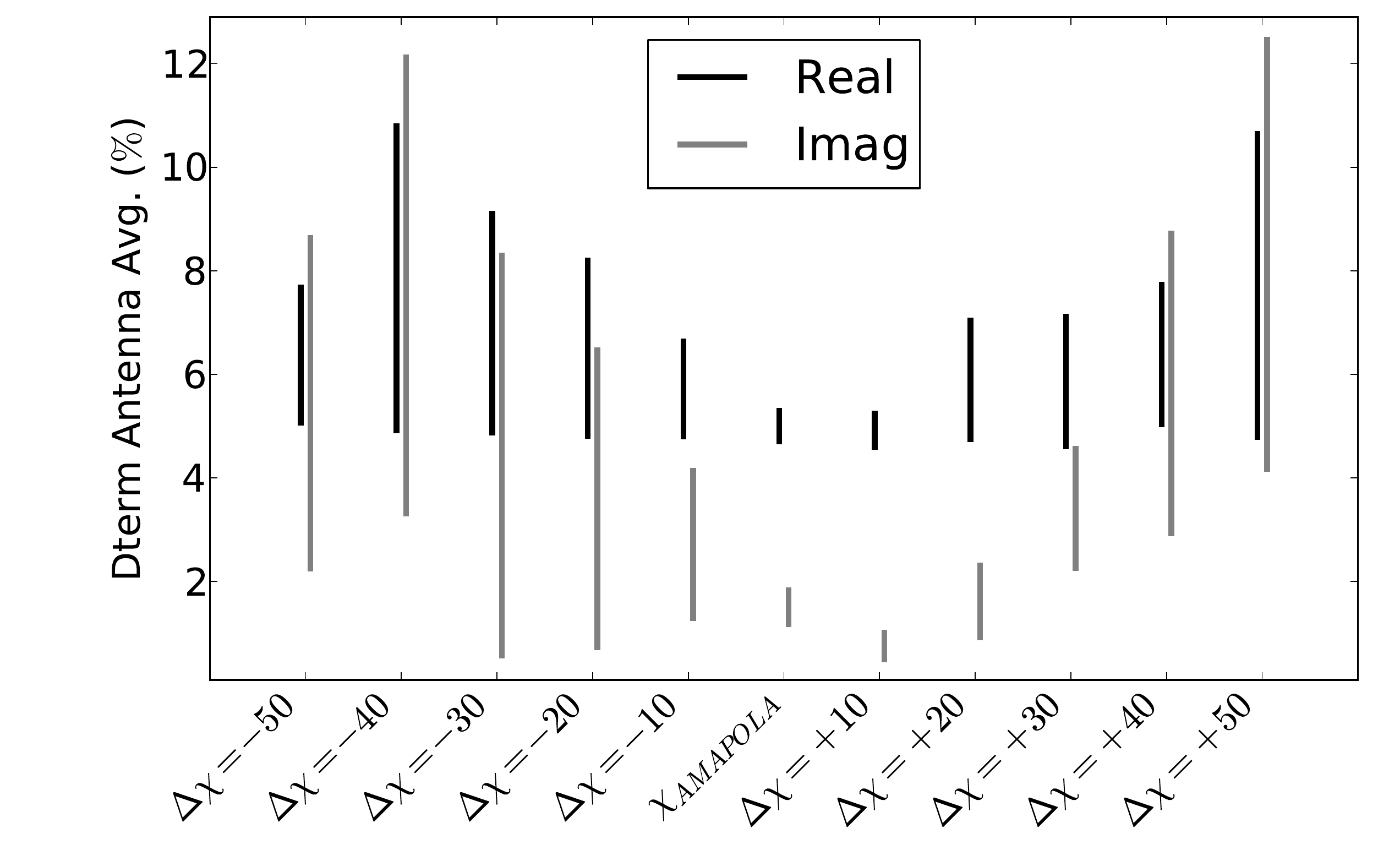} 
\caption{ 
Left panel: Impact of the deviations from the J0510+1800 AMAPOLA model on the D-terms of the phased-antennas. Models including deviations in the polarized flux: 
$P = \Delta P \times  P_{\rm amapola}$ (where $P_{\rm amapola}=0.16$~Jy). 
Right panel: Models including deviations in the  polarization position angle: 
 $\chi =  \chi_{\rm amapola} \pm \Delta {\chi}$ (where $\chi_{\rm amapola}=75^{\circ}$).
}
\label{fig:polcalmodel}
\end{figure*}

 We then assessed the impact of different models on the D-terms of the phased-antennas.
To this end, we  computed  
 the D-terms ``channel dispersion", a parameter defined by the range (minimum to maximum) of the average of the D-terms (absolute) values computed across all antennas for each channel. 
Figure~\ref{fig:polcalmodel} summarizes the values of this parameter for the class of models presented above. 
 We found that deviations from the AMAPOLA model always resulted in a deterioration of the D-terms of the phased-antennas, 
with a monotonically increase in the D-term dispersions (typically a factor of few higher than the original model).
This simple analysis provides an a-posteriori validation of the  polarization model assumed to perform the interferometric data calibration.

\section{Concerning Phasing Efficiency}
\label{app:efficiency}
Here we provide some remarks on the characterization of phasing efficiency, which
is used as a key metric in the current paper for evaluating the APS performance 
in Band~7.

As described in \cite{APPPaper},
for an idealized phased array,
a phasing efficiency, $\eta_{p}$,  can be defined
as a function of the cross-correlation between the summed signal and that of a comparison antenna, 
divided by the averaged cross-correlations between the comparison antenna and the individual phased elements: 

\begin{equation}
\eta_p \equiv \frac{\left<V_{\rm sum}V_c\right>}{\left< V_i V_c\right>} \frac{1}{\sqrt{N_{\rm phased}}}.
\label{eqn:etap}
\end{equation}

\noindent Here  $N_{\rm phased}$ is the number of phased antennas.

In principle, monitoring $\eta_{p}$ during an observation can provide 
a fundamental figure-of-merit to characterize the performance of the
phasing system.
To meet the need for a readily computable efficiency metric, the APS was designed to include the ability to
 substitute one antenna
input to the ALMA Baseline Correlator with a copy of the phased-sum signal.
The correlator then produces visibilities on baselines to this ``sum
antenna'', allowing it to be analyzed by the online system in a manner analogous to
the real antennas comprising the ALMA array.
This system was modeled after the 
PhRInGES software previously
used at the Submillimeter Array (SMA; J. Weintroub, private communication).\footnote{%
PhRInGES was replaced starting in 2015 by the SMA Wideband
Astronomical ROACH2 Machine (SWARM, \cite{SWARM}).} 
Figure~\ref{fig:amp+pha_vs_time}, discussed in the main text,
shows examples of plots of phase and correlated amplitude, respectively, as a function of
time for baselines to this sum antenna.
As described in the main text, such plots
allow useful qualitative assessment of the phasing performance.

By definition, 
a perfectly phased array will have
$\eta_{p}=1$ and deviations from this ideal can then be used to characterize efficiency losses.
In a real-world system, such efficiency losses occur from
multiple effects, including imperfections in the phasing solutions (e.g., due to rapid atmospheric
phase fluctuations and/or time lags in the application of the phasing solutions), 
non-optimized antenna weighting in the phased sum, and
inherent losses due to quantization effects and other properties of signal chain and hardware \citep{ALMAPHTR}.

As discussed in \cite{APPPaper}, under optimal weather conditions the end-to-end efficiency of the APS 
as defined by Eq.~\ref{eqn:etap} has been found to be $\sim$60\% of an ideal system  in Bands~3 and 6 ($\eta_{p}\sim$0.6). Portions of these efficiency losses result from
known effects, including the two-bit quantization of the phased sum signal, residual delay errors, and the neglect of
antenna-based weighting factors in computing the phased sum. However, approximately 20\% of this efficiency loss remained
unaccounted for. Subsequent investigation has now uncovered the source of this additional loss;
specifically, because
the logic that computes the
sum signal in the APS is significantly slower than originally designed, it arrives with a
24-lag delay (192~ns). In a 128 lag design, this results in a  loss factor of
[$(128 - 24)/128 = 0.8125$], i.e., $\sim$20\% in correlated amplitude. 

Because certain efficiency losses in any phasing system, including the aforementioned 24-lag delay, are essentially
fixed and unvarying, it is useful to decouple these
from losses in efficiency that result from time-varying phenomena (e.g., weather; baseline length, etc.), as the latter may be used to inform real-time adjustments to
the observing strategy and/or phased array parameters (e.g., changes in 
selection of antennas within the phased array or termination of a phased array observation because of poor phasing stability).
This need has led us to define and compute (in ALMA's TelCal software) an alternate phasing efficiency metric, $\eta_{v}$:
\begin{equation}
\label{eqn:sumcosines}
    \eta_v ~ = ~{ { \sum_{ij}{ v_{ij} } }
      \over { \sum_{ij}{ | v_{ij} | } } }
    \label{eqn:etav}
\end{equation}
Here $v_{ij}$ is the visibility on the baseline between
antennas $i$ and $j$ of the $N$ phased array antennas. This manner of defining phasing efficiency is similar to the approach 
adopted in the SMA SWARM correlator \citep{SWARMASS}. During observations,
$\eta_{v}$ provides a useful quantitative, real-time metric
of phasing efficiency. Because this definition is not computed based on the phased sum signal, it is free from
losses due to quantization and the 24-lag delay.  We therefore quote $\eta_{v}$ as a metric of phasing efficiency  throughout the discussions
of phasing performance in the main text. In contrast, we limit the use of
the phasing efficiency, $\eta_{p}$ defined by Eq.~\ref{eqn:etap} (computed using baselines to the APS
phased sum ``antenna'') only for {\it qualitative}
evaluation (e.g., Figure~\ref{fig:amp+pha_vs_time} of the main text).  We note that the original design requirements of the APS specified
that under band-appropriate observing conditions, the system should routinely achieve
$\eta_{v}\ge$0.9 in Bands 3 and 6 \citep{APPPaper}. However, in practice, under  excellent weather conditions we find
$\eta_{v}\rightarrow$1, even in Band~7.

During VLBI observations,
the calculation in Eq.~\ref{eqn:etav} is repeated during the polarization conversion ({\sc{PolConvert}})
process that converts ALMA's linearly polarized data products to a circular basis (see Section~7 in \citealt{APPPaper}).
This allows recovery of the correct amplitude scaling of the VLBI
products for use in the ANTAB tables \citep{QA2Paper}.

%
%


\
\bibliography{app2_band7_ms}{}
\bibliographystyle{aasjournal}

\end{document}